\documentclass[12pt]{article}
\usepackage{amsmath}
\allowdisplaybreaks
\usepackage{amsfonts}
\usepackage{amssymb}
\usepackage{graphicx,psfrag,epsf}
\usepackage{enumerate}
\usepackage{natbib}
\usepackage{url} 
\usepackage{lscape}
\usepackage{pdflscape}
\newcommand\independent{\protect\mathpalette{\protect\independenT}{\perp}}
\def\independenT#1#2{\mathrel{\rlap{$#1#2$}\mkern2mu{#1#2}}}
\usepackage{mathtools}
\usepackage{xfrac}
\usepackage{booktabs}
\usepackage{multirow}
\usepackage{enumitem}
\usepackage{xcolor}
\usepackage{subcaption}

\usepackage{theorem}
\usepackage{ulem}

\newtheorem{assumption}{Assumption}

\newtheorem{corollary}{Corollary}

\newtheorem{lemma}{Lemma}

\newtheorem{proposition}{Proposition}

\newcommand{\blind}{0}

\begin{document}

\def\spacingset#1{\renewcommand{\baselinestretch}%
{#1}\small\normalsize} \spacingset{1}


\if0\blind
{
  \title{\bf Probability of Causation with Sample Selection: A Reanalysis of the Impacts of \textit{Jóvenes en Acción} on Formality}
  \author{Vitor Possebom\thanks{vitor.possebom@fgv.br}\hspace{.2cm}\\
    Sao Paulo School of Economics - FGV\\
    and \\
    Flavio Riva\thanks{flaviorussoriva@gmail.com} \\
    Instituto Mobilidade e Desenvolvimento Social - Imds}
  \maketitle
} \fi

\if1\blind
{
  \bigskip
  \bigskip
  \bigskip
  \begin{center}
    {\LARGE\bf Probability of Causation with Sample Selection: A Reanalysis of the Impacts of \textit{Jóvenes en Acción} on Formality}
\end{center}
  \medskip
} \fi

\bigskip
\begin{abstract}
This paper identifies the probability of causation when there is sample selection. We show that the probability of causation is partially identified for individuals who are always observed regardless of treatment status and derive sharp bounds under three increasingly restrictive sets of assumptions. The first set imposes an exogenous treatment and a monotone sample selection mechanism. To tighten these bounds, the second set also imposes the monotone treatment response assumption, while the third set additionally imposes a stochastic dominance assumption. Finally, we use experimental data from the Colombian job training program \textit{Jóvenes en Acción} to empirically illustrate our approach's usefulness. We find that, among always-employed women, at least 10.2\% and at most 13.4\% transitioned to the formal labor market because of the program. However, our 90\%-confidence region does not reject the null hypothesis that the lower bound is equal to zero.
\end{abstract}

\noindent%
\vfill

\newpage
\spacingset{1.8} 

\section{Introduction}\label{Sintro}
\newsavebox{\tablebox} \newlength{\tableboxwidth}

{ Many policy evaluation questions involve two simultaneous identification challenges: the causal parameter of interest depends on the joint distribution of potential outcomes {\citep{Heckman1997,Pearl1999,Tian2000,Jun2019,Cinelli2021}}, and sample selection is present {\citep{Lee2009,Chen2015,Bartalotti2021}}. For example, when evaluating the effects of job training programs {\citep{Heckman1999a,Attanasio2011,Attanasio2017,Blanco2018}}, the researcher may be interested in learning to what extent the transition from informal to formal employment can be attributed to the policy. Still, she only observes formality status among those who are employed. This double identification challenge also arises when researchers analyze the effects of a political campaign on agents' opinions {\citep{DellaVigna2007,DellaVigna2010}} if agents may not reply to the researchers' survey.}

In this paper, we derive novel sharp bounds around the probability of causation parameter \citep{Pearl1999,Tian2000,Jun2019,Cinelli2021} for individuals who self-select into the sample regardless of their treatment assignment. The probability of causation parameter summarizes one crucial aspect of the effects of treatments on binary outcomes: the proportion of individuals who benefit from being treated within the subgroup who would, counterfactually, experience a negative untreated outcome. Thus, our target parameter helps researchers gauge to what extent the transition from one state to another can be attributed to the treatment in a relevant latent sub-population.

Our partial identification strategies are based on three increasingly restrictive sets of assumptions. They extend the identification of probabilities of causation to scenarios with endogenous sample selection. In our model, treatment effects can be related to the sample selection mechanism even though treatment take-up is exogenous. We also discuss when our assumptions have identification power and how to test them through necessary observable conditions.

Our first identification result relies on a monotone sample selection mechanism. This condition imposes that treatment has a non-negative effect on the sample selection indicator for all individuals. In the job training example, this restriction implies that the treatment can move workers into employment but never out of employment.

Our second result further assumes a monotone treatment response to tighten the identified bounds. This condition imposes that treatment has a non-negative effect on the potential outcomes for all individuals. In the job training example, this restriction implies that the treatment can move workers into formal jobs but never into informal jobs.

Our final result additionally relies on a stochastic dominance assumption to further reduce the identified set. This condition imposes that the sub-population that self-selects into the sample regardless of the treatment status has higher treated potential outcomes than the sub-population that self-selects into the sample only when treated. In the job training example, this restriction implies that the agents who are always employed are more likely to have a formal job if treated than those who are employed only when treated.

Additionally, we propose parametric estimators for all these bounds. We also combine the precision-corrected bounds proposed by \cite{Chernozhukov2013} with a Bonferroni-style correction to derive confidence regions that contain the identified region with a pre-specified confidence level.

To empirically illustrate the usefulness of our approach, we provide bounds for the probability of causation of an intensive training program: \textit{Jóvenes en Acción}. This program aimed to improve the labor market prospects and, in particular, the quality of jobs held by disadvantaged youths in seven large cities in Colombia. It offered in-classroom intensive training in occupational skills to qualify unemployed individuals for locally demanded jobs. Additionally, it focused on socioemotional development and offered on-the-job internships with formal employers.

Previous research \citep{Attanasio2011,Attanasio2017} finds that this program positively affects employment and unconditional formality. However, less is known about whether the program achieves its goal of improving job quality conditioning on having a job. We study its effects on the job quality margin by considering the share of women that transitioned to the formal labor market \textit{because} they participated in the training program. We find that incorporating selection and bounding the probability of causation leads to a pessimistic view of the program’s impacts. More precisely, we find that at most 13.4\% of the always-employed women switched their formality status because they were assigned to the \textit{Jóvenes en Acción} training program. Moreover, our 90\%-confidence region includes the zero, implying that we cannot reject the null hypothesis that our target parameter's lower bound is equal to zero.

Concerning its theoretical contribution, our work is inserted in two research areas: identification of probabilities of causation and identification in the presence of sample selection.

\cite{Heckman1997} motivate the focus on a parameter closely connected to the probability of causation based on the political economy of policy evaluation. They argue that a program would only be adopted in a democracy if it benefited most people in the population. {They either make strong probabilistic assumptions or impose model restrictions on treatment take-up decisions} to point-identify this parameter, while we focus entirely on partial identification strategies based on a menu of easily interpretable assumptions.

\cite{Pearl1999} and \cite{Tian2000} discuss how to interpret and partially identify probabilities of causation in a single population where agents are always observed. \cite{Cinelli2021} extend their work by combining experimental results from multiple trials to extrapolate probabilities of causation from one population to a different population. Moreover, \cite{Jun2019} extend their work by considering endogenous selection into treatment.

We extend the work by \cite{Pearl1999} and \cite{Tian2000} in a different direction. We identify probabilities of causation when the agents' realized outcomes may not be observed due to endogenous sample selection. To do so, we combine the tools developed in the literature about probabilities of causation with the trimming bounds developed in the sample selection literature \citep{Horowitz1995,Lee2009,Chen2015,Bartalotti2021}.

Concerning its empirical contribution, our work is inserted in the literature about job training programs. \cite{Attanasio2011} and \cite{Attanasio2017} analyze the average treatment effect (ATE) of \textit{Jóvenes en Acción} on short and long-term outcomes associated with labor force attachment. We extend their work by analyzing a treatment effect parameter that focuses on job quality instead of labor force attachment. Importantly, \cite{Blanco2018} also analyze the impact of a job training program on job quality using partial identification strategies. However, we focus on different contexts (\textit{Job Corps} v. \textit{Jóvenes en Acción}) and different target parameters (Quantile Treatment Effects v. Probabilities of Causation).

{This paper is organized as follows. Section \ref{Sframework} presents our structural model, sample selection mechanism, and identifying assumptions. It also discusses the testable restrictions imposed by our model. Section \ref{Sident} describes our main identification results, while Section \ref{Sestimation} proposes a parametric estimator for our bounds and discusses an inferential method for the identified region. Moreover, Section \ref{Sapplication} discusses the results of our empirical application. In the end, Section \ref{Sconclusion} concludes.

Moreover, we also have an online appendix with additional details and results. Appendix \ref{AppProofs} presents the proofs of all our identification results, while Appendix \ref{Sexample} intuitively explains them using a numerical example. Moreover, Appendix \ref{Stestable} brings a detailed discussion about the testable restrictions of our identifying assumptions, while Appendix \ref{Spoc} compares our target parameter against other causal parameters. Furthermore, Appendix \ref{AppDetailEstimationInference} detailedly explains our estimator and inferential method. Finally, Appendix \ref{AppAddEmpirical} presents additional empirical results.}

\section{Analytical Framework}\label{Sframework}

We aim to identify the probability of causation \citep{Pearl1999,Tian2000,Jun2019,Cinelli2021} within the always-observed subsample. To do so, we consider the generalized sample selection model \citep{Lee2009}, described in the potential outcomes framework:
\begin{equation}\label{EQoutcomes}
	\left\{ \begin{array}{lcl}
		Y^*&=&Y^{*}_{1} \cdot D+Y^{*}_{0} \cdot (1-D)\\
		S&=&S_1 \cdot D+S_0 \cdot (1-D)\\
		Y&=&Y^{*} \cdot S
	\end{array} \right.
\end{equation}
where $D$ is the treatment status indicator (in our application, being selected to enroll in the \textit{Jóvenes in Acción} training program). The variable $Y^{*}$ is the possibly censored realized outcome variable (indicator for whether the agent has a formal or informal job) with support $\mathcal Y = \left\lbrace 0, 1 \right\rbrace$, while $Y_{0}^{*}$ and $Y_{1}^{*}$ are the possibly censored potential outcomes when the person is untreated and treated, respectively. Similarly, $S$ is the realized sample selection indicator (indicator for whether the agent holds a job), and $S_{0}$ and $S_{1}$ are potential sample selection indicators when individuals are untreated and treated. Moreover, $Y$ is the uncensored observed outcome. {Finally, $X$ is a set of exogenous covariates (indicator variables for each course-city pair in the \textit{Jóvenes in Acción} training program) whose support is denoted by $\mathcal{X}$. The researcher observes only the vector $\left(Y,D,S,X\right)$,} while $Y^{*}_{1}$, $Y^{*}_{0}$, $S_{1}$ and $S_{0}$ are latent variables.

In the setting analyzed here, learning about the probability of causation \citep{Pearl1999,Tian2000,Jun2019,Cinelli2021} is further complicated by the potential for nonrandom sample selection. As pointed out by \cite{Lee2009}, even in the simpler case of the average treatment effect (ATE), point identification is no longer possible, leading him to derive bounds for the ATE.

This paper combines the insights of these literatures to develop sharp bounds for the probability of causation under sample selection. To do so, we define four latent groups based on the potential sample selection indicators. The sub-populations are defined as: always-observed ($S_{0} = 1, S_{1} = 1$), observed-only-when-treated ($S_{0} = 0, S_{1} = 1$), observed-only-when-untreated ($S_{0} = 1, S_{1} = 0$), and never-observed ($S_{0} = 0, S_{1} = 0$). They are denoted by $OO$, $NO$, $ON$ and $NN$ respectively.

Following \cite{Zhang2008} and \cite{Lee2009}, we focus on the always-observed sub-population $\left(S_{0} = 1, S_{1} = 1\right)$. Importantly, this sub-population is the only group with censored potential outcomes observed in both treatment arms. For the other three sub-populations, treatment effect parameters are not point-identified or bounded in a non-trivial way without further parametric assumptions because at least one of the potential outcomes ($Y_{0}^{*}$ or $Y_{1}^{*}$) is never observed. Since we focus on a fully non-parametric identification strategy, we do not discuss parametric identification of unconditional treatment effect parameters or treatment effect parameters associated with the latent groups $ON$, $NO$ and $NN$.

Our target parameter is the probability of causation within the sub-population that is always observed:
\begin{equation}\label{EQtarget}
	\theta^{OO} = \mathbb{P}\left[\left. Y_{1}^{*} = 1 \right\vert Y_{0}^{*} = 0, S_{0} = 1, S_{1} = 1\right]
\end{equation}
and depends on the joint distribution of potential outcomes $\left(Y_{0}^{*}, Y_{1}^{*}\right)$.

The unconditional probability of causation $\left(\mathbb{P}\left[\left. Y_{1}^{*} = 1 \right\vert Y_{0}^{*} = 0\right]\right)$ captures, within the sub-population whose untreated potential outcome is equal to zero, the share whose treated potential outcome is equal to one. Intuitively, it measures the share of agents who benefited from the treatment within the subgroup with a negative untreated outcome. In our empirical application, the unconditional probability of causation captures, within the population with an informal job if untreated, the share of workers with a formal job if treated. (In Appendix \ref{Spoc}, we compare the probability of causation parameter against other treatment effect parameters frequently discussed in the literature. In particular, we discuss the concepts of ``persuasion effect'' proposed by \cite{Jun2019}, of ``distribution of gains at selected base state values'' and ``probability of employed with treatment, not employed without treatment'' proposed by \cite{Heckman1997}, and of the average treatment effect.)

Our target parameter in Equation \eqref{EQtarget} focuses on the probability of causation for the always-observed latent group. In our empirical application, our target parameter captures, within the population who is employed regardless of treatment status and has an informal job if untreated, the share of workers with a formal job if treated. Intuitively, we focus on the population who is always employed and found a job of higher observable quality because they were assigned to the \textit{Jóvenes in Acción} training program.

Analogously to \cite{Heckman1997}, \cite{Jun2019} and \cite{Cinelli2021}, identification of $\theta^{OO}$ is complicated because it depends on the joint distribution of the potential outcomes $\left(Y_{0}^{*}, Y_{1}^{*}\right)$ while, even in a randomized controlled trial, we can only identify the marginal distributions of the potential outcomes. Analogously to \cite{Lee2009}, identification of $\theta^{OO}$ is complex because sample selection is nonrandom and possibly impacted by the treatment.

To simultaneously address these issues, we follow a layered policy analysis approach \citep{Manski2011} and consider three sets of assumptions to partially identify our target parameter. The identified set {weakly} shrinks when stronger assumptions are used. Assumptions \ref{ASexogeneity}-\ref{ASmonotonicity} are sufficient {to derive sharp bounds around $\theta^{OO}$.}

\begin{assumption}[Random Assignment]\label{ASexogeneity}
	{Treatment $D$ is randomly assigned after conditioning on the covariates, i.e., $\left. D \independent (Y^{*}_{0},Y^{*}_{1},S_{0},S_{1}) \right\vert X$.}
\end{assumption}

Assumption \ref{ASexogeneity} modifies the standard independence assumption \citep{Imbens2009a} to account for sample selection. Instead of assuming that the treatment variable is independent of the potential outcomes only, we also assume independence between the treatment variable and the potential sample selection indicators similarly to \cite{Lee2009}. In our empirical application, it holds conditionally on course indicators because the possibility of enrolling in the \textit{Jóvenes in Acción} training program was randomly allocated within oversubscribed courses.

\begin{assumption}[Positive Mass]\label{ASpositive}
	{Both treatment groups and the always-observed sub-population who chooses $Y_{0}^{*} = 0$ exist after conditioning on the covariates, i.e., $0 < \mathbb{P}\left[\left. D = 1 \right\vert X = x\right] < 1$ and  $\mathbb{P}\left[\left. Y_{0}^{*} = 0, S_{0} = 1, S_{1} = 1 \right\vert X = x \right] > 0$ for every value $x \in \mathcal{X}$.}
\end{assumption}

Assumption \ref{ASpositive} is crucial for the identification results because it ensures that our sub-population of interest exists. In our empirical application, it requires that oversubscribed courses are the only ones to exist and that there are always-employed individuals who have an informal job when untreated for every course-city pair.

\begin{assumption}[Monotone Sample Selection]\label{ASmonotonicity}
	Treatment has a non-negative effect on the sample selection indicator for all individuals, i.e., $S_{1} \geq S_{0}$.
\end{assumption}

Assumption \ref{ASmonotonicity} is a monotonicity restriction that rules out the existence of the observed-only-when-untreated sub-population and is commonly used in the literature about sample selection \citep{Lee2009, Chen2015,Bartalotti2021}. In our empirical application, it imposes that the \textit{Jóvenes in Acción} training program can only move agents into employment. {This assumption is plausible if the training program improves the workers' social skills, boosting their performance in job interviews. However, this assumption is implausible if the training program stimulates them to pursue further education.}

Assumptions \ref{ASexogeneity}-\ref{ASmonotonicity} form our first set of assumptions required to {derive sharp bounds around} the probability of causation within the always-observed individuals. Importantly, this set of assumptions has a testable implication, as discussed in Lemma \ref{LErestriction1}.

Even though these assumptions are sufficient to {derive sharp bounds around} $\theta^{OO}$, the identified set {may} be substantially tightened by additionally imposing that the treatment can only increase the possibly censored potential outcome.

\begin{assumption}[Monotone Treatment Response]\label{ASmonotonicityY}
	Treatment has a non-negative effect on the censored outcome variable for all individuals, i.e., $Y_{1}^{*} \geq Y_{0}^{*}$.
\end{assumption}

Assumption \ref{ASmonotonicityY} is a monotonicity restriction common in the partial identification literature \citep{Manski1997,Manski2000,Jun2019}. In our empirical application, it imposes that the \textit{Jóvenes in Acción} training program can only move agents from informal jobs to formal ones. {This assumption is plausible if the training program increases the workers' productivity. However, this assumption is implausible if the training program stimulates them to open their own informal firms.}

Assumptions \ref{ASexogeneity}-\ref{ASmonotonicityY} form our second set of assumptions required to {derive sharp bounds around} the probability of causation within the always-observed individuals. Importantly, this set of assumptions has an extra testable implication, as discussed in Proposition \ref{PROPrestriction2}.

We {may} further shrink the identified set around $\theta^{OO}$ by adding Assumption \ref{ASstochastic} and completing our final set of identifying assumptions.

\begin{assumption}[Stochastic Dominance]\label{ASstochastic}
	 {After conditioning on the covariates, the treated counterfactual for the always-observed group stochastically dominates the treated counterfactual for the observed-only-when-treated group, i.e., $$\mathbb{P}\left[\left. Y_{1}^{*} = 1 \right\vert S_{0} = 1, S_{1} = 1, X = x \right] \geq \mathbb{P}\left[\left. Y_{1}^{*} = 1 \right\vert S_{0} =0, S_{1} = 1, X = x \right]$$ for every value $x \in \mathcal{X}$.}
\end{assumption}

Assumption \ref{ASstochastic} is a stochastic dominance restriction that imposes that the always-observed sub-population has higher potential treated outcomes than the observed-only-when-treated group. This type of assumption is common in the literature \citep{Imai2008,Blanco2013,Huber2015,Huber2017,Bartalotti2021} and is intuitively based on the argument that some sub-groups have more favorable underlying characteristics than others. In our empirical application, it imposes that the always-employed sub-population has higher potential formality when treated than the employed-only-when-treated sub-population. {This assumption is plausible if individuals with better employment status are more likely to have better (i.e., formal) jobs because they are more productive or skillful. However, this assumption will be invalid if always-employed individuals have jobs because they are willing to accept any working opportunity, even if it is an informal job.}

\subsection{Testable Restrictions}\label{Srestriction}

This subsection discusses testable restrictions implied by the assumptions described in Section \ref{Sframework}.

First, the testable restriction implied by Assumptions \ref{ASexogeneity}-\ref{ASmonotonicity} was already derived by \cite{Lee2009}. We state it here for completeness.
\begin{lemma}\label{LErestriction1}
	Under Assumptions \ref{ASexogeneity}-\ref{ASmonotonicity}, the following inequality holds:
	\begin{equation*}
		\mathbb{P}\left[\left. S = 1 \right\vert D = 1, X\right] - \mathbb{P}\left[\left. S = 1 \right\vert D = 0, X\right] \geq 0.
	\end{equation*}
\end{lemma}

Second, we derive a set of testable restrictions implied by Assumptions \ref{ASexogeneity}-\ref{ASmonotonicityY} as detailed in Proposition \ref{PROPrestriction2}. Its proof is in Appendix \ref{PROOFrestriction2}.

\begin{proposition}\label{PROPrestriction2}
	Under Assumptions \ref{ASexogeneity}-\ref{ASmonotonicityY}, the following inequalities hold:
	\begin{align}
		\label{EQrestriction1} \mathbb{P}\left[\left. S = 1 \right\vert D = 1, X\right] - \mathbb{P}\left[\left. S = 1 \right\vert D = 0, X\right] & \geq 0, \\
		\label{EQrestriction2} \mathbb{P}\left[\left. Y = 1 \right\vert D = 1, X \right] - \mathbb{P}\left[\left. Y = 1 \right\vert D = 0, X\right] & \geq 0.
	\end{align}
\end{proposition}

Intuitively, the monotonicity of the sample selection indicator and the censored potential outcome implies that treatment positively affects the uncensored potential outcome.

These restrictions can be easily tested using two one-sided tests of mean differences. {In Appendix \ref{Stestable}, we discuss the relationship between these testable restrictions and the bounds proposed in Section \ref{Sident}.}

\section{Identification Results}\label{Sident}

{

In this section, we partially identify the probability of causation within the always-observed sub-population (Equation \eqref{EQtarget}). To do so, we start by identifying the conditional probability of causation within the always-observed sub-population, $$\theta^{OO}\left(x\right) \coloneqq \mathbb{P}\left[\left. Y_{1}^{*} = 1 \right\vert Y_{0}^{*} = 0, S_{0} = 1, S_{1} = 1, X = x\right],$$ and, then, integrate over the distribution of the covariates for the always-observed sub-population with a zero untreated potential outcome, $ \left. X \right\vert Y_{0}^{*} = 0, S_{0} = 1, S_{1} = 1,$ to identify our target parameter $\theta^{OO}$ (Equation \eqref{EQtarget}).

First, we identify $\theta^{OO}\left(x\right)$ under our three sets of assumptions and discuss the identifying power of our assumptions.

Combining Assumptions \ref{ASexogeneity}-\ref{ASmonotonicity}, we derive sharp bounds around the conditional probability of causation within the always-observed sub-population as detailed in Proposition \ref{PROPmonotonicity}. Its proof is in Appendix \ref{PROOFmonotonicity}.

\begin{proposition}\label{PROPmonotonicity}
	Under Assumptions \ref{ASexogeneity}-\ref{ASmonotonicity}, the conditional probability of causation is partially identified for the always-observed subgroup, i.e.,
	\begin{equation*}
		LB_{1}\left(x\right) \leq \theta^{OO}\left(x\right) \leq UB_{1}\left(x\right),
	\end{equation*}
	where $$LB_{1}\left(x\right) \coloneqq \max\left\lbrace \dfrac{ \left[B\left(x\right) - \left(1 - A\left(x\right) \right)\right] \cdot \left[A\left(x\right)\right]^{-1} + C\left(x\right) - 1}{C\left(x\right)} , 0 \right\rbrace,$$ $$UB_{1}\left(x\right) \coloneqq \min \left\lbrace \dfrac{B\left(x\right) \cdot \left[A\left(x\right)\right]^{-1}}{C\left(x\right)} , 1 \right\rbrace,$$ $A\left(x\right) \coloneqq \dfrac{\mathbb{P}\left[\left.S = 1 \right\vert D = 0, X = x \right]}{\mathbb{P}\left[\left.S = 1 \right\vert D = 1, X = x \right]},$ $B\left(x\right) \coloneqq \mathbb{P}\left[\left.Y = 1 \right\vert S = 1, D = 1, X = x \right],$ and $C\left(x\right) \coloneqq \mathbb{P}\left[\left.Y = 0 \right\vert S = 1, D = 0, X = x \right]$ for every value $x \in \mathcal{X}$.

	Moreover, these bounds are sharp.
\end{proposition}

{

Corollary \ref{CorIDprop2} describes when Assumptions \ref{ASexogeneity}-\ref{ASmonotonicity} have identifying power, i.e., the identified set in Proposition \ref{PROPmonotonicity} is strictly smaller than the unit interval. Its proof is in Appendix \ref{PROOFidProp2}.

\begin{corollary}\label{CorIDprop2}
	If Assumptions \ref{ASexogeneity}-\ref{ASmonotonicity} hold and
	\begin{align}
	& \mathbb{P}\left[\left. Y_{0}^{*} = 0, S_{0} = 1\right\vert X = x\right] \nonumber \\
	& \hspace{20pt} > \max \left\lbrace \mathbb{P}\left[\left. Y_{1}^{*} = 0, S_{1} = 1 \right\vert X = x\right], \mathbb{P}\left[\left. Y_{1}^{*} = 1, S_{1} = 1 \right\vert X = x\right]  \right\rbrace \label{EQrestrictionID1}
	\end{align}
	for every value $x \in \mathcal{X}$, then $LB_{1}\left(x\right) > 0$ and $UB_{1}\left(x\right) < 1$.
\end{corollary}

Intuitively, Assumptions \ref{ASexogeneity}-\ref{ASmonotonicity} have identifying power if the group who is informally employed when untreated is sufficiently large.

}

In practice, the bounds in Proposition \ref{PROPmonotonicity} may be {wide} even though they are sharp. To derive tighter bounds, researchers can add increasingly stronger assumptions. Even though the credibility of these assumptions depends on their empirical contexts, applied researchers frequently have some prior about the direction of the treatment effect. Using this prior, the researcher can impose the monotone treatment response condition.

Formally, combining Assumptions \ref{ASexogeneity}-\ref{ASmonotonicityY}, we derive sharp bounds around $\theta^{OO}\left(x\right)$ as detailed in Proposition \ref{PROPmonotonicityY}. Its proof is in Appendix \ref{PROOFmonotonicityY}.

\begin{proposition}\label{PROPmonotonicityY}
	Under Assumptions \ref{ASexogeneity}-\ref{ASmonotonicityY}, the conditional probability of causation is partially identified for the always-observed subgroup, i.e.,
	\begin{equation*}
		LB_{1}\left(x\right) \leq \theta^{OO}\left(x\right) \leq UB_{2}\left(x\right),
	\end{equation*}
	where $$UB_{2}\left(x\right) \coloneqq \min \left\lbrace \dfrac{B\left(x\right) \cdot \left[A\left(x\right)\right]^{-1} + C\left(x\right) - 1}{C\left(x\right)} , 1 \right\rbrace$$  for every value $x \in \mathcal{X}$.

	Moreover, these bounds are sharp.
\end{proposition}

{

	Corollary \ref{CorIDprop3} describes when Assumption \ref{ASmonotonicityY} has additional identifying power, i.e., the identified set in Proposition \ref{PROPmonotonicityY} is strictly smaller than the identified set in Proposition \ref{PROPmonotonicity}. Its proof is in Appendix \ref{PROOFidProp3}.

	\begin{corollary}\label{CorIDprop3}
		If Assumptions \ref{ASexogeneity}-\ref{ASmonotonicityY} hold, Inequality \eqref{EQrestrictionID1} holds, and
		\begin{equation}\label{EQrestrictionID3}
		\mathbb{P}\left[\left. Y_{0}^{*} = 1, Y_{1}^{*} = 1 \right\vert S_{0} = 1, S_{1} = 1, X = x\right] > 0
		\end{equation}
		for every value $x \in \mathcal{X}$, then $LB_{1}\left(x\right) > 0$ and $UB_{2}\left(x\right) < UB_{1}\left(x\right) < 1$.
	\end{corollary}

Note that the identifying power of Assumption \ref{ASmonotonicityY} is illustrated by a strictly smaller upper bound in Proposition \ref{PROPmonotonicityY} in comparison with Proposition \ref{PROPmonotonicity}. Intuitively, Assumption \ref{ASmonotonicityY} has additional identifying power if some always-employed individuals have a formal job regardless of their treatment status.
}

To achieve even tighter bounds, researchers can impose the stochastic dominance condition. Formally, combining Assumptions \ref{ASexogeneity}-\ref{ASstochastic}, we derive sharp bounds around the conditional probability of causation within the always-observed sub-population as detailed in Proposition \ref{PROPstochastic}. Its proof is in Appendix \ref{PROOFstochastic}.

\begin{proposition}\label{PROPstochastic}
	Under Assumptions \ref{ASexogeneity}-\ref{ASstochastic}, the conditional probability of causation is partially identified for the always-observed subgroup, i.e.,
	\begin{equation*}
		LB_{3}\left(x\right) \leq \theta^{OO}\left(x\right) \leq UB_{2}\left(x\right),
	\end{equation*}
	where $$LB_{3}\left(x\right) \coloneqq \max\left\lbrace \dfrac{B\left(x\right) + C\left(x\right) - 1}{C\left(x\right)} , 0 \right\rbrace$$  for every value $x \in \mathcal{X}$.

	Moreover, these bounds are sharp.
\end{proposition}

{

	Corollary \ref{CorIDprop4} describes when Assumption \ref{ASstochastic} has additional identifying power, i.e., the identified set in Proposition \ref{PROPstochastic} is strictly smaller than the identified set in Proposition \ref{PROPmonotonicityY}. Its proof is in Appendix \ref{PROOFidProp4}.

	\begin{corollary}\label{CorIDprop4}
		If Assumptions \ref{ASexogeneity}-\ref{ASstochastic} hold, Inequalities \eqref{EQrestrictionID1} and \eqref{EQrestrictionID3} hold, $\mathbb{P}\left[\left. S_{0} = 0, S_{1} = 1 \right\vert X = x\right] > 0$ and $\mathbb{P}\left[\left. Y_{0}^{*} = 0, Y_{1}^{*} = 0 \right\vert S_{1} = 1, X = x\right] > 0$ for every value $x \in \mathcal{X}$, then $LB_{3}\left(x\right) > LB_{1}\left(x\right) > 0$ and $UB_{2}\left(x\right) < UB_{1}\left(x\right) < 1$.
	\end{corollary}

	Note that the identifying power of Assumption \ref{ASstochastic} is illustrated by a strictly larger lower bound in Proposition \ref{PROPstochastic} in comparison with Proposition \ref{PROPmonotonicityY}. Intuitively, Assumption \ref{ASstochastic} has additional identifying power if there are employed-only-when-treated individuals and if some employed-when-treated individuals never have a formal job.

}

Second, we identify the distribution of the covariates for the always-observed sub-population with a zero untreated potential outcome, $ \left. X \right\vert Y_{0}^{*} = 0, S_{0} = 1, S_{1} = 1,$ in Lemma \ref{LEcovariates}. For ease of notation, we assume that all covariates $X$ are discrete, as in our empirical application. This lemma's proof is in Appendix \ref{PROOFcovariates}.

\begin{lemma}\label{LEcovariates}
	Under Assumptions \ref{ASexogeneity}-\ref{ASmonotonicity}, the distribution of the covariates for the always-observed sub-population with a zero untreated potential outcome is point identified, i.e.,
	\begin{align*}
	\omega\left(x\right) & \coloneqq \mathbb{P}\left[\left. X = x \right\vert Y_{0}^{*} = 0, S_{0} = 1, S_{1} = 1\right] \\
	& = \dfrac{\mathbb{P}\left[\left. Y = 0, S = 1 \right\vert D = 0, X = x \right] \cdot \mathbb{P}\left[X = x\right]}{\sum_{x^{\prime} \in \mathcal{X}} \mathbb{P}\left[\left. Y = 0, S = 1 \right\vert D = 0, X = x^{\prime} \right] \cdot \mathbb{P}\left[X = x^{\prime}\right]}
	\end{align*}
	for every $x \in \mathcal{X}$.
\end{lemma}

Finally, we can combine Propositions \ref{PROPmonotonicity}-\ref{PROPstochastic} and Lemma \ref{LEcovariates} to partially identify our target parameter $\theta^{OO}$ (Equation \eqref{EQtarget})  as detailed in Corollary \ref{CorTarget}.

\begin{corollary}\label{CorTarget}
	The probability of causation is partially identified for the always-observed subgroup, i.e., $$\sum_{x \in \mathcal{X}} LB_{1}\left(x\right) \cdot \omega\left(x\right) \leq \theta^{OO} \leq \sum_{x \in \mathcal{X}} UB_{1}\left(x\right) \cdot \omega\left(x\right)$$ under Assumptions \ref{ASexogeneity}-\ref{ASmonotonicity}, $$\sum_{x \in \mathcal{X}} LB_{1}\left(x\right) \cdot \omega\left(x\right) \leq \theta^{OO} \leq \sum_{x \in \mathcal{X}} UB_{2}\left(x\right) \cdot \omega\left(x\right)$$ under Assumptions \ref{ASexogeneity}-\ref{ASmonotonicityY}, and $$\sum_{x \in \mathcal{X}} LB_{3}\left(x\right) \cdot \omega\left(x\right) \leq \theta^{OO} \leq \sum_{x \in \mathcal{X}} UB_{2}\left(x\right) \cdot \omega\left(x\right)$$ under Assumptions \ref{ASexogeneity}-\ref{ASstochastic}.
\end{corollary}

Furthermore, in Appendix \ref{Sexample}, we illustrate this section's results with a numerical example that captures the intuition behind them.

}


\section{Estimation and Inference}\label{Sestimation}

This section is divided in two parts. In the first part, we discuss how to estimate the bounds proposed in Section \ref{Sident}. In the second part, we propose estimators for 90\%-confidence regions that contain the identified sets described in Corollary \ref{CorTarget}.

Importantly, in Section \ref{Sinference}, we do not discuss how to conduct inference around the target parameter in Equation \eqref{EQtarget}. Our choice of conducting inference around the target parameter's identified region may have a cost in terms of statistical power and may explain our null results in Section \ref{Sapplication}. However, our chosen procedure has the advantage of being simpler and more intuitive.

\subsection{Estimation}\label{Sestimator}

{
	In this section, we propose estimators for the bounds described in Propositions \ref{PROPmonotonicity}-\ref{PROPstochastic} and Corollary \ref{CorTarget}, and the weights in Lemma \ref{LEcovariates}. To do so, we need to estimate $\mathbb{P}\left[\left. S = 1\right\vert D = d, X = x \right]$, $\mathbb{P}\left[\left. Y = y\right\vert S = 1, D = d, X = x \right]$, $\mathbb{P}\left[\left. Y = 0, S = 1\right\vert D = 0, X = x \right]$ and $\mathbb{P}\left[X = x\right]$ for any $y \in \left\lbrace 0,1 \right\rbrace$, $d \in \left\lbrace 0,1 \right\rbrace$ and $x \in \mathcal{X}$.

	We estimate these objects parametrically using maximum likelihood estimators. {To simplify our notation, we follow our empirical application and impose that the covariates $X$ are stratum (course-city pair) fixed effects (417 strata). Moreover, to ensure that the first part of Assumption \ref{ASpositive} holds, we delete non-oversubscribed strata (327 strata remain). Finally, to estimate $B\left(x\right)$ and $C\left(x\right)$, we delete strata without post-treatment employed individuals (246 strata remain).}

	Let $\lambda\left(\cdot\right)$ be a link function, such as the logistic link function or the normal link function. Our parametric regression models are given by:
	\begin{enumerate}
		\item $\mathbb{P}\left[\left. S = 1\right\vert D = d, X = x \right] = \lambda\left(\alpha_{0} + \alpha_{1} \cdot d + \alpha_{x}\right)$,

		\item $\mathbb{P}\left[\left. Y = 1\right\vert S = 1, D = d, X = x \right] = \lambda\left(\beta_{0} + \beta_{1} \cdot d + \beta_{x}\right)$, where we only use the employed subsample to estimate $\beta_{0}$, $\beta_{1}$ and $\beta_{x}$, and

		\item $\mathbb{P}\left[\left. W = 1 \right\vert D = d, X = x \right] = \lambda\left(\gamma_{0} + \gamma_{1} \cdot d + \gamma_{x}\right)$, where $W \coloneqq \mathbf{1}\left\lbrace Y = 0, S = 1 \right\rbrace$.
	\end{enumerate}

	Denoting our coefficients' estimators with the hat notation, the bounds in Propositions \ref{PROPmonotonicity}-\ref{PROPstochastic} can be estimated using the following objects:
	\begin{enumerate}
		\item $\hat{A}\left(x\right) = \dfrac{\lambda\left(\hat{\alpha}_{0} + \hat{\alpha}_{x}\right)}{\lambda\left(\hat{\alpha}_{0} + \hat{\alpha}_{1} + \hat{\alpha}_{x}\right)}$,

		\item $\hat{B}\left(x\right) = \lambda\left(\hat{\beta}_{0} + \hat{\beta}_{1} + \hat{\beta}_{x}\right)$, and

		\item $\hat{C}\left(x\right) = 1 - \lambda\left(\hat{\beta}_{0} + \hat{\beta}_{x}\right)$.
	\end{enumerate}

	Furthermore, the weights in Lemma \ref{LEcovariates} can be estimated by $$\hat{\omega}\left(x\right) = \dfrac{\lambda\left(\hat{\gamma}_{0} + \hat{\gamma}_{x}\right) \cdot \sum_{i = 1}^{N} \mathbf{1}\left\lbrace X_{i} = x \right\rbrace}{\sum_{x^{\prime} \in \mathcal{X}} \lambda\left(\hat{\gamma}_{0} + \hat{\gamma}_{x^{\prime}}\right) \cdot \sum_{i = 1}^{N} \mathbf{1}\left\lbrace X_{i} = x^{\prime} \right\rbrace}.$$

	In Appendix \ref{AppDetailEstimation}, we present the full formulas of our estimators for the bounds in Propositions \ref{PROPmonotonicity}-\ref{PROPstochastic} and Corollary \ref{CorTarget}.

	We must also test the restrictions in Proposition \ref{PROPrestriction2}. The first restriction is equivalent to testing the null hypothesis that $\alpha_{1} \geq 0$. The second restriction is equivalent to testing the null hypothesis that $\delta_{1} \geq 0$ in the following model: $$\mathbb{P}\left[\left. Y = 1\right\vert D = d, X = x \right] = \lambda\left(\delta_{0} + \delta_{1} \cdot d + \delta_{x}\right).$$ To control size appropriately, we use a Bonferroni correction for the p-values of both tests. {When using either a Probit Model or a Logit Model for the link function $\lambda\left(\cdot\right)$, we find Bonferroni corrected p-values equal to 1.00 for $H_{0}: \alpha_{1} \geq 0$ and $H_{0}: \delta_{1} \geq 0$.} These results suggest, based on Proposition \ref{PROPrestriction2}, that our identifying assumptions are not refuted.

}

\subsection{Inference}\label{Sinference}
In this section, we suggest possible estimators for 90\%-confidence regions that contain the identified sets described in Corollary \ref{CorTarget}. To fix ideas, we will focus on the bounds under Assumptions \ref{ASexogeneity}-\ref{ASstochastic}, but all the ideas here extend to the bounds under our other sets of assumptions.

Imposing Assumptions \ref{ASexogeneity}-\ref{ASstochastic}, we have that $\theta^{OO} \in \left[\sum_{x \in \mathcal{X}} LB_{3}\left(x\right) \cdot \omega\left(x\right), \sum_{x \in \mathcal{X}} UB_{2}\left(x\right) \cdot \omega\left(x\right) \right]$ and $\theta^{OO}\left(x\right) \in \left[ LB_{3}\left(x\right), UB_{2}\left(x\right) \right]$ for any $x \in \mathcal{X}$. We want to find random sets $\widehat{Q}_{N}\left(x\right)$ and $\widehat{R}_{N}$ such that
\begin{equation}
\label{EQplimQ}
\mathbb{P}\left[ \left[ LB_{3}\left(x\right), UB_{2}\left(x\right) \right] \subseteq \widehat{Q}_{N}\left(x\right) \right] \geq p_{Q} - o\left(1\right)
\end{equation}
for any $x \in \mathcal{X}$ and
\begin{equation}
\label{EQplimR}
\mathbb{P}\left[ \left[ \sum_{x \in \mathcal{X}} LB_{3}\left(x\right) \cdot \omega\left(x\right), \sum_{x \in \mathcal{X}} UB_{2}\left(x\right) \cdot \omega\left(x\right) \right] \subseteq \widehat{R}_{N} \right] \geq p - o\left(1\right),
\end{equation}
where $N$ is the sample size, $p_{Q} \in \left(\sfrac{1}{2}, 1\right)$ and $p = 0.9$.

The $p_{Q}$-confidence region $\widehat{Q}_{N}\left(x\right)$ is given by the precision-corrected estimator proposed by \cite{Chernozhukov2013}. The $p$-confidence region $\widehat{R}_{N}$ is given by a set that combines the precision-corrected estimator proposed by \cite{Chernozhukov2013} with a Bonferroni-style correction.

For any $x \in \mathcal{X}$, let $\widehat{Q}_{N}\left(x\right) \coloneqq \left[ \widehat{LB}_{3,N}^{CLR}\left(x,\sfrac{\left(1 + p_{Q}\right)}{2}\right), \widehat{UB}^{CLR}_{2,N}\left(x,\sfrac{\left(1 + p_{Q}\right)}{2}\right) \right]$, where $\widehat{LB}_{3,N}^{CLR}\left(x,\sfrac{\left(1 + p_{Q}\right)}{2}\right)$ and $\widehat{UB}^{CLR}_{2,N}\left(x,\sfrac{\left(1 + p_{Q}\right)}{2}\right)$ are the precision-corrected estimators proposed by \cite{Chernozhukov2013} for the bounds $LB_{3}\left(x\right)$ and $UB_{2}\left(x\right)$. These estimators satisfy $$\mathbb{P}\left[ \widehat{LB}_{3,N}^{CLR}\left(x,\sfrac{\left(1 + p_{Q}\right)}{2}\right) \leq LB_{3}\left(x\right) \right] \geq \dfrac{1 + p_{Q}}{2} - o\left(1\right)$$ and $$\mathbb{P}\left[ UB_{2}\left(x\right) \leq \widehat{UB}^{CLR}_{2,N}\left(x,\sfrac{\left(1 + p_{Q}\right)}{2}\right) \right] \geq \dfrac{1 + p_{Q}}{2} - o\left(1\right),$$ implying that Equation \eqref{EQplimQ} holds. We formally prove this result in Appendix \ref{AppDetailInference}.

Now, we define
\begin{equation}
\label{EqConfRegion}
\widehat{R}_{N} \coloneqq \left[ \sum_{x \in \mathcal{X}} \widehat{LB}_{3,N}^{CLR}\left(x,\sfrac{\left(1 + p_{Q}\right)}{2}\right) \cdot \hat{\omega}\left(x\right), \sum_{x \in \mathcal{X}} \widehat{UB}^{CLR}_{2,N}\left(x,\sfrac{\left(1 + p_{Q}\right)}{2}\right) \cdot \hat{\omega}\left(x\right) \right].
\end{equation}

This choice of estimator for a feasible $p-$confidence region is inspired by the unfeasible set given by
\begin{equation}
	\label{EqConfRegionUnfeasible}
	{R}_{N} \coloneqq \left[ \sum_{x \in \mathcal{X}} \widehat{LB}_{3,N}^{CLR}\left(x,\sfrac{\left(1 + p_{Q}\right)}{2}\right) \cdot {\omega}\left(x\right), \sum_{x \in \mathcal{X}} \widehat{UB}^{CLR}_{2,N}\left(x,\sfrac{\left(1 + p_{Q}\right)}{2}\right) \cdot {\omega}\left(x\right) \right],
\end{equation}
which assumes we know the true population weights ${\omega}\left(\cdot\right)$ instead of using the estimated weights $\hat{\omega}\left(\cdot\right)$ proposed in Section \ref{Sestimator}.

In Appendix \ref{AppDetailInference}, we show that the unfeasible set ${R}_{N}$ is a valid $p$-confidence region around the identified set $\left[\sum_{x \in \mathcal{X}} LB_{3}\left(x\right) \cdot \omega\left(x\right), \sum_{x \in \mathcal{X}} UB_{2}\left(x\right) \cdot \omega\left(x\right) \right]$. In particular, a Bonferroni-style correction implies that $p = 90\%$ if $p_{Q} = 99.96\%$. Additionally, if our goal was to derive half-median unbiased estimators, we could use $p_{Q} = 99.8\%$.

Appendix \ref{AppDetailInference} also contain details on how to implement the precision-corrected estimators proposed by \cite{Chernozhukov2013}. This appendix relies heavily on the work done by \cite{Flores2013}, who intuitively explain the method proposed by \cite{Chernozhukov2013}.

As a caveat, we highlight that we do not show that the feasible set $\widehat{R}_{N}$ is a valid $p$-confidence region around the identified set. We believe that taking into consideration the uncertainty behind the estimation of $\omega\left(\cdot\right)$ is beyond the scope of this paper and emphasize that a rigorous treatment of feasible inference around the identified set $\left[\sum_{x \in \mathcal{X}} LB_{3}\left(x\right) \cdot \omega\left(x\right), \sum_{x \in \mathcal{X}} UB_{2}\left(x\right) \cdot \omega\left(x\right) \right]$ is an interesting area for future work.

Despite the absence of a formal proof, Appendix \ref{AppMC} describes a Monte Carlo Simulation that illustrates the finite sample properties of the feasible inference procedure proposed in this section. We find that, in our simulated data-generating process, $\widehat{R}_{N}$ covers the identified set more frequently than its nominal confidence level of 90\%. This result suggests that using the feasible set $\widehat{R}_{N}$ in place of the unfeasible set ${R}_{N}$ may work appropriately, suggesting the importance of developing formal results related to this inference procedure in the future.


\section{Empirical Application: Transition into Formality in the \textit{Jóvenes in Acción} Training Program}\label{Sapplication}

{

Our empirical application uses experimental data on a large job training program called \textit{Jóvenes en Acción}, implemented in Colombia's seven largest cities between 2002 and 2005. The program’s main goals were to increase the labor market attachment and the quality of jobs that disadvantaged young individuals (between 18 and 25 years old) held. To this end, \textit{Jóvenes en Acción} combined three main components:
(i) three months of classroom training on occupational-specific skills in private training centers, with an additional focus on building ``soft'' skills, such as proactive behavior, resourcefulness, openness to feedback and teamwork;
(ii) three months of on-the-job training provided by legally registered companies in the form of an unpaid internship;
(iii) elaboration of a project of life, orienting youth towards a positive visualization of their abilities and work perspectives.

An additional key feature of \textit{Jóvenes en Acción} was that the payment structure of training centers incentivized them to help their trainees complete the program and secure jobs after the program.
Specifically, training centers received a large fraction of their payment conditional on the student completing the course and obtaining an internship. More importantly, they were awarded an additional bonus if the firm hired the trainee on a formal contract.
This tight incentive structure and curricula encompassing a large set of potentially productive skills allows one to consider \textit{Jóvenes en Acción} as an intensive program with high potential to improve the employability and the quality of jobs held by its beneficiaries.

The short-run experimental effects of the program have been described in \cite{Attanasio2011} and point to improvements along the employability and job quality margins. We follow \cite{Attanasio2011} and \cite{Attanasio2017} in analyzing effects separately by gender, focusing on women since there was a significant differential sample selection into employment in this sub-sample in the short run. Specifically, women selected to participate in \textit{Jóvenes en Acción} were 6.1 percentage points (or 9.6\%) more likely to be employed between 13 and 15 months after exiting the program according to \cite{Attanasio2011}. Moreover, they also document that women selected to participate in \textit{Jóvenes en Acción} were 7.1 percentage points (or 36\%) more likely to be formally employed approximately one year after exiting the program.

Differently from \cite{Attanasio2011}, we are interested in learning more about the effects of \textit{Jóvenes en Acción} on job quality \emph{after accounting for sample selection}. Distinguishing between effects on the job quality margin that would occur irrespective of the movements towards employment is important to understand better whether the program led to more favorable labor market outcomes. We focus on formality, which, in most developing countries, is strongly associated with employer compliance with labor market statutes (minimum wage and firing regulations), higher productivity and pay, and social security contributions \citep{Meghir2015wages, Attanasio2017}.

We use our partial identification results to learn about the share of women who became formal \textit{because} they were selected to participate in the program.
As explained in Section \ref{Sframework}, our target parameter is the probability of causation for the latent group that would be employed regardless of treatment assignment.
We compute bounds around this probability of causation by considering assignment to the program as the treatment indicator, employment (either in the formal or the informal sector) as the selection indicator, and an indicator that equals one if the person has a formal job and zero if the person has an informal job as our variable of interest.

We start by providing descriptive statistics on the size of our latent groups of interest, i.e., the share of the female population who would be employed regardless of being assigned to the \textit{Jóvenes en Acción} training program and, within this group, the share of women who would have an informal job if they were assigned to the control group.
Since both objects are point-identified under Assumptions \ref{ASexogeneity}-\ref{ASmonotonicity}, we focus on our first set of assumptions when estimating them.
We find that 71.9\% of the women are always-employed using either a Probit or Logit model as the link function $\lambda\left(\cdot\right)$. Within this subgroup, we also estimate the probability of having an informal job when untreated as 49.7\% using either a Probit or Logit model as the link function $\lambda\left(\cdot\right)$. Thus, our latent group of interest represents a non-negligible share (approximately 35.7\%) of the program's pool of potential female participants.

Our main results are presented in Figure \ref{FigPoC}. The intervals in this figure represent estimated lower and upper bounds on the probability of causation for the always-employed women (Corollary \ref{CorTarget}) using data from the job training program \textit{Jóvenes en Acción} and the estimator proposed in Section \ref{Sestimator}. The black estimated intervals are based on Assumptions \ref{ASexogeneity}-\ref{ASmonotonicity}. The dark gray estimated intervals are based on Assumptions \ref{ASexogeneity}-\ref{ASmonotonicityY}. The light gray estimated intervals are based on Assumptions \ref{ASexogeneity}-\ref{ASstochastic}. Subfigure \ref{FigPoCprobit} uses a Probit Model as the link function $\lambda\left(\cdot\right)$ while Subfigure \ref{FigPoClogit} uses a Logit Model. The dots represent the lower and upper bounds of 90\%-confidence regions around the identified sets. These confidence regions are based on the inferential method proposed by \cite{Chernozhukov2013} and explained in Section \ref{Sinference}. Since the bounds with a Probit or a Logit link function are very similar, we focus our discussion on the former.

\begin{figure}[!htbp]\caption{Estimated Bounds on the Probability of Causation in the \textit{Jóvenes in Acción}}\label{FigPoC}
	\begin{subfigure}[t]{0.47\textwidth}
			\centering
			\includegraphics[width = \textwidth]{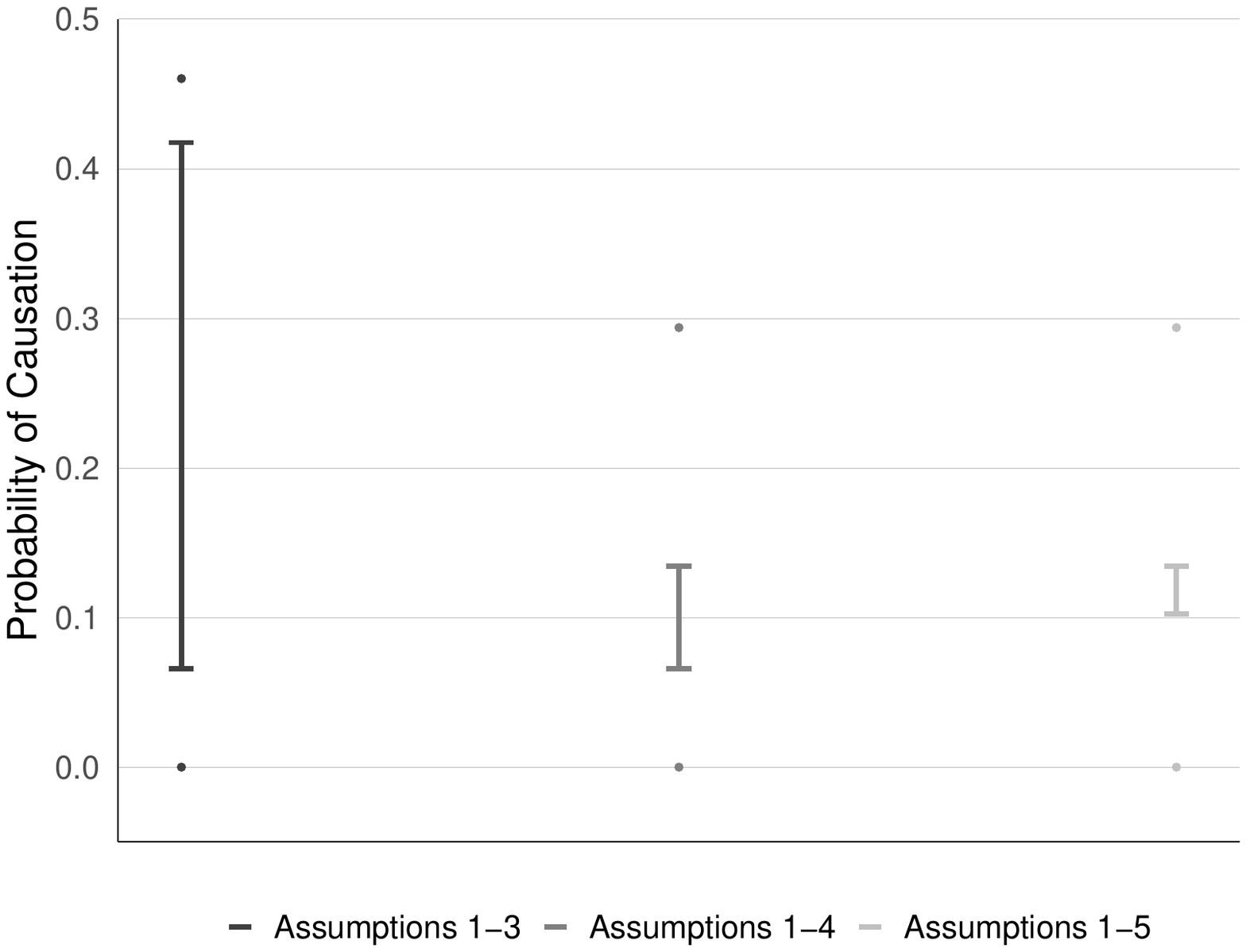}
			\caption{Probit Model as Link Function}
			\label{FigPoCprobit}
		\end{subfigure}
        \hfill
        \begin{subfigure}[t]{0.47\textwidth}
			\centering
			\includegraphics[width = \textwidth]{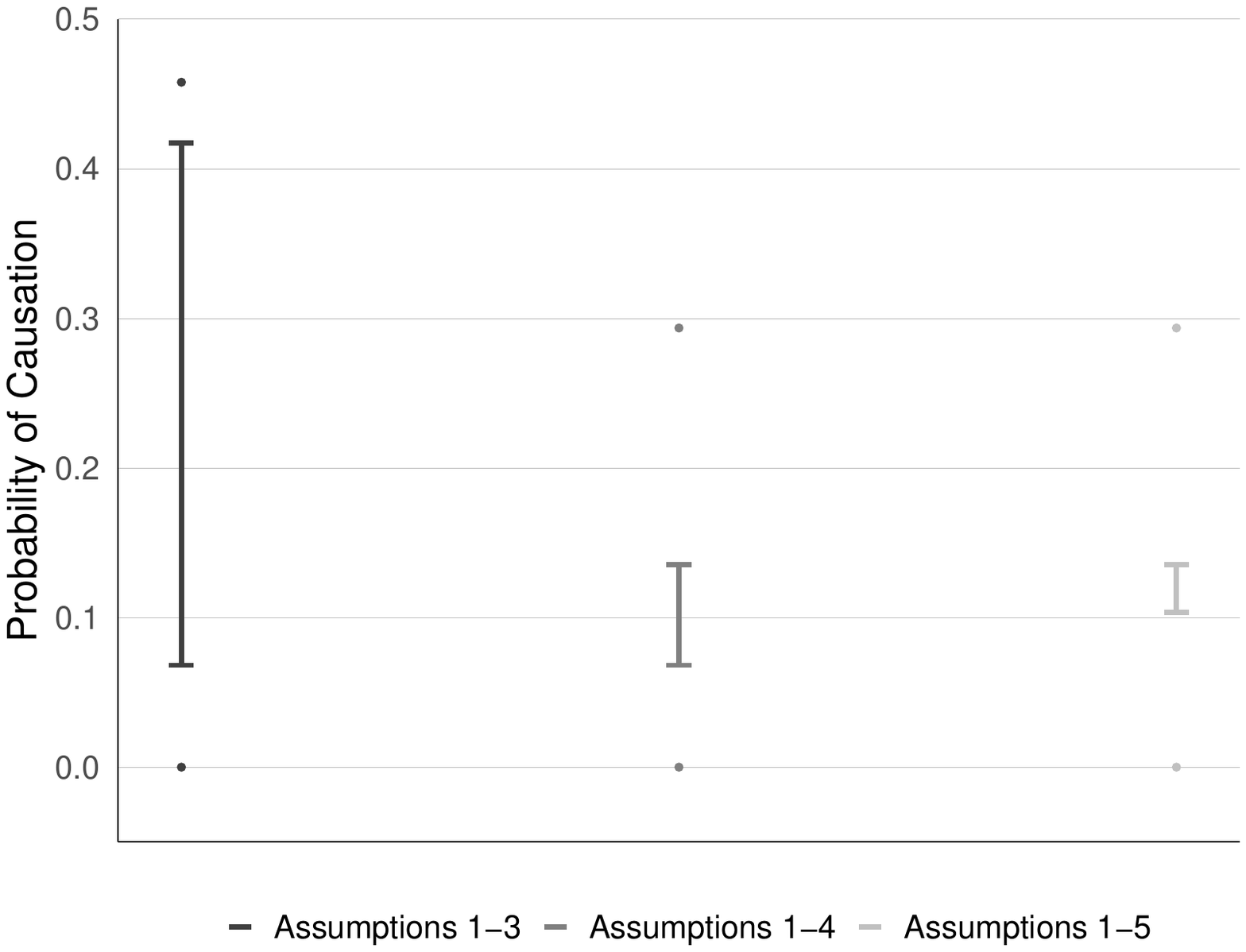}
			\caption{Logit Model as Link Function}
			\label{FigPoClogit}
		\end{subfigure}
	\textit{Notes}: The intervals in this figure represent estimated lower and upper bounds on the probability of causation for the always-employed women (Corollary \ref{CorTarget}) using data from the job training program \textit{Jóvenes en Acción} and the estimator proposed in Section \ref{Sestimator}. The outcome of interest is formal employment one year after the training program, the selection indicator is employment, and the treatment is a randomized assignment indicator. The black estimated intervals are based on Assumptions \ref{ASexogeneity}-\ref{ASmonotonicity}. The dark gray estimated intervals are based on Assumptions \ref{ASexogeneity}-\ref{ASmonotonicityY}. The light gray estimated intervals are based on Assumptions \ref{ASexogeneity}-\ref{ASstochastic}. Subfigure \ref{FigPoCprobit} uses a Probit Model as the link function $\lambda\left(\cdot\right)$ while Subfigure \ref{FigPoClogit} uses a Logit Model. The dots represent the lower and upper bounds of 90\%-confidence regions around the identified sets. These confidence regions are based on the inferential method proposed by \cite{Chernozhukov2013} and explained in Section \ref{Sinference}.
\end{figure}

We start by presenting the bounds on the probability of causation for the always-employed women (Corollary \ref{CorTarget}) under Assumptions \ref{ASexogeneity}-\ref{ASmonotonicity}. In this case, we only impose, beyond the random assignment and positive mass assumptions, that participation in the program does not deter employment (monotone sample selection).

Assumption \ref{ASmonotonicity} is plausible in the \textit{Jóvenes in Acción} context. First, the training program's focus on ``soft skills'' is likely to boost the workers' performance in job interviews, improving their employment prospects. Second, as discussed in Section \ref{Sestimation}, the test proposed in Lemma \ref{LErestriction1} does not reject the null hypothesis that is implied by Assumptions \ref{ASexogeneity}-\ref{ASmonotonicity}.

We find that the estimated bounds are very wide. They imply that our estimates are consistent with a large variety of values for the probability of causation for the always-employed women ($[6.6\%, 41.8\%]$). It implies that the \textit{Jóvenes in Acción} training program formalized, at least, 6.6\% of the women who are always-employed and would have an informal job if untreated. Moreover, the 90\%-confidence region includes the zero, implying that we cannot reject the null hypothesis that our target parameter's lower bound is equal to zero.

To tighten the estimated intervals, we now discuss the bounds obtained by additionally imposing Assumption \ref{ASmonotonicityY}. In this case, we assume that participation in the program can only move agents from informal jobs to formal ones.

Assumption \ref{ASmonotonicityY} is plausible in the \textit{Jóvenes in Acción} context. First, the program's occupational-specific classes and on-the-job training are likely to increase the workers' productivity, helping them find better (i.e., formal) jobs. Second, training centers are incentivized to help their trainees secure a formal job in the firm where they interned. Furthermore, as discussed in Section \ref{Sestimation}, the test proposed in Proposition \ref{PROPrestriction2} does not reject the null hypotheses that are implied by Assumptions \ref{ASexogeneity}-\ref{ASmonotonicityY}.

We find that imposing a monotone treatment response decreases the upper bound substantially. The dark gray interval in Figure \ref{FigPoCprobit} suggests that \textit{Jóvenes en Acción} formalized at most 13.4\% of the women who are always-employed and would have an informal job if untreated. Furthermore, the upper bound of the 90\%-confidence region decreases to 29.4\%.

To further tighten the estimated intervals, we discuss the bounds obtained by additionally imposing Assumption \ref{ASstochastic}. In this case, we assume that the always-employed sub-population has higher potential formality when treated than the employed-only-when-treated sub-population. This assumption is plausible because individuals with better employment status are more likely to be more skillful, increasing their chances of having a better (i.e., formal) job.

We find that imposing this stochastic dominance assumption increases the lower bound. The light gray interval in Figure \ref{FigPoCprobit} suggests that \textit{Jóvenes en Acción} formalized at least 10.2\% of the women who are always-employed and would have an informal job if untreated. Importantly, the 90\%-confidence region includes zero, implying that we cannot reject the null hypothesis that the lower bound of the probability of causation for the always-employed women is zero.

Finally, in Appendix \ref{AppAddEmpirical}, we present additional results focusing on the heterogeneity generated by different course-city pairs.

}


\section{Conclusion}\label{Sconclusion}

This paper partially identifies the probability of causation for the always-observed subgroup when sample selection occurs. This parameter is important for researchers aiming to describe treatment effects in a way that is relevant to policy-makers. Intuitively, it describes the share of the population induced by the treatment to switch from a negative to a positive state. We derive sharp bounds around this parameter under three increasingly restrictive sets of assumptions.

To illustrate the usefulness of our partial identification strategy, we use experimental data from the Colombian job training program \textit{Jóvenes en Acción}. { Contradicting the positive effects on the share of women employed in the formal labor market \citep{Attanasio2011}, we find that incorporating selection and bounding the probability of causation leads to a pessimistic view of the program’s impacts. More precisely, we find that at most 13.4\% of the always-employed women switched their formality status because they were assigned to the \textit{Jóvenes en Acción} training program. Moreover, even our tightest 90\%-confidence region includes zero, implying that we cannot reject the null hypothesis that our lower bound is equal to zero.}

{Beyond the analysis of job training programs, our partial identification strategy can be useful for researchers interested in assessing the impacts of interventions in the presence of sample selection. For example, when analyzing the effects of a political campaign {\citep{DellaVigna2007,DellaVigna2010}}, the researcher may be interested in identifying the share of the population who supports policy A when treated, given that they would support policy B if untreated. In this case, the researcher only observes the agents' opinions if they reply to a survey. This double identification challenge also arises when researchers consider the effects of health interventions on health quality {\citep{Health2004}} if agents may pass away, or the effects of educational interventions on learning {\citep{Angrist2006,Chetty2011,Dobbie2015}} if there is selection into test-taking.}


\section{Acknowledgment}\label{Sacknowledgment}

We thank Donald Andrews, Xiaohong Chen, Fernanda Estevan, Bruno Ferman, Sergio Firpo, John Eric Humphries, Helena Laneuville, Guilherme Lichand, Yusuke Narita, Cormac O'Dea, Giovanni Di Pietra, Rudi Rocha, Edward Vytlacil, Siu Yuat Wong, and seminar participants at Yale University, EPGE Brazilian School of Economics and Finance, Sao Paulo School of Economics, Federal University of Paraiba and State University of New York (Albany) for helpful suggestions. We thank Joana Getlinger for providing excellent research assistance.

\bibliographystyle{chicago}

\bibliography{Bibliography-MM-MC}

\newpage


\appendix

\begin{center}
	\huge
	Supporting Information

	(Online Appendix)

\end{center}

\section{Proofs}\label{AppProofs}

\setcounter{table}{0}
\renewcommand\thetable{A.\arabic{table}}

\setcounter{figure}{0}
\renewcommand\thefigure{A.\arabic{figure}}

\setcounter{equation}{0}
\renewcommand\theequation{A.\arabic{equation}}

\setcounter{theorem}{0}
\renewcommand\thetheorem{A.\arabic{theorem}}

\setcounter{lemma}{0}
\renewcommand\thelemma{A.\arabic{lemma}}

\setcounter{proposition}{0}
\renewcommand\theproposition{A.\arabic{proposition}}

\setcounter{corollary}{0}
\renewcommand\thecorollary{A.\arabic{corollary}}

\setcounter{assumption}{0}
\renewcommand\theassumption{A.\arabic{assumption}}

\subsection{Proof of Proposition \ref{PROPrestriction2}}\label{PROOFrestriction2}

{For ease of notation, we omit from the proof that all probabilities are conditional on covariates $X$.}

To prove Proposition \ref{PROPrestriction2}, we must prove that Inequalities \eqref{EQrestriction1} and \eqref{EQrestriction2} hold. Since the validity of Inequality \eqref{EQrestriction1} is a direct consequence of Lemma \ref{LErestriction1}, we focus on proving Inequality \eqref{EQrestriction2}. Note that
\begin{align*}
	& \mathbb{P}\left[\left. Y = 1 \right\vert D = 1\right] - \mathbb{P}\left[\left. Y = 1 \right\vert D = 0\right] \\
	& \hspace{20pt} = \mathbb{P}\left[\left. Y_{1}^{*} \cdot S_{1} = 1 \right\vert D = 1\right] - \mathbb{P}\left[\left. Y_{0}^{*} \cdot S_{0} = 1 \right\vert D = 0\right] \\
	& \hspace{40pt} \text{by Equation \eqref{EQoutcomes}} \\
	& \hspace{20pt} = \mathbb{P}\left[Y_{1}^{*} \cdot S_{1} = 1 \right] - \mathbb{P}\left[ Y_{0}^{*} \cdot S_{0} = 1 \right] \\
	& \hspace{40pt} \text{by Assumption \ref{ASexogeneity}} \\
	& \hspace{20pt} = \mathbb{P}\left[Y_{1}^{*} = 1, S_{1} = 1 \right] - \mathbb{P}\left[ Y_{0}^{*} = 1, S_{0} = 1 \right] \\
	& \hspace{20pt} \geq \mathbb{P}\left[Y_{1}^{*} = 1, S_{0} = 1 \right] - \mathbb{P}\left[ Y_{0}^{*} = 1, S_{0} = 1 \right] \\
	& \hspace{40pt} \text{by Assumption \ref{ASmonotonicity}} \\
	& \hspace{20pt} \geq \mathbb{P}\left[Y_{0}^{*} = 1, S_{0} = 1 \right] - \mathbb{P}\left[ Y_{0}^{*} = 1, S_{0} = 1 \right] \\
	& \hspace{40pt} \text{by Assumption \ref{ASmonotonicityY}} \\
	& \hspace{20pt} = 0.
\end{align*}

\subsection{Proof of Proposition \ref{PROPmonotonicity}}\label{PROOFmonotonicity}

{For ease of notation, we omit from the proof that all probabilities are conditional on covariates $X$.}

To prove Proposition \ref{PROPmonotonicity}, we first show that $LB_{1} \leq \theta^{OO}$ and $\theta^{OO} \leq UB_{1}$. Then, we show that $LB_{1}$ and $UB_{1}$ are sharp bounds. For completeness, we state four lemmas previously derived in the literature and used in our proofs. We prove them in Appendix \ref{PROOFlemmas}.

\begin{lemma}\label{LEboole}
	Boole-Frechet Bounds \citep{Imai2008}: We have that
	\begin{align*}
		& \mathbb{P}\left[\left. Y_{1}^{*} = 1 \right\vert S_{0} = 1, S_{1} = 1 \right] + \mathbb{P}\left[\left. Y_{0}^{*} = 0 \right\vert S_{0} = 1, S_{1} = 1 \right] - 1 \\
		& \hspace{40pt} \leq \mathbb{P}\left[\left. Y_{1}^{*} = 1, Y_{0}^{*} = 0  \right\vert S_{0} = 1, S_{1} = 1 \right] \\
		& \hspace{40pt} \leq \min \left\lbrace \mathbb{P}\left[\left. Y_{1}^{*} = 1 \right\vert S_{0} = 1, S_{1} = 1 \right], \mathbb{P}\left[\left. Y_{0}^{*} = 0 \right\vert S_{0} = 1, S_{1} = 1 \right] \right\rbrace.
	\end{align*}
\end{lemma}

\begin{lemma}\label{LEhorowitz}
	\citet[Corollary 1.2]{Horowitz1995}: Under Assumptions \ref{ASexogeneity} and \ref{ASpositive}, we have that
	\begin{align*}
		\dfrac{\mathbb{P}\left[\left. Y = 1 \right\vert S = 1, D = 1 \right] - \left(1 - \mathbb{P}\left[\left. S_{0} = 1, S_{1} = 1 \right\vert S_{1} = 1 \right] \right)}{\mathbb{P}\left[\left. S_{0} = 1, S_{1} = 1 \right\vert S_{1} = 1 \right]}
		& \leq \mathbb{P}\left[\left. Y_{1}^{*} = 1 \right\vert S_{0} = 1, S_{1} = 1 \right] \\
		& \leq \dfrac{\mathbb{P}\left[\left. Y = 1 \right\vert S = 1, D = 1 \right]}{\mathbb{P}\left[\left. S_{0} = 1, S_{1} = 1 \right\vert S_{1} = 1 \right]}.
	\end{align*}
\end{lemma}

\begin{lemma}\label{LEleealpha}
	\citet{Lee2009}: Under Assumptions \ref{ASexogeneity}-\ref{ASmonotonicity}, we have that $$\mathbb{P}\left[\left. S_{0} = 1, S_{1} = 1 \right\vert S_{1} = 1 \right] = \dfrac{\mathbb{P}\left[\left.S = 1 \right\vert D = 0 \right]}{\mathbb{P}\left[\left.S = 1 \right\vert D = 1 \right]}.$$
\end{lemma}

\begin{lemma}\label{LEleeuntreated}
	\cite{Lee2009}: Under Assumptions \ref{ASexogeneity}-\ref{ASmonotonicity}, we have that $$\mathbb{P}\left[\left. Y_{0}^{*} = 0 \right\vert S_{0} = 1, S_{1} = 1\right] = \mathbb{P}\left[\left.Y = 0 \right\vert S = 1, D = 0 \right].$$
\end{lemma}

\subsubsection{Lower Bound: $LB_{1} \leq \theta^{OO}$}

Note that
\begin{align*}
	\theta^{OO} & \coloneqq \mathbb{P}\left[\left. Y_{1}^{*} = 1 \right\vert Y_{0}^{*} = 0, S_{0} = 1, S_{1} = 1 \right] \\
	& = \dfrac{\mathbb{P}\left[\left. Y_{1}^{*} = 1, Y_{0}^{*} = 0 \right\vert S_{0} = 1, S_{1} = 1\right]}{\mathbb{P}\left[\left. Y_{0}^{*} = 0 \right\vert S_{0} = 1, S_{1} = 1\right]} \\
	& \geq \dfrac{\mathbb{P}\left[\left. Y_{1}^{*} = 1 \right\vert S_{0} = 1, S_{1} = 1 \right] + \mathbb{P}\left[\left. Y_{0}^{*} = 0 \right\vert S_{0} = 1, S_{1} = 1 \right] - 1}{\mathbb{P}\left[\left. Y_{0}^{*} = 0 \right\vert S_{0} = 1, S_{1} = 1\right]} \\
	& \hspace{40pt} \text{ by Lemma \ref{LEboole}} \\
	& \geq \dfrac{\dfrac{\mathbb{P}\left[\left. Y = 1 \right\vert S = 1, D = 1 \right] - \left(1 - \mathbb{P}\left[\left. S_{0} = 1, S_{1} = 1 \right\vert S_{1} = 1 \right] \right)}{\mathbb{P}\left[\left. S_{0} = 1, S_{1} = 1 \right\vert S_{1} = 1 \right]} + \mathbb{P}\left[\left. Y_{0}^{*} = 0 \right\vert S_{0} = 1, S_{1} = 1 \right] - 1}{\mathbb{P}\left[\left. Y_{0}^{*} = 0 \right\vert S_{0} = 1, S_{1} = 1\right]} \\
	& \hspace{40pt} \text{ by Lemma \ref{LEhorowitz}} \\
	& = \dfrac{\dfrac{\mathbb{P}\left[\left. Y = 1 \right\vert S = 1, D = 1 \right] - \left(1 - \dfrac{\mathbb{P}\left[\left.S = 1 \right\vert D = 0 \right]}{\mathbb{P}\left[\left.S = 1 \right\vert D = 1 \right]} \right)}{\dfrac{\mathbb{P}\left[\left.S = 1 \right\vert D = 0 \right]}{\mathbb{P}\left[\left.S = 1 \right\vert D = 1 \right]}} + \mathbb{P}\left[\left. Y_{0}^{*} = 0 \right\vert S_{0} = 1, S_{1} = 1 \right] - 1}{\mathbb{P}\left[\left. Y_{0}^{*} = 0 \right\vert S_{0} = 1, S_{1} = 1\right]} \\
	& \hspace{40pt} \text{ by Lemma \ref{LEleealpha}} \\
	& = \dfrac{\dfrac{\mathbb{P}\left[\left. Y = 1 \right\vert S = 1, D = 1 \right] - \left(1 - \dfrac{\mathbb{P}\left[\left.S = 1 \right\vert D = 0 \right]}{\mathbb{P}\left[\left.S = 1 \right\vert D = 1 \right]} \right)}{\dfrac{\mathbb{P}\left[\left.S = 1 \right\vert D = 0 \right]}{\mathbb{P}\left[\left.S = 1 \right\vert D = 1 \right]}} + \mathbb{P}\left[\left.Y = 0 \right\vert S = 1, D = 0 \right] - 1}{\mathbb{P}\left[\left.Y = 0 \right\vert S = 1, D = 0 \right]} \\
	& \hspace{40pt} \text{ by Lemma \ref{LEleeuntreated}.}
\end{align*}

Moreover, $\theta^{OO} \geq 0$ by definition.

\subsubsection{Upper Bound: $\theta^{OO} \leq UB_{1}$}

Note that
\begin{align*}
	\theta^{OO} & \coloneqq \mathbb{P}\left[\left. Y_{1}^{*} = 1 \right\vert Y_{0}^{*} = 0, S_{0} = 1, S_{1} = 1 \right] \\
	& = \dfrac{\mathbb{P}\left[\left. Y_{1}^{*} = 1, Y_{0}^{*} = 0 \right\vert S_{0} = 1, S_{1} = 1\right]}{\mathbb{P}\left[\left. Y_{0}^{*} = 0 \right\vert S_{0} = 1, S_{1} = 1\right]} \\
	& \leq \dfrac{\min \left\lbrace \mathbb{P}\left[\left. Y_{1}^{*} = 1 \right\vert S_{0} = 1, S_{1} = 1 \right], \mathbb{P}\left[\left. Y_{0}^{*} = 0 \right\vert S_{0} = 1, S_{1} = 1 \right] \right\rbrace}{\mathbb{P}\left[\left. Y_{0}^{*} = 0 \right\vert S_{0} = 1, S_{1} = 1\right]} \\
	& \hspace{40pt} \text{ by Lemma \ref{LEboole}} \\
	& = \min \left\lbrace \dfrac{\mathbb{P}\left[\left. Y_{1}^{*} = 1 \right\vert S_{0} = 1, S_{1} = 1 \right]}{\mathbb{P}\left[\left. Y_{0}^{*} = 0 \right\vert S_{0} = 1, S_{1} = 1\right]}, 1 \right\rbrace \\
	& \leq \min \left\lbrace \dfrac{\dfrac{\mathbb{P}\left[\left. Y = 1 \right\vert S = 1, D = 1 \right]}{\mathbb{P}\left[\left. S_{0} = 1, S_{1} = 1 \right\vert S_{1} = 1 \right]}}{\mathbb{P}\left[\left. Y_{0}^{*} = 0 \right\vert S_{0} = 1, S_{1} = 1\right]}, 1 \right\rbrace \\
	& \hspace{40pt} \text{ by Lemma \ref{LEhorowitz}} \\
	& = \min \left\lbrace \dfrac{\mathbb{P}\left[\left. Y = 1 \right\vert S = 1, D = 1 \right] \cdot \dfrac{\mathbb{P}\left[\left.S = 1 \right\vert D = 1 \right]}{\mathbb{P}\left[\left.S = 1 \right\vert D = 0 \right]}}{\mathbb{P}\left[\left. Y_{0}^{*} = 0 \right\vert S_{0} = 1, S_{1} = 1\right]}, 1 \right\rbrace \\
	& \hspace{40pt} \text{ by Lemma \ref{LEleealpha}} \\
	& = \min \left\lbrace \dfrac{\mathbb{P}\left[\left. Y = 1 \right\vert S = 1, D = 1 \right] \cdot \dfrac{\mathbb{P}\left[\left.S = 1 \right\vert D = 1 \right]}{\mathbb{P}\left[\left.S = 1 \right\vert D = 0 \right]}}{\mathbb{P}\left[\left.Y = 0 \right\vert S = 1, D = 0 \right]}, 1 \right\rbrace \\
	& \hspace{40pt} \text{ by Lemma \ref{LEleeuntreated}.}
\end{align*}

\subsubsection{$LB_{1}$ and $UB_{1}$ are sharp bounds}

To show that $LB_{1}$ and $UB_{1}$ are sharp bounds, we have to show that, for any $\tilde{\theta} \in \left[LB_{1}, UB_{1}\right]$, there exist candidate random variables $\left(\tilde{Y}^{*}_{0}, \tilde{Y}^{*}_{1}, \tilde{S}_{0}, \tilde{S}_{1}, \tilde{D}\right)$ that satisfy the following conditions:\footnote{Intuitively, the definition of sharpness says that there exist candidate random variables $\left(\tilde{Y}^{*}_{0}, \tilde{Y}^{*}_{1}, \tilde{S}_{0}, \tilde{S}_{1}, \tilde{D}\right)$ that attain the candidate target parameter $\tilde{\theta}$, satisfy the model restrictions and are indistinguishable from the true latent variables $\left(Y^{*}_{0}, Y^{*}_{1}, S_{0}, S_{1}, D\right)$ in the sense that they generate the same distribution of the observable data $\left(\tilde{Y}, \tilde{S}, \tilde{D}\right)$ as the distribution of the data that is actually observed, i.e., $\left(Y,S,D\right)$.}
\begin{enumerate}[label=(\Alph*)]
	\item The model restrictions hold, i.e., $\left(\tilde{Y}^{*}_{0}, \tilde{Y}^{*}_{1}, \tilde{S}_{0}, \tilde{S}_{1}, \tilde{D}\right)$ satisfy Assumptions \ref{ASexogeneity}-\ref{ASmonotonicity}.

	\item The data restrictions hold, i.e., $\mathbb{P}\left[\left. \tilde{Y} = 1 \right\vert \tilde{S} = 1, \tilde{D} = d \right] = \mathbb{P}\left[\left. Y = 1 \right\vert S = 1, D = d \right]$, $\mathbb{P}\left[\left. \tilde{S} = 1 \right\vert \tilde{D} = d \right] = \mathbb{P}\left[\left. S = 1 \right\vert D = d \right]$ for any $d \in \left\lbrace 0, 1 \right\rbrace$ and $\mathbb{P}\left[\tilde{D} = 1\right] = \mathbb{P}\left[D = 1\right]$, where $\tilde{Y}^* = \tilde{Y}^{*}_{1} \cdot \tilde{D} + \tilde{Y}^{*}_{0} \cdot (1 - \tilde{D})$, $\tilde{S} = \tilde{S}_{1} \cdot \tilde{D} + \tilde{S}_{0} \cdot (1 - \tilde{D})$ and $\tilde{Y} = \tilde{Y}^{*} \cdot \tilde{S}$.\footnote{From the observable data, one can estimate: \begin{enumerate}
			\item The joint distribution of $(S,D)$, which is equivalent to estimating $P\left[\left. S=1 \right\vert D=d\right]$ for all $d \in \{0,1\}$ and $P\left[D=1\right]$ given that $S$ and $D$ are binary;
			\item The joint distribution of $(Y,D)\vert S=1$, which is equivalent to estimating $P\left[\left. Y=1 \right\vert S=1, D=d \right]$ for all $d\in \{0,1\}$ and $P\left[D=1\right]$ because $Y$ and $D$ are binary.
		\end{enumerate}
		Hence, the data restrictions guarantee that the proposed latent variables are indistinguishable from the real latent variables in the data.}

	\item $\tilde{\theta}$ is attained, i.e., $\mathbb{P}\left[\left. \tilde{Y}_{1}^{*} = 1 \right\vert \tilde{Y}_{0}^{*} = 0, \tilde{S}_{0} = 1, \tilde{S}_{1} = 1 \right] = \tilde{\theta}$.
\end{enumerate}

To do so, we construct random variables $(\tilde{Y}_0^*, \tilde{Y}_1^*, \tilde{S}_0,\tilde{S}_1, \tilde{D}$) by:
\begin{itemize}
	\item[Part 1.] imposing a joint distribution that satisfies Assumptions \ref{ASexogeneity}-\ref{ASpositive} and ensures that the marginal distribution of $\tilde{D}$ is the same as the marginal distribution of $D$;

	\item[Part 2.] imposing a joint distribution of $(\tilde{S}_0, \tilde{S}_1)$ that satisfies Assumptions \ref{ASpositive}-\ref{ASmonotonicity} and ensures that the conditional distribution of $\left. \tilde{S} \right\vert \tilde{D}$ is the same as the conditional distribution of $\left. S \right\vert D$;

	\item[Part 3.] constructing a conditional distribution $\left. \left(\tilde{Y}^*_0, \tilde{Y}_1^*\right) \right\vert \left(\tilde{S}_0^*, \tilde{S}_1^* \right)$ that is a probability distribution, satisfies the data restrictions, and generates a probability of causation parameter $\tilde{\theta}$ respectively equal to:
	\begin{enumerate}
		\item[(3.a)] the lower bound;
		\item[(3.b)] the upper bound;
		\item[(3.c)] any value in the interval $(LB_1, UB_1)$.
	\end{enumerate}
\end{itemize}

\

\fbox{\textbf{Part 1: The distribution of $\mathbf{\tilde{D}}$ and  Assumptions \ref{ASexogeneity}-\ref{ASpositive}}}

Fix $\left(y_{0}, y_{1}, s_{0}, s_{1}, d\right) \in \left\lbrace 0, 1 \right\rbrace^{5}$ arbitrarily.

To ensure that Assumption \ref{ASexogeneity} holds, we impose that $\mathbb{P}\left[\tilde{Y}^{*}_{0} = y_{0}, \tilde{Y}^{*}_{1} = y_{1}, \tilde{S}_{0} = s_{0}, \tilde{S}_{1} = s_{1}, \tilde{D} = d\right] = \mathbb{P}\left[\tilde{Y}^{*}_{0} = y_{0}, \tilde{Y}^{*}_{1} = y_{1}, \tilde{S}_{0} = s_{0}, \tilde{S}_{1} = s_{1}\right] \cdot \mathbb{P}\left[\tilde{D} = d\right].$

We set
\begin{equation}\label{D1}
	\mathbb{P}\left[\tilde{D}=1\right]=\mathbb{P}\left[D=1\right].
\end{equation}
Note that Assumption \ref{ASpositive} holds because $\mathbb{P}\left[D=1\right] \in \left(0, 1\right)$ according to Assumption \ref{ASpositive} for the true variable $D$.

We also impose that
\begin{equation}
	\mathbb{P}\left[\tilde{D} = 0\right] = 1 - \mathbb{P}\left[\tilde{D} = 1\right],
\end{equation}
so that $\tilde{D}$ has a probability distribution.

\

\fbox{\textbf{Part 2: The distribution of $\mathbf{(\tilde{S}_0, \tilde{S}_1)}$ and  Assumptions \ref{ASpositive}-\ref{ASmonotonicity}}} \\

Since we have defined $\mathbb{P}\left[\tilde{D} = d\right]$ in Part 1, it remains to define $$\mathbb{P}\left[\tilde{Y}^{*}_{0} = y_{0}, \tilde{Y}^{*}_{1} = y_{1}, \tilde{S}_{0} = s_{0}, \tilde{S}_{1} = s_{1}\right].$$ Since $\mathbb{P}\left[\tilde{Y}^{*}_{0} = y_{0}, \tilde{Y}^{*}_{1} = y_{1}, \tilde{S}_{0} = s_{0}, \tilde{S}_{1} = s_{1}\right] = \mathbb{P}\left[\left. \tilde{Y}^{*}_{0} = y_{0}, \tilde{Y}^{*}_{1} = y_{1} \right\vert \tilde{S}_{0} = s_{0}, \tilde{S}_{1} = s_{1}\right] \cdot \mathbb{P}\left[\tilde{S}_{0} = s_{0}, \tilde{S}_{1} = s_{1}\right]$, we define $\mathbb{P}\left[\tilde{S}_{0} = s_{0}, \tilde{S}_{1} = s_{1}\right]$ here and $\mathbb{P}\left[\left. \tilde{Y}^{*}_{0} = y_{0}, \tilde{Y}^{*}_{1} = y_{1} \right\vert \tilde{S}_{0} = s_{0}, \tilde{S}_{1} = s_{1}\right]$ in Part 3.

We set
\begin{equation}\label{S11}
	\mathbb{P}\left[\tilde{S}_0=1, \tilde{S}_1=1 \right] = \mathbb{P}\left[\left. S=1 \right\vert D=0 \right],
\end{equation}
implying that  Assumption \ref{ASpositive} holds because $\mathbb{P}\left[\left. S=1 \right\vert D=0 \right] = \mathbb{P}\left[S_0 = 1 \right] = \mathbb{P}\left[S_0=1, S_1=1 \right] > 0$ according to Assumption \ref{ASexogeneity}-\ref{ASmonotonicity} for the true latent variables.

To ensure that Assumption \ref{ASmonotonicity} holds, we set $\mathbb{P}\left[\tilde{S}_0=1, \tilde{S}_1=0 \right]=0$.

To finish defining the distribution of $(\tilde{S}_0, \tilde{S}_1)$, let

\begin{equation}\label{S01}
	\mathbb{P}\left[\tilde{S}_0=0, \tilde{S}_1=1 \right] = \mathbb{P}\left[\left. S=1 \right\vert D=1 \right] - \mathbb{P}\left[\left. S=1 \right\vert D=0 \right]
\end{equation}
and
\begin{equation}\label{S00}
	\mathbb{P}\left[\tilde{S}_0=0, \tilde{S}_1=0 \right] = 1 - \mathbb{P}\left[\left. S=1 \right\vert D=1 \right].
\end{equation}

To see that what we have indeed defined a probability distribution for $(\tilde{S}_0, \tilde{S}_1)$, note that $$\mathbb{P}\left[\tilde{S}_{0} = 0, \tilde{S}_{1} = 1\right] = \mathbb{P}\left[ S_1 = 1 \right] - \mathbb{P}\left[ S_0 = 1 \right] = \mathbb{P}\left[S_{0} = 0, S_{1} = 1\right] \geq 0$$ by Assumptions \ref{ASexogeneity} and \ref{ASmonotonicity} for the true latent variables, and $$\mathbb{P}\left[\tilde{S}_{0} = 0, \tilde{S}_{1} = 0\right] + \mathbb{P}\left[\tilde{S}_{0} = 1, \tilde{S}_{1} = 0\right] + \mathbb{P}\left[\tilde{S}_{0} = 0, \tilde{S}_{1} = 1\right] + \mathbb{P}\left[\tilde{S}_{0} = 1, \tilde{S}_{1} = 1\right] = 1$$ by construction.

We conclude this part by showing that the distribution of $\tilde{S}|\tilde{D}$ is the same as that of $S|D$.  Note that
\begin{equation*}
	\mathbb{P}\left[\left. \tilde{S} = 1 \right\vert \tilde{D} = 0 \right] = \mathbb{P}\left[\tilde{S}_{0} = 1 \right] = \mathbb{P}\left[\tilde{S}_{0} = 1, \tilde{S}_{1} = 1 \right] = \mathbb{P}\left[\left. S = 1 \right\vert D = 0\right]
\end{equation*}
and that
\begin{align*}
	\mathbb{P}\left[\left. \tilde{S} = 1 \right\vert \tilde{D} = 1 \right] & = \mathbb{P}\left[\tilde{S}_{1} = 1 \right] = \mathbb{P}\left[\tilde{S}_{0} = 1, \tilde{S}_{1} = 1 \right] + \mathbb{P}\left[\tilde{S}_{0} = 0, \tilde{S}_{1} = 1 \right] \\
	& = \mathbb{P}\left[\left. S = 1 \right\vert D = 0\right] + \mathbb{P}\left[\left. S = 1 \right\vert D = 1\right] - \mathbb{P}\left[\left. S = 1 \right\vert D = 0\right] \\
	& = \mathbb{P}\left[\left. S = 1 \right\vert D = 1\right].
\end{align*}

\

\fbox{\textbf{Part 3: The distribution of $\mathbf{(\tilde{Y}^*_1, \tilde{Y}^*_0)|(\tilde{S}_1, \tilde{S}_0)}$}}\\

Since we have defined $\mathbb{P}\left[\tilde{D} = d\right]$ in Part 1 and $\mathbb{P}\left[\tilde{S}_{0} = s_{0}, \tilde{S}_{1} = s_{1}\right]$ in Part 2, it remains to define $\mathbb{P}\left[\left. \tilde{Y}^{*}_{0} = y_{0}, \tilde{Y}^{*}_{1} = y_{1} \right\vert \tilde{S}_{0} = s_{0}, \tilde{S}_{1} = s_{1}\right]$.

We will define $(\tilde{Y}^*_1, \tilde{Y}^*_0)|(\tilde{S}_1, \tilde{S}_0)$ in three different ways so that $\tilde{\theta}$ attains each value in the identified interval $\left[LB_1, UB_1 \right]$ and $\left. \tilde{Y} \right\vert \tilde{S}=1, \tilde{D}$ has the same distribution as $\left. Y \right\vert S=1, D$.

\textbf{(Part 3.a) Constructing a conditional distribution such that $\mathbf{\tilde{\theta} = LB_{1}}$}

Since $\mathbb{P}\left[\tilde{S}_{0} = 1, \tilde{S}_{1} = 0\right] = 0$, we do not need to define $\mathbb{P}\left[\left. \tilde{Y}^{*}_{0} = y_{0}, \tilde{Y}^{*}_{1} = y_{1} \right\vert \tilde{S}_{0} = 1, \tilde{S}_{1} = 0\right]$. We define $\mathbb{P}\left[\left. \tilde{Y}^{*}_{0} = y_{0}, \tilde{Y}^{*}_{1} = y_{1} \right\vert \tilde{S}_{0} = 0, \tilde{S}_{1} = 0\right] = \sfrac{1}{4}$ for any $\left(y_0, y_1\right) \in \{0,1\}^{2}$. We also define the constant

\begin{equation*}
	\blacklozenge = \max \left\lbrace \dfrac{\mathbb{P}\left[\left.Y = 1 \right\vert S = 1, D = 1 \right] - \left(1 - \dfrac{\mathbb{P}\left[\left.S = 1 \right\vert D = 0 \right]}{\mathbb{P}\left[\left.S = 1 \right\vert D = 1 \right]} \right)}{\dfrac{\mathbb{P}\left[\left.S = 1 \right\vert D = 0 \right]}{\mathbb{P}\left[\left.S = 1 \right\vert D = 1 \right]}}, 0 \right\rbrace,
\end{equation*}
and the conditional probabilities
\begin{align}
	& \mathbb{P}\left[\left. \tilde{Y}^{*}_{0} = 0, \tilde{Y}^{*}_{1} = 1 \right\vert \tilde{S}_{0} = 1, \tilde{S}_{1} = 1\right] = \max \{ \blacklozenge  + \mathbb{P}\left[\left.Y = 0 \right\vert S = 1, D = 0 \right] - 1, 0 \} \label{EQlb1} \\
	& \mathbb{P}\left[\left. \tilde{Y}^{*}_{0} = 1, \tilde{Y}^{*}_{1} = 1 \right\vert \tilde{S}_{0} = 1, \tilde{S}_{1} = 1\right]= \min\{ 1- \mathbb{P}\left[\left. Y=0 \right\vert S=1, D=0 \right], \blacklozenge \} \label{EQlb2} \\
	& \mathbb{P}\left[\left. \tilde{Y}^{*}_{0} = 0, \tilde{Y}^{*}_{1} = 0 \right\vert \tilde{S}_{0} = 1, \tilde{S}_{1} = 1\right]  \label{EQlb3}\\
	& \hspace{20pt} = \mathbb{P}\left[\left.Y = 0 \right\vert S = 1, D = 0 \right] - \mathbb{P}\left[\left. \tilde{Y}^{*}_{0} = 0, \tilde{Y}^{*}_{1} = 1 \right\vert \tilde{S}_{0} = 1, \tilde{S}_{1} = 1\right], \nonumber \\
	& \mathbb{P}\left[\left. \tilde{Y}^{*}_{0} = 1, \tilde{Y}^{*}_{1} = 0 \right\vert \tilde{S}_{0} = 1, \tilde{S}_{1} = 1\right] & \label{EQlb4}\\
	& \hspace{20pt} = \mathbb{P}\left[\left.Y = 1 \right\vert S = 1, D = 0 \right] - \mathbb{P}\left[\left. \tilde{Y}^{*}_{0} = 1, \tilde{Y}^{*}_{1} = 1 \right\vert \tilde{S}_{0} = 1, \tilde{S}_{1} = 1\right], \nonumber \\
	& \mathbb{P}\left[\left. \tilde{Y}^{*}_{0} = 0, \tilde{Y}^{*}_{1} = 1 \right\vert \tilde{S}_{0} = 0, \tilde{S}_{1} = 1\right] \label{EQlb5}\\
	& \hspace{20pt} = \dfrac{\mathbb{P}\left[\left. Y = 1, \right\vert S = 1, D = 1\right] - \mathbb{P}\left[\left. \tilde{Y}^{*}_{1} = 1 \right\vert \tilde{S}_{0} = 1, \tilde{S}_{1} = 1\right] \cdot \dfrac{\mathbb{P}\left[\left.S = 1 \right\vert D = 0 \right]}{\mathbb{P}\left[\left.S = 1 \right\vert D = 1 \right]}}{1 - \dfrac{\mathbb{P}\left[\left.S = 1 \right\vert D = 0 \right]}{\mathbb{P}\left[\left.S = 1 \right\vert D = 1 \right]}}, \nonumber \\
	& \mathbb{P}\left[\left. \tilde{Y}^{*}_{0} = 1, \tilde{Y}^{*}_{1} = 1 \right\vert \tilde{S}_{0} = 0, \tilde{S}_{1} = 1\right] = 0, \label{EQlb6} \\
	& \mathbb{P}\left[\left. \tilde{Y}^{*}_{0} = 0, \tilde{Y}^{*}_{1} = 0 \right\vert \tilde{S}_{0} = 0, \tilde{S}_{1} = 1\right] = 1 - \mathbb{P}\left[\left. \tilde{Y}^{*}_{0} = 0, \tilde{Y}^{*}_{1} = 1 \right\vert \tilde{S}_{0} = 0, \tilde{S}_{1} = 1\right], \label{EQlb7}\\
	& \mathbb{P}\left[\left. \tilde{Y}^{*}_{0} = 1, \tilde{Y}^{*}_{1} = 0 \right\vert \tilde{S}_{0} = 0, \tilde{S}_{1} = 1\right] = 0. \label{EQlb8}
\end{align}

\textbf{(Part 3.a.1) The candidate conditional distribution is a probability distribution}

Now, we want to show that the functions described by equations \eqref{EQlb1}-\eqref{EQlb8} are a probability mass function. First, note that:
\begin{align*}
	& \mathbb{P}\left[\left. \tilde{Y}^{*}_{0} = 0, \tilde{Y}^{*}_{1} = 1 \right\vert \tilde{S}_{0} = 1, \tilde{S}_{1} = 1\right] + \mathbb{P}\left[\left. \tilde{Y}^{*}_{0} = 1, \tilde{Y}^{*}_{1} = 1 \right\vert \tilde{S}_{0} = 1, \tilde{S}_{1} = 1\right] \\
	& \hspace{20pt} + \mathbb{P}\left[\left. \tilde{Y}^{*}_{0} = 0, \tilde{Y}^{*}_{1} = 0 \right\vert \tilde{S}_{0} = 1, \tilde{S}_{1} = 1\right] + \mathbb{P}\left[\left. \tilde{Y}^{*}_{0} = 1, \tilde{Y}^{*}_{1} = 0 \right\vert \tilde{S}_{0} = 1, \tilde{S}_{1} = 1\right] = 1
\end{align*}
and
\begin{align*}
	& \mathbb{P}\left[\left. \tilde{Y}^{*}_{0} = 0, \tilde{Y}^{*}_{1} = 1 \right\vert \tilde{S}_{0} = 0, \tilde{S}_{1} = 1\right] + \mathbb{P}\left[\left. \tilde{Y}^{*}_{0} = 1, \tilde{Y}^{*}_{1} = 1 \right\vert \tilde{S}_{0} = 0, \tilde{S}_{1} = 1\right] \\
	& \hspace{20pt} + \mathbb{P}\left[\left. \tilde{Y}^{*}_{0} = 0, \tilde{Y}^{*}_{1} = 0 \right\vert \tilde{S}_{0} = 0, \tilde{S}_{1} = 1\right] + \mathbb{P}\left[\left. \tilde{Y}^{*}_{0} = 1, \tilde{Y}^{*}_{1} = 0 \right\vert \tilde{S}_{0} = 0, \tilde{S}_{1} = 1\right] = 1.
\end{align*}

We must show that all values in \eqref{EQlb1}-\eqref{EQlb8} are in the interval $[0,1]$.

Note that
\begin{equation*}
	\blacklozenge \in [0,1]
\end{equation*}
because $\blacklozenge \geq 0$ by construction, and, using Lemma \ref{LEleealpha}, the expression in the definition of $\blacklozenge$ becomes the expression on the left hand side of Lemma \ref{LEhorowitz} and, therefore, $\blacklozenge \le \mathbb{P}\left[ Y^*_1|S_0=1, S_1=1\right] \le 1$.

Furthermore, by construction, we have that:
\begin{equation}\label{Cond0111}
	\max\{0, \blacklozenge -1+ \mathbb{P}\left[ Y=0| S=1, D=0 \right]\} = \mathbb{P}\left[\tilde{Y}^*_0=0,\tilde{Y}^*_1=1 | \tilde{S}_0=1, \tilde{S}_1=1\right] \le  \blacklozenge
\end{equation}
\begin{equation}\label{Cond1111}
	0 \le \mathbb{P}\left[\tilde{Y}^*_0=1,\tilde{Y}^*_1=1 | \tilde{S}_0=1, \tilde{S}_1=1\right] \le  1-\mathbb{P}\left[Y=0|S=1,D=0\right] \le 1
\end{equation}

Given Equation \eqref{Cond1111} and the fact that $1-\mathbb{P}\left[ Y=0| S=1, D=0 \right]=\mathbb{P}\left[ Y=1| S=1, D=0 \right]$, Equation \eqref{EQlb4} implies that
\begin{equation}\label{Cond1011}
	0 \le \mathbb{P}\left[\tilde{Y}^*_0=1,\tilde{Y}^*_1=0 | \tilde{S}_0=1, \tilde{S}_1=1\right] \le 1.
\end{equation}

Given Equations \eqref{EQlb2} and \eqref{Cond0111}, Equation \eqref{EQlb3} implies that
\begin{equation}\label{Cond0011}
	1-\blacklozenge \le \mathbb{P}\left[ \tilde{Y}^*_0=0, \tilde{Y}_1^*=0| \tilde{S}_0=1, \tilde{S}_1=1 \right] \le  \mathbb{P}\left[\left.  Y=0 \right\vert S=1, D=0 \right] \le 1.
\end{equation}

In order to bound $\mathbb{P}\left[\left. \tilde{Y}^*_0=0, \tilde{Y}_1^*=1 \right\vert  \tilde{S}_0=0, \tilde{S}_1=1 \right]$, consider three cases:
\begin{itemize}
	\item[Case 1)] $\blacklozenge=0$:

	In this case, using Equations \eqref{EQlb1} and \eqref{EQlb2}, we get that:
	\begin{align*}
		& \mathbb{P}\left[\tilde{Y}_1^*=1| \tilde{S}_0=1, \tilde{S}_1=1 \right] \\
		& \hspace{20pt} = \mathbb{P}\left[\tilde{Y}_0^*=1, \tilde{Y}_1^*=1| \tilde{S}_0=1, \tilde{S}_1=1 \right]+\mathbb{P}\left[\tilde{Y}_0^*=0, \tilde{Y}_1^*=1| \tilde{S}_0=1, \tilde{S}_1=1 \right] \\
		& \hspace{20pt}=0
	\end{align*}

	Also, by the definition of $\blacklozenge$, it is the case that:

	\begin{equation*}
		\mathbb{P}\left[\left. Y=1 \right\vert S=1, D=1 \right] \le 1-\frac{\mathbb{P}\left[S=1|D=0\right]}{\mathbb{P}\left[ S=1|D=1\right]},
	\end{equation*}
	implying, by Equation \eqref{EQlb5}, that
	\begin{equation*}
		0 \le \frac{\mathbb{P}\left[\left. Y=1 \right\vert S=1, D=1 \right]}{1- \frac{\mathbb{P}\left[S=1|D=0\right]}{\mathbb{P}\left[ S=1|D=1\right]}}= \mathbb{P}\left[\left. \tilde{Y}_0^*=0, \tilde{Y}_1^*=1 \right\vert \tilde{S}_0=0, \tilde{S}_1=1\right] \le 1.
	\end{equation*}

	\item[Case 2)] $\blacklozenge > 1-\mathbb{P}\left[\left. Y=0 \right\vert S=1, D=0 \right]$.

	In this case, Equations \eqref{EQlb1} and \eqref{EQlb2} imply that
	\begin{align*}
		& \mathbb{P}\left[\left. \tilde{Y}^*_1=1 \right\vert \tilde{S}_0=1, \tilde{S}_1=1 \right] \\
		& \hspace{20pt} = \mathbb{P}\left[\left. \tilde{Y}^*_0=1, \tilde{Y}^*_1=1 \right\vert \tilde{S}_0=1, \tilde{S}_1 =1\right]+\mathbb{P}\left[\left. \tilde{Y}^*_0=0, \tilde{Y}^*_1=1 \right\vert \tilde{S}_0=1, \tilde{S}_1=1\right] \\
		& \hspace{20pt} = 1-\mathbb{P}\left[\left.  Y=0 \right\vert  S=1, D=0 \right] + \blacklozenge - \left(   1-\mathbb{P}\left[\left.  Y=0 \right\vert S=1, D=0 \right]\right) \\
		& \hspace{20pt} =\blacklozenge.
	\end{align*}

	\item[Case 3)] $\blacklozenge \in (0, 1-\mathbb{P}\left[ Y=0| S=1, D=0\right]]$.

	In this case, we have that $\mathbb{P}\left[\left. \tilde{Y}_0^*=0, \tilde{Y}_1^*=1 \right\vert \tilde{S}_0=1, \tilde{S}_1=1 \right]=0$ by Equation \eqref{EQlb1} and $\mathbb{P} \left[\left. \tilde{Y}_0^*=1, \tilde{Y}_1^*=1 \right\vert \tilde{S}_0=1, \tilde{S}_1=1 \right]=\blacklozenge$ by Equation \eqref{EQlb2}, implying that \begin{equation*}
		\mathbb{P}\left[\left. \tilde{Y}_1^*=1 \right\vert \tilde{S}_0=1, \tilde{S}_1=1 \right]=\blacklozenge.
	\end{equation*}

	In Cases 2 and 3,  we can use Equation \eqref{EQlb5} to see that
	\begin{equation*}
		\mathbb{P}\left[\tilde{Y}^*_0=0, \tilde{Y}^*_1=1| \tilde{S}_0=0, \tilde{S}_1=1\right]= \frac{\mathbb{P}\left[ Y=1| S=1, D=1\right]- \blacklozenge \cdot  \frac{\mathbb{P}\left[\left. S=1 \right\vert D=0 \right]}{\mathbb{P}\left[\left. S=1 \right\vert D=1 \right]}}{1-\frac{\mathbb{P}\left[\left. S=1 \right\vert D=0 \right]}{\mathbb{P}\left[\left. S=1 \right\vert D=1 \right]}},
	\end{equation*}
	implying, by the definition of $\blacklozenge$, that
	\begin{align}
		& \mathbb{P}\left[\left. \tilde{Y}^*_0=0, \tilde{Y}^*_1=1 \right\vert \tilde{S}_0=0, \tilde{S}_1=1\right] \label{Square} \\
		& \hspace{20pt} = \frac{\mathbb{P}\left[ Y=1| S=1, D=1\right]- \left(\mathbb{P}\left[ Y=1| S=1, D=1\right] - \left( 1- \frac{\mathbb{P}\left[S=1|D=0\right]}{\mathbb{P}\left[S=1|D=1\right]}\right) \right)}{1-\frac{\mathbb{P}\left[S=1|D=0\right]}{\mathbb{P}\left[S=1|D=1\right]}} \nonumber \\
		& \hspace{20pt}=1. \nonumber
	\end{align}

\end{itemize}

Since $ \mathbb{P}\left[\tilde{Y}^*_0=0, \tilde{Y}^*_1=1| \tilde{S}_0=0, \tilde{S}_1=1\right] \in [0,1]$, Equation \eqref{EQlb7} ensures that
\begin{equation*}
	\mathbb{P}\left[\tilde{Y}^*_0=0, \tilde{Y}^*_1=0| \tilde{S}_0=0, \tilde{S}_1=1\right] \in [0,1].
\end{equation*}

\textbf{(Part 3.a.2) The candidate conditional distribution satisfies its data restrictions}

The data restrictions for $\left. \tilde{Y} \right\vert \tilde{S} = 1, \tilde{D}$ are satisfied because:

\begin{itemize}
	\item $\mathbb{P}\left[\left. \tilde{Y}=1 \right\vert \tilde{S}=1, \tilde{D}=0 \right]=\mathbb{P}\left[ Y=1| S=1, D=0 \right]$;

	To see that,  use Equations \eqref{EQlb2} and \eqref{EQlb4} and the fact that $\mathbb{P}\left[\tilde{S}_0=1,\tilde{S}_1=0\right]=0$ to write:
	\begin{align*}
		& \mathbb{P}\left[\left. \tilde{Y} = 1 \right\vert \tilde{S} = 1, \tilde{D} = 0 \right] \\
		& \hspace{20pt} = \mathbb{P}\left[\left. \tilde{Y}_{0}^{*} = 1 \right\vert \tilde{S}_{0} = 1\right] \\
		& \hspace{20pt} = \mathbb{P}\left[\left. \tilde{Y}_{0}^{*} = 1 \right\vert \tilde{S}_{0} = 1, \tilde{S}_{1} = 1\right] \\
		& \hspace{20pt} = \mathbb{P}\left[\left. \tilde{Y}_{0}^{*} = 1, \tilde{Y}_{1}^{*} = 1 \right\vert \tilde{S}_{0} = 1, \tilde{S}_{1} = 1\right] + \mathbb{P}\left[\left. \tilde{Y}_{0}^{*} = 1, \tilde{Y}_{1}^{*} = 0 \right\vert \tilde{S}_{0} = 1, \tilde{S}_{1} = 1\right] \\
		& \hspace{20pt} = \mathbb{P}\left[\left.Y = 1 \right\vert S = 1, D = 0 \right].
	\end{align*}

	\item $\mathbb{P}\left[\left. \tilde{Y} \right\vert \tilde{S}=1, \tilde{D}=1 \right]=\mathbb{P}\left[Y|S=1, D=1\right]$.

	To see that, note that we can write:
	\begin{align*}
		& \mathbb{P}\left[\left. \tilde{Y} = 1 \right\vert \tilde{S} = 1, \tilde{D} = 1 \right] \\
		& \hspace{20pt} = \mathbb{P}\left[\left. \tilde{Y}_{1}^{*} = 1 \right\vert \tilde{S}_{1} = 1\right] \\
		& \hspace{20pt} = \mathbb{P}\left[\left. \tilde{Y}_{1}^{*} = 1 \right\vert \tilde{S}_{0} = 1, \tilde{S}_{1} = 1 \right] \cdot \mathbb{P}\left[\left. \tilde{S}_{0} = 1, \tilde{S}_{1} = 1 \right\vert \tilde{S}_{1} = 1 \right] \\
		& \hspace{40pt} + \mathbb{P}\left[\left. \tilde{Y}_{1}^{*} = 1 \right\vert \tilde{S}_{0} = 0, \tilde{S}_{1} = 1 \right] \cdot \left( 1 - \mathbb{P}\left[\left. \tilde{S}_{0} = 1, \tilde{S}_{1} = 1 \right\vert \tilde{S}_{1} = 1 \right] \right)
	\end{align*}

	Now, note that we can sum Equations \eqref{EQlb5} and \eqref{EQlb6} and find that
	\begin{equation}\label{plug1}
		\mathbb{P}\left[\tilde{Y}^*_1=1|\tilde{S}_0=0, \tilde{S}_1=1 \right]=\frac{\mathbb{P}\left[Y=1| S=1, D=1\right]-\mathbb{P}\left[ \tilde{Y}^*_1| \tilde{S}_0=1, \tilde{S}_1=1\right] \cdot \frac{\mathbb{P}\left[\left. S=1 \right\vert D=0 \right]}{\mathbb{P}\left[\left. S=1 \right\vert D=1 \right]}}{1-\frac{\mathbb{P}\left[\left. S=1 \right\vert D=0 \right]}{\mathbb{P}\left[\left. S=1 \right\vert D=1 \right]}}
	\end{equation}

	Using Equations \eqref{S01} and \eqref{S11} from Part 1, we get:

	\begin{equation}\label{plug2}
		\mathbb{P}\left[\tilde{S}_1=1, \tilde{S}_0=1| \tilde{S}_1=1\right]=\frac{\mathbb{P}\left[\tilde{S}_1=1, \tilde{S}_0=1\right]}{\mathbb{P}\left[\tilde{S}_0=1, \tilde{S}_1=1\right]+\mathbb{P}\left[\tilde{S}_0=0, \tilde{S}_1=1\right]}=\frac{\mathbb{P}\left[\left. S=1 \right\vert D=0 \right]}{\mathbb{P}\left[\left. S=1 \right\vert D=1 \right]}
	\end{equation}

	Plugging \eqref{plug1} and \eqref{plug2} in the expression above, we get:
	\begin{align*}
		& \mathbb{P}\left[\left. \tilde{Y} = 1 \right\vert \tilde{S} = 1, \tilde{D} = 1 \right] \\
		& \hspace{20pt} = \mathbb{P}\left[\tilde{Y}^*_1=1|\tilde{S}_0=1, \tilde{S}_1=1 \right]\cdot \frac{\mathbb{P}\left[\left. S=1 \right\vert D=0 \right]}{\mathbb{P}\left[\left. S=1 \right\vert D=1 \right]} \\
		& \hspace{40pt} + \frac{\mathbb{P}\left[Y=1| S=1, D=1\right]-\mathbb{P}\left[ \tilde{Y}^*_1| \tilde{S}_0=1, \tilde{S}_1=1\right]\frac{\mathbb{P}\left[\left. S=1 \right\vert D=0 \right]}{\mathbb{P}\left[\left. S=1 \right\vert D=1 \right]}}{1-\frac{\mathbb{P}\left[\left. S=1 \right\vert D=0 \right]}{\mathbb{P}\left[\left. S=1 \right\vert D=1 \right]}} \cdot \left( 1- \frac{\mathbb{P}\left[\left. S=1 \right\vert D=0 \right]}{\mathbb{P}\left[\left. S=1 \right\vert D=1 \right]}    \right) \\
		& \hspace{20pt} = \mathbb{P}\left[\left. Y=1 \right\vert S=1, D=1 \right]
	\end{align*}

\end{itemize}

\textbf{(Part 3.a.3)} \textbf{The probability of causation $\mathbb{\tilde{\theta}}$ reaches the lower bound $\mathbf{LB_1}$}

Finally, note that the lower bound $LB_{1}$ is attained because
\begin{align*}
	& \mathbb{P}\left[\left. \tilde{Y}_{1}^{*} = 1 \right\vert \tilde{Y}_{0}^{*} = 0, \tilde{S}_{0} = 1, \tilde{S}_{1} = 1 \right] \\
	& \hspace{5pt} = \dfrac{\mathbb{P}\left[\left. \tilde{Y}_{0}^{*} = 0, \tilde{Y}_{1}^{*} = 1 \right\vert \tilde{S}_{0} = 1, \tilde{S}_{1} = 1 \right]}{\mathbb{P}\left[\left. \tilde{Y}_{0}^{*} = 0 \right\vert \tilde{S}_{0} = 1, \tilde{S}_{1} = 1 \right]} \\
	& \hspace{5pt} = \dfrac{\mathbb{P}\left[\left. \tilde{Y}_{0}^{*} = 0, \tilde{Y}_{1}^{*} = 1 \right\vert \tilde{S}_{0} = 1, \tilde{S}_{1} = 1 \right]}{\mathbb{P}\left[\left. \tilde{Y}_{0}^{*} = 0, \tilde{Y}_{1}^{*} = 1 \right\vert \tilde{S}_{0} = 1, \tilde{S}_{1} = 1 \right] + \mathbb{P}\left[\left. \tilde{Y}_{0}^{*} = 0, \tilde{Y}_{1}^{*} = 0 \right\vert \tilde{S}_{0} = 1, \tilde{S}_{1} = 1 \right]} \\
	& \hspace{5pt} = \dfrac{\mathbb{P}\left[\left. \tilde{Y}_{0}^{*} = 0, \tilde{Y}_{1}^{*} = 1 \right\vert \tilde{S}_{0} = 1, \tilde{S}_{1} = 1 \right]}{\mathbb{P}\left[\left. \tilde{Y}_{0}^{*} = 0, \tilde{Y}_{1}^{*} = 1 \right\vert \tilde{S}_{0} = 1, \tilde{S}_{1} = 1 \right] + \mathbb{P}\left[\left.Y = 0 \right\vert S = 1, D = 0 \right] - \mathbb{P}\left[\left. \tilde{Y}^{*}_{0} = 0, \tilde{Y}^{*}_{1} = 1 \right\vert \tilde{S}_{0} = 1, \tilde{S}_{1} = 1\right]} \\
	& \hspace{5pt} = \dfrac{\max \left\lbrace \dfrac{\mathbb{P}\left[\left.Y = 1 \right\vert S = 1, D = 1 \right] - \left(1 - \dfrac{\mathbb{P}\left[\left.S = 1 \right\vert D = 0 \right]}{\mathbb{P}\left[\left.S = 1 \right\vert D = 1 \right]} \right)}{\dfrac{\mathbb{P}\left[\left.S = 1 \right\vert D = 0 \right]}{\mathbb{P}\left[\left.S = 1 \right\vert D = 1 \right]}} + \mathbb{P}\left[\left.Y = 0 \right\vert S = 1, D = 0 \right] - 1, 0 \right\rbrace}{\mathbb{P}\left[\left.Y = 0 \right\vert S = 1, D = 0 \right]} \\
	& \hspace{5pt} = LB_{1}.
\end{align*}

\textbf{(Part 3.b) Constructing a conditional distribution such that} $\mathbf{\tilde{\theta} = UB_{1}}$\\

Since $\mathbb{P}\left[\tilde{S}_{0} = 1, \tilde{S}_{1} = 0\right] = 0$, we do not need to define $\mathbb{P}\left[\left. \tilde{Y}^{*}_{0} = y_{0}, \tilde{Y}^{*}_{1} = y_{1} \right\vert \tilde{S}_{0} = 1, \tilde{S}_{1} = 0\right]$. We define $\mathbb{P}\left[\left. \tilde{Y}^{*}_{0} = y_{0}, \tilde{Y}^{*}_{1} = y_{1} \right\vert \tilde{S}_{0} = 0, \tilde{S}_{1} = 0\right] = \sfrac{1}{4}$ for any $\left(y_0, y_1\right) \in \{0,1\}^{2}$. We also define:
\begin{align}
	& \mathbb{P}\left[\left. \tilde{Y}^{*}_{0} = 0, \tilde{Y}^{*}_{1} = 1 \right\vert \tilde{S}_{0} = 1, \tilde{S}_{1} = 1\right] \label{EQub1}\\
	& \hspace{20pt} = \min \left\lbrace \mathbb{P}\left[\left. Y = 1 \right\vert S = 1, D = 1\right] \cdot \dfrac{\mathbb{P}\left[\left. S = 1 \right\vert D = 1\right]}{\mathbb{P}\left[\left. S = 1 \right\vert D = 0\right]}, \mathbb{P}\left[\left. Y = 0 \right\vert S = 1, D = 0\right] \right\rbrace, \nonumber \\
	& \mathbb{P}\left[\left. \tilde{Y}^{*}_{0} = 1, \tilde{Y}^{*}_{1} = 1 \right\vert \tilde{S}_{0} = 1, \tilde{S}_{1} = 1\right] \\
	& \hspace{20pt} = \max \left\lbrace \min\left\lbrace \mathbb{P}\left[\left. Y = 1 \right\vert S = 1, D = 1\right] \cdot \dfrac{\mathbb{P}\left[\left. S = 1 \right\vert D = 1\right]}{\mathbb{P}\left[\left. S = 1 \right\vert D = 0\right]}, 1 \right\rbrace - \mathbb{P}\left[\left. Y = 0 \right\vert S = 1, D = 0\right], 0 \right\rbrace, \nonumber \\
	& \mathbb{P}\left[\left. \tilde{Y}^{*}_{0} = 0, \tilde{Y}^{*}_{1} = 0 \right\vert \tilde{S}_{0} = 1, \tilde{S}_{1} = 1\right] \\
	& \hspace{20pt} = \mathbb{P}\left[\left.Y = 0 \right\vert S = 1, D = 0 \right] - \mathbb{P}\left[\left. \tilde{Y}^{*}_{0} = 0, \tilde{Y}^{*}_{1} = 1 \right\vert \tilde{S}_{0} = 1, \tilde{S}_{1} = 1\right], \nonumber \\
	& \mathbb{P}\left[\left. \tilde{Y}^{*}_{0} = 1, \tilde{Y}^{*}_{1} = 0 \right\vert \tilde{S}_{0} = 1, \tilde{S}_{1} = 1\right] & \\
	& \hspace{20pt} = \mathbb{P}\left[\left.Y = 1 \right\vert S = 1, D = 0 \right] - \mathbb{P}\left[\left. \tilde{Y}^{*}_{0} = 1, \tilde{Y}^{*}_{1} = 1 \right\vert \tilde{S}_{0} = 1, \tilde{S}_{1} = 1\right], \nonumber \\
	& \mathbb{P}\left[\left. \tilde{Y}^{*}_{0} = 0, \tilde{Y}^{*}_{1} = 1 \right\vert \tilde{S}_{0} = 0, \tilde{S}_{1} = 1\right] \\
	& \hspace{20pt} = \max\left\lbrace \dfrac{\mathbb{P}\left[\left. Y = 1 \right\vert S = 1, D = 1\right] - \dfrac{\mathbb{P}\left[\left. S = 1 \right\vert D = 0\right]}{\mathbb{P}\left[\left. S = 1 \right\vert D = 1\right]}}{1 - \dfrac{\mathbb{P}\left[\left. S = 1 \right\vert D = 0\right]}{\mathbb{P}\left[\left. S = 1 \right\vert D = 1\right]}} , 0 \right\rbrace, \nonumber \\
	& \mathbb{P}\left[\left. \tilde{Y}^{*}_{0} = 1, \tilde{Y}^{*}_{1} = 1 \right\vert \tilde{S}_{0} = 0, \tilde{S}_{1} = 1\right] = 0, \\
	& \mathbb{P}\left[\left. \tilde{Y}^{*}_{0} = 0, \tilde{Y}^{*}_{1} = 0 \right\vert \tilde{S}_{0} = 0, \tilde{S}_{1} = 1\right] = 1 - \mathbb{P}\left[\left. \tilde{Y}^{*}_{0} = 0, \tilde{Y}^{*}_{1} = 1 \right\vert \tilde{S}_{0} = 0, \tilde{S}_{1} = 1\right], \\
	& \mathbb{P}\left[\left. \tilde{Y}^{*}_{0} = 1, \tilde{Y}^{*}_{1} = 0 \right\vert \tilde{S}_{0} = 0, \tilde{S}_{1} = 1\right] = 0. \label{EQub8}
\end{align}

Observe that
\begin{align*}
	& \mathbb{P}\left[\left. \tilde{Y}^{*}_{0} = 0, \tilde{Y}^{*}_{1} = 0 \right\vert \tilde{S}_{0} = 1, \tilde{S}_{1} = 1\right] \\
	& \hspace{20pt} \geq \mathbb{P}\left[\left.Y = 0 \right\vert S = 1, D = 0 \right] - \mathbb{P}\left[\left.Y = 0 \right\vert S = 1, D = 0 \right] \\
	& \hspace{20pt} \geq 0,
\end{align*}
and
\begin{align*}
	& \mathbb{P}\left[\left. \tilde{Y}^{*}_{0} = 1, \tilde{Y}^{*}_{1} = 0 \right\vert \tilde{S}_{0} = 1, \tilde{S}_{1} = 1\right] \\
	& \hspace{20pt} \geq \mathbb{P}\left[\left.Y = 1 \right\vert S = 1, D = 0 \right] - 1 + \mathbb{P}\left[\left.Y = 0 \right\vert S = 1, D = 0 \right] \\
	& \hspace{20pt} = 0
\end{align*}
and
\begin{align*}
	& \mathbb{P}\left[\left. \tilde{Y}^{*}_{0} = 0, \tilde{Y}^{*}_{1} = 1 \right\vert \tilde{S}_{0} = 1, \tilde{S}_{1} = 1\right] + \mathbb{P}\left[\left. \tilde{Y}^{*}_{0} = 1, \tilde{Y}^{*}_{1} = 1 \right\vert \tilde{S}_{0} = 1, \tilde{S}_{1} = 1\right] \\
	& \hspace{20pt} + \mathbb{P}\left[\left. \tilde{Y}^{*}_{0} = 0, \tilde{Y}^{*}_{1} = 0 \right\vert \tilde{S}_{0} = 1, \tilde{S}_{1} = 1\right] + \mathbb{P}\left[\left. \tilde{Y}^{*}_{0} = 1, \tilde{Y}^{*}_{1} = 0 \right\vert \tilde{S}_{0} = 1, \tilde{S}_{1} = 1\right] = 1.
\end{align*}
Moreover, note that $\mathbb{P}\left[\left. \tilde{Y}^{*}_{0} = 0, \tilde{Y}^{*}_{1} = 1 \right\vert \tilde{S}_{0} = 0, \tilde{S}_{1} = 1\right] \in \left[0, 1 \right)$ by construction.

Notice also that the data restrictions are satisfied because
\begin{align*}
	& \mathbb{P}\left[\left. \tilde{Y} = 1 \right\vert \tilde{S} = 1, \tilde{D} = 0 \right] \\
	& \hspace{20pt} = \mathbb{P}\left[\left. \tilde{Y}_{0}^{*} = 1 \right\vert \tilde{S}_{0} = 1\right] \\
	& \hspace{20pt} = \mathbb{P}\left[\left. \tilde{Y}_{0}^{*} = 1 \right\vert \tilde{S}_{0} = 1, S_{1} = 1\right] \\
	& \hspace{20pt} = \mathbb{P}\left[\left. \tilde{Y}_{0}^{*} = 1, \tilde{Y}_{1}^{*} = 1 \right\vert \tilde{S}_{0} = 1, S_{1} = 1\right] + \mathbb{P}\left[\left. \tilde{Y}_{0}^{*} = 1, \tilde{Y}_{1}^{*} = 0 \right\vert \tilde{S}_{0} = 1, S_{1} = 1\right] \\
	& \hspace{20pt} = \mathbb{P}\left[\left.Y = 1 \right\vert S = 1, D = 0 \right]
\end{align*}
and
\begin{align*}
	& \mathbb{P}\left[\left. \tilde{Y} = 1 \right\vert \tilde{S} = 1, \tilde{D} = 1 \right] \\
	& \hspace{20pt} = \mathbb{P}\left[\left. \tilde{Y}_{1}^{*} = 1 \right\vert \tilde{S}_{1} = 1\right] \\
	& \hspace{20pt} = \mathbb{P}\left[\left. \tilde{Y}_{1}^{*} = 1 \right\vert \tilde{S}_{0} = 1, \tilde{S}_{1} = 1 \right] \cdot \mathbb{P}\left[\left. \tilde{S}_{0} = 1, \tilde{S}_{1} = 1 \right\vert \tilde{S}_{1} = 1 \right] \\
	& \hspace{40pt} + \mathbb{P}\left[\left. \tilde{Y}_{1}^{*} = 1 \right\vert \tilde{S}_{0} = 0, \tilde{S}_{1} = 1 \right] \cdot \left( 1 - \mathbb{P}\left[\left. \tilde{S}_{0} = 1, \tilde{S}_{1} = 1 \right\vert \tilde{S}_{1} = 1 \right] \right), \\
	& \hspace{20pt} = \left(\mathbb{P}\left[\left. \tilde{Y}_{0}^{*} = 0, \tilde{Y}_{1}^{*} = 1 \right\vert \tilde{S}_{0} = 1, \tilde{S}_{1} = 1 \right] + \mathbb{P}\left[\left. \tilde{Y}_{0}^{*} = 1, \tilde{Y}_{1}^{*} = 1 \right\vert \tilde{S}_{0} = 1, \tilde{S}_{1} = 1 \right]\right) \\
	& \hspace{60pt} \cdot \mathbb{P}\left[\left. \tilde{S}_{0} = 1, \tilde{S}_{1} = 1 \right\vert \tilde{S}_{1} = 1 \right] \\
	& \hspace{40pt} + \left(\mathbb{P}\left[\left. \tilde{Y}_{0}^{*} = 0, \tilde{Y}_{1}^{*} = 1 \right\vert \tilde{S}_{0} = 0, \tilde{S}_{1} = 1 \right] + \mathbb{P}\left[\left. \tilde{Y}_{0}^{*} = 1, \tilde{Y}_{1}^{*} = 1 \right\vert \tilde{S}_{0} = 0, \tilde{S}_{1} = 1 \right]\right) \\
	& \hspace{60pt} \cdot \left( 1 - \mathbb{P}\left[\left. \tilde{S}_{0} = 1, \tilde{S}_{1} = 1 \right\vert \tilde{S}_{1} = 1 \right] \right) \\
	& \hspace{20pt} = \min\left\lbrace \mathbb{P}\left[\left. Y = 1 \right\vert S = 1, D = 1\right] \cdot \dfrac{\mathbb{P}\left[\left. S = 1 \right\vert D = 1\right]}{\mathbb{P}\left[\left. S = 1 \right\vert D = 0\right]} , 1 \right\rbrace \cdot \dfrac{\mathbb{P}\left[\left. S = 1 \right\vert D = 0\right]}{\mathbb{P}\left[\left. S = 1 \right\vert D = 1\right]} \\
	& \hspace{40pt} + \max\left\lbrace \dfrac{\mathbb{P}\left[\left. Y = 1 \right\vert S = 1, D = 1\right] - \dfrac{\mathbb{P}\left[\left. S = 1 \right\vert D = 0\right]}{\mathbb{P}\left[\left. S = 1 \right\vert D = 1\right]}}{1 - \dfrac{\mathbb{P}\left[\left. S = 1 \right\vert D = 0\right]}{\mathbb{P}\left[\left. S = 1 \right\vert D = 1\right]}} , 0 \right\rbrace \cdot \left(1 - \dfrac{\mathbb{P}\left[\left. S = 1 \right\vert D = 0\right]}{\mathbb{P}\left[\left. S = 1 \right\vert D = 1\right]}\right) \\
	& \hspace{20pt} = \min\left\lbrace \mathbb{P}\left[\left. Y = 1 \right\vert S = 1, D = 1\right], \dfrac{\mathbb{P}\left[\left. S = 1 \right\vert D = 0\right]}{\mathbb{P}\left[\left. S = 1 \right\vert D = 1\right]} \right\rbrace \\
	& \hspace{40pt} + \max\left\lbrace \mathbb{P}\left[\left. Y = 1 \right\vert S = 1, D = 1\right] - \dfrac{\mathbb{P}\left[\left. S = 1 \right\vert D = 0\right]}{\mathbb{P}\left[\left. S = 1 \right\vert D = 1\right]} , 0 \right\rbrace \\
	& \hspace{20pt} = \mathbb{P}\left[\left. Y = 1 \right\vert S = 1, D = 1\right].
\end{align*}

Finally, note that
\begin{align*}
	& \mathbb{P}\left[\left. \tilde{Y}_{1}^{*} = 1 \right\vert \tilde{Y}_{0}^{*} = 0, \tilde{S}_{0} = 1, \tilde{S}_{1} = 1 \right] \\
	& \hspace{5pt} = \dfrac{\mathbb{P}\left[\left. \tilde{Y}_{0}^{*} = 0, \tilde{Y}_{1}^{*} = 1 \right\vert \tilde{S}_{0} = 1, \tilde{S}_{1} = 1 \right]}{\mathbb{P}\left[\left. \tilde{Y}_{0}^{*} = 0 \right\vert \tilde{S}_{0} = 1, \tilde{S}_{1} = 1 \right]} \\
	& \hspace{5pt} = \dfrac{\mathbb{P}\left[\left. \tilde{Y}_{0}^{*} = 0, \tilde{Y}_{1}^{*} = 1 \right\vert \tilde{S}_{0} = 1, \tilde{S}_{1} = 1 \right]}{\mathbb{P}\left[\left. \tilde{Y}_{0}^{*} = 0, \tilde{Y}_{1}^{*} = 1 \right\vert \tilde{S}_{0} = 1, \tilde{S}_{1} = 1 \right] + \mathbb{P}\left[\left. \tilde{Y}_{0}^{*} = 0, \tilde{Y}_{1}^{*} = 0 \right\vert \tilde{S}_{0} = 1, \tilde{S}_{1} = 1 \right]} \\
	& \hspace{5pt} = \dfrac{\mathbb{P}\left[\left. \tilde{Y}_{0}^{*} = 0, \tilde{Y}_{1}^{*} = 1 \right\vert \tilde{S}_{0} = 1, \tilde{S}_{1} = 1 \right]}{\mathbb{P}\left[\left. \tilde{Y}_{0}^{*} = 0, \tilde{Y}_{1}^{*} = 1 \right\vert \tilde{S}_{0} = 1, \tilde{S}_{1} = 1 \right] + \mathbb{P}\left[\left.Y = 0 \right\vert S = 1, D = 0 \right] - \mathbb{P}\left[\left. \tilde{Y}^{*}_{0} = 0, \tilde{Y}^{*}_{1} = 1 \right\vert \tilde{S}_{0} = 1, \tilde{S}_{1} = 1\right]} \\
	& \hspace{5pt} = \dfrac{\min \left\lbrace \mathbb{P}\left[\left. Y = 1 \right\vert S = 1, D = 1\right] \cdot \dfrac{\mathbb{P}\left[\left. S = 1 \right\vert D = 1\right]}{\mathbb{P}\left[\left. S = 1 \right\vert D = 0\right]}, \mathbb{P}\left[\left. Y = 0 \right\vert S = 1, D = 0\right] \right\rbrace}{\mathbb{P}\left[\left.Y = 0 \right\vert S = 1, D = 0 \right]} \\
	& \hspace{5pt} = UB_{1}.
\end{align*}

\textbf{(Part 3.c) Constructing a conditional distribution that attains any $\mathbf{\tilde{\theta} \in \left(LB_{1}, UB_{1}\right)}$}

Since $\tilde{\theta} \in \left(LB_{1}, UB_{1}\right)$, there exists $\omega \in \left(0, 1\right)$ such that $\tilde{\theta} = \omega \cdot LB_{1} + \left(1 - \omega\right) UB_{1}$.

Since $\mathbb{P}\left[\tilde{S}_{0} = 1, \tilde{S}_{1} = 0\right] = 0$, we do not need to define $\mathbb{P}\left[\left. \tilde{Y}^{*}_{0} = y_{0}, \tilde{Y}^{*}_{1} = y_{1} \right\vert \tilde{S}_{0} = 1, \tilde{S}_{1} = 0\right]$. We define $\mathbb{P}\left[\left. \tilde{Y}^{*}_{0} = y_{0}, \tilde{Y}^{*}_{1} = y_{1} \right\vert \tilde{S}_{0} = 0, \tilde{S}_{1} = 0\right] = \sfrac{1}{4}$  for any $\left(y_0, y_1\right) \in \{0,1\}^{2}$. We also define
\begin{align}
	& \mathbb{P}\left[\left. \tilde{Y}^{*}_{0} = 0, \tilde{Y}^{*}_{1} = 1 \right\vert \tilde{S}_{0} = 1, \tilde{S}_{1} = 1\right] \\
	& \hspace{20pt} = \omega \cdot \mathbb{P}_{L}\left[\left. \tilde{Y}^{*}_{0} = 0, \tilde{Y}^{*}_{1} = 1 \right\vert \tilde{S}_{0} = 1, \tilde{S}_{1} = 1\right] + \left(1 - \omega\right) \cdot \mathbb{P}_{U}\left[\left. \tilde{Y}^{*}_{0} = 0, \tilde{Y}^{*}_{1} = 1 \right\vert \tilde{S}_{0} = 1, \tilde{S}_{1} = 1\right], \nonumber \\
	& \mathbb{P}\left[\left. \tilde{Y}^{*}_{0} = 1, \tilde{Y}^{*}_{1} = 1 \right\vert \tilde{S}_{0} = 1, \tilde{S}_{1} = 1\right] \\
	& \hspace{20pt} = \omega \cdot \mathbb{P}_{L}\left[\left. \tilde{Y}^{*}_{0} = 1, \tilde{Y}^{*}_{1} = 1 \right\vert \tilde{S}_{0} = 1, \tilde{S}_{1} = 1\right] + \left(1 - \omega\right) \cdot \mathbb{P}_{U}\left[\left. \tilde{Y}^{*}_{0} = 1, \tilde{Y}^{*}_{1} = 1 \right\vert \tilde{S}_{0} = 1, \tilde{S}_{1} = 1\right], \nonumber \\
	& \mathbb{P}\left[\left. \tilde{Y}^{*}_{0} = 0, \tilde{Y}^{*}_{1} = 0 \right\vert \tilde{S}_{0} = 1, \tilde{S}_{1} = 1\right] \\
	& \hspace{20pt} = \omega \cdot \mathbb{P}_{L}\left[\left. \tilde{Y}^{*}_{0} = 0, \tilde{Y}^{*}_{1} = 0 \right\vert \tilde{S}_{0} = 1, \tilde{S}_{1} = 1\right] + \left(1 - \omega\right) \cdot \mathbb{P}_{U}\left[\left. \tilde{Y}^{*}_{0} = 0, \tilde{Y}^{*}_{1} = 0 \right\vert \tilde{S}_{0} = 1, \tilde{S}_{1} = 1\right], \nonumber \\
	& \mathbb{P}\left[\left. \tilde{Y}^{*}_{0} = 1, \tilde{Y}^{*}_{1} = 0 \right\vert \tilde{S}_{0} = 1, \tilde{S}_{1} = 1\right] \\
	& \hspace{20pt} = \omega \cdot \mathbb{P}_{L}\left[\left. \tilde{Y}^{*}_{0} = 1, \tilde{Y}^{*}_{1} = 0 \right\vert \tilde{S}_{0} = 1, \tilde{S}_{1} = 1\right] + \left(1 - \omega\right) \cdot \mathbb{P}_{U}\left[\left. \tilde{Y}^{*}_{0} = 1, \tilde{Y}^{*}_{1} = 0 \right\vert \tilde{S}_{0} = 1, \tilde{S}_{1} = 1\right], \nonumber \\
	& \mathbb{P}\left[\left. \tilde{Y}^{*}_{0} = 0, \tilde{Y}^{*}_{1} = 1 \right\vert \tilde{S}_{0} = 0, \tilde{S}_{1} = 1\right] \\
	& \hspace{20pt} = \omega \cdot \mathbb{P}_{L}\left[\left. \tilde{Y}^{*}_{0} = 0, \tilde{Y}^{*}_{1} = 1 \right\vert \tilde{S}_{0} = 0, \tilde{S}_{1} = 1\right] + \left(1 - \omega\right) \cdot \mathbb{P}_{U}\left[\left. \tilde{Y}^{*}_{0} = 0, \tilde{Y}^{*}_{1} = 1 \right\vert \tilde{S}_{0} = 0, \tilde{S}_{1} = 1\right], \nonumber \\
	& \mathbb{P}\left[\left. \tilde{Y}^{*}_{0} = 1, \tilde{Y}^{*}_{1} = 1 \right\vert \tilde{S}_{0} = 0, \tilde{S}_{1} = 1\right] \\
	& \hspace{20pt} = \omega \cdot \mathbb{P}_{L}\left[\left. \tilde{Y}^{*}_{0} = 1, \tilde{Y}^{*}_{1} = 1 \right\vert \tilde{S}_{0} = 0, \tilde{S}_{1} = 1\right] + \left(1 - \omega\right) \cdot \mathbb{P}_{U}\left[\left. \tilde{Y}^{*}_{0} = 1, \tilde{Y}^{*}_{1} = 1 \right\vert \tilde{S}_{0} = 0, \tilde{S}_{1} = 1\right], \nonumber \\
	& \mathbb{P}\left[\left. \tilde{Y}^{*}_{0} = 0, \tilde{Y}^{*}_{1} = 0 \right\vert \tilde{S}_{0} = 0, \tilde{S}_{1} = 1\right] \\
	& \hspace{20pt} = \omega \cdot \mathbb{P}_{L}\left[\left. \tilde{Y}^{*}_{0} = 0, \tilde{Y}^{*}_{1} = 0 \right\vert \tilde{S}_{0} = 0, \tilde{S}_{1} = 1\right] + \left(1 - \omega\right) \cdot \mathbb{P}_{U}\left[\left. \tilde{Y}^{*}_{0} = 0, \tilde{Y}^{*}_{1} = 0 \right\vert \tilde{S}_{0} = 0, \tilde{S}_{1} = 1\right], \nonumber \\
	& \mathbb{P}\left[\left. \tilde{Y}^{*}_{0} = 1, \tilde{Y}^{*}_{1} = 0 \right\vert \tilde{S}_{0} = 0, \tilde{S}_{1} = 1\right] \\
	& \hspace{20pt} = \omega \cdot \mathbb{P}_{L}\left[\left. \tilde{Y}^{*}_{0} = 1, \tilde{Y}^{*}_{1} = 0 \right\vert \tilde{S}_{0} = 0, \tilde{S}_{1} = 1\right] + \left(1 - \omega\right) \cdot \mathbb{P}_{U}\left[\left. \tilde{Y}^{*}_{0} = 1, \tilde{Y}^{*}_{1} = 0 \right\vert \tilde{S}_{0} = 0, \tilde{S}_{1} = 1\right]. \nonumber \\
\end{align}
where the sub-index $L$ denotes the conditional probabilities defined for the lower bound (Equations \eqref{EQlb1}-\eqref{EQlb8}) and the sub-index $U$ denotes the conditional probabilities defined for the upper bound (Equations \eqref{EQub1}-\eqref{EQub8}).

Notice that the data restrictions are satisfied because
\begin{align*}
	& \mathbb{P}\left[\left. \tilde{Y} = 1 \right\vert \tilde{S} = 1, \tilde{D} = d \right] \\
	& \hspace{20pt} = \omega \cdot \mathbb{P}_{L}\left[\left. \tilde{Y} = 1 \right\vert \tilde{S} = 1, \tilde{D} = d \right] + \left(1 - \omega\right) \cdot \mathbb{P}_{U}\left[\left. \tilde{Y} = 1 \right\vert \tilde{S} = 1, \tilde{D} = d \right] \\
	& \hspace{20pt} = \mathbb{P}\left[\left.Y = 1 \right\vert S = 1, D = d \right].
\end{align*}

Finally, note that
\begin{align*}
	& \mathbb{P}\left[\left. \tilde{Y}_{1}^{*} = 1 \right\vert \tilde{Y}_{0}^{*} = 0, \tilde{S}_{0} = 1, \tilde{S}_{1} = 1 \right] \\
	& \hspace{20pt} = \dfrac{\mathbb{P}\left[\left. \tilde{Y}_{0}^{*} = 0, \tilde{Y}_{1}^{*} = 1 \right\vert \tilde{S}_{0} = 1, \tilde{S}_{1} = 1 \right]}{\mathbb{P}\left[\left. \tilde{Y}_{0}^{*} = 0 \right\vert \tilde{S}_{0} = 1, \tilde{S}_{1} = 1 \right]} \\
	& \hspace{20pt} = \dfrac{\mathbb{P}\left[\left. \tilde{Y}_{0}^{*} = 0, \tilde{Y}_{1}^{*} = 1 \right\vert \tilde{S}_{0} = 1, \tilde{S}_{1} = 1 \right]}{\mathbb{P}\left[\left. \tilde{Y}_{0}^{*} = 0, \tilde{Y}_{1}^{*} = 1 \right\vert \tilde{S}_{0} = 1, \tilde{S}_{1} = 1 \right] + \mathbb{P}\left[\left. \tilde{Y}_{0}^{*} = 0, \tilde{Y}_{1}^{*} = 0 \right\vert \tilde{S}_{0} = 1, \tilde{S}_{1} = 1 \right]} \\
	& \hspace{20pt} = \dfrac{\omega \cdot \mathbb{P}_{L}\left[\left. \tilde{Y}_{0}^{*} = 0, \tilde{Y}_{1}^{*} = 1 \right\vert \tilde{S}_{0} = 1, \tilde{S}_{1} = 1 \right] + \left(1 - \omega\right) \cdot \mathbb{P}_{U}\left[\left. \tilde{Y}_{0}^{*} = 0, \tilde{Y}_{1}^{*} = 1 \right\vert \tilde{S}_{0} = 1, \tilde{S}_{1} = 1 \right]}{ \mathbb{P}\left[\left.Y = 0 \right\vert S = 1, D = 0 \right]} \\
	& \hspace{20pt} = \omega \cdot LB_{1} + \left(1 - \omega \right) \cdot UB_{1} \\
	& \hspace{20pt} = \tilde{\theta}.
\end{align*}

\subsection{Proofs of Lemmas \ref{LEboole}-\ref{LEleeuntreated}}\label{PROOFlemmas}

\subsubsection{Lemma \ref{LEboole}}

{For ease of notation, we omit from the proof that all probabilities are conditional on covariates $X$.}

For the upper bound, note that
\begin{align*}
	& \mathbb{P}\left[\left. Y_{1}^{*} = 1, Y_{0}^{*} = 0  \right\vert S_{0} = 1, S_{1} = 1 \right] \\
	& \hspace{40pt} \leq \mathbb{P}\left[\left. Y_{1}^{*} = 1, Y_{0}^{*} = 0  \right\vert S_{0} = 1, S_{1} = 1 \right] + \mathbb{P}\left[\left. Y_{1}^{*} = 1, Y_{0}^{*} = 1  \right\vert S_{0} = 1, S_{1} = 1 \right] \\
	& \hspace{40pt} = \mathbb{P}\left[\left. Y_{1}^{*} = 1 \right\vert S_{0} = 1, S_{1} = 1 \right]
\end{align*}
and
\begin{align*}
	& \mathbb{P}\left[\left. Y_{1}^{*} = 1, Y_{0}^{*} = 0  \right\vert S_{0} = 1, S_{1} = 1 \right] \\
	& \hspace{40pt} \leq \mathbb{P}\left[\left. Y_{1}^{*} = 1, Y_{0}^{*} = 0  \right\vert S_{0} = 1, S_{1} = 1 \right] + \mathbb{P}\left[\left. Y_{1}^{*} = 0, Y_{0}^{*} = 0  \right\vert S_{0} = 1, S_{1} = 1 \right] \\
	& \hspace{40pt} = \mathbb{P}\left[\left. Y_{0}^{*} = 0 \right\vert S_{0} = 1, S_{1} = 1 \right].
\end{align*}

For the lower bound, observe that
\begin{align*}
	& \mathbb{P}\left[\left. Y_{1}^{*} = 1, Y_{0}^{*} = 0  \right\vert S_{0} = 1, S_{1} = 1 \right] \\
	& \hspace{20pt} = \mathbb{P}\left[\left. Y_{1}^{*} = 1 \right\vert S_{0} = 1, S_{1} = 1 \right] + \mathbb{P}\left[\left. Y_{0}^{*} = 0 \right\vert S_{0} = 1, S_{1} = 1 \right] - \mathbb{P}\left[\left. Y_{1}^{*} = 1 \text{ or } Y_{0}^{*} = 0  \right\vert S_{0} = 1, S_{1} = 1 \right] \\
	& \hspace{20pt} \geq \mathbb{P}\left[\left. Y_{1}^{*} = 1 \right\vert S_{0} = 1, S_{1} = 1 \right] + \mathbb{P}\left[\left. Y_{0}^{*} = 0 \right\vert S_{0} = 1, S_{1} = 1 \right] - 1.
\end{align*}

\subsubsection{Lemma \ref{LEhorowitz}}

{For ease of notation, we omit from the proof that all probabilities are conditional on covariates $X$.}

Note that
\begin{align*}
	\mathbb{P}\left[\left. Y = 1 \right\vert S = 1, D = 1 \right] & = \mathbb{P}\left[\left. Y_{1}^{*} = 1 \right\vert S_{1} = 1, D = 1 \right] \\
	& = \dfrac{\mathbb{P}\left[\left. Y_{1}^{*} = 1, S_{1} = 1 \right\vert D = 1 \right]}{\mathbb{P}\left[\left. S_{1} = 1 \right\vert D = 1 \right]} \\
	& = \dfrac{\mathbb{P}\left[Y_{1}^{*} = 1, S_{1} = 1 \right]}{\mathbb{P}\left[S_{1} = 1 \right]} \text{ by Assumption \ref{ASexogeneity}} \\
	& = \mathbb{P}\left[\left. Y_{1}^{*} = 1 \right\vert S_{1} = 1 \right] \\
	& = \mathbb{P}\left[\left. Y_{1}^{*} = 1 \right\vert S_{0} = 1, S_{1} = 1 \right] \cdot \mathbb{P}\left[\left. S_{0} = 1, S_{1} = 1 \right\vert S_{1} = 1 \right] \\
	& \hspace{20pt} + \mathbb{P}\left[\left. Y_{1}^{*} = 1 \right\vert S_{0} = 0, S_{1} = 1 \right] \cdot \left( 1 - \mathbb{P}\left[\left. S_{0} = 1, S_{1} = 1 \right\vert S_{1} = 1 \right] \right),
\end{align*}
implying that
\begin{align*}
	& \mathbb{P}\left[\left. Y_{1}^{*} = 1 \right\vert S_{0} = 1, S_{1} = 1 \right] \\
	& \hspace{20pt} = \dfrac{\mathbb{P}\left[\left. Y = 1 \right\vert S = 1, D = 1 \right] - \mathbb{P}\left[\left. Y_{1}^{*} = 1 \right\vert S_{0} = 0, S_{1} = 1 \right] \cdot \left(1 - \mathbb{P}\left[\left. S_{0} = 1, S_{1} = 1 \right\vert S_{1} = 1 \right] \right)}{\mathbb{P}\left[\left. S_{0} = 1, S_{1} = 1 \right\vert S_{1} = 1 \right]}.
\end{align*}
Since $\mathbb{P}\left[\left. Y_{1}^{*} = 1 \right\vert S_{0} = 0, S_{1} = 1 \right] \in \left[0, 1\right]$, we can conclude that the bounds above hold.

\subsubsection{Lemma \ref{LEleealpha}}

{For ease of notation, we omit from the proof that all probabilities are conditional on covariates $X$.}

Note that
\begin{align*}
	\mathbb{P}\left[\left. S_{0} = 1, S_{1} = 1 \right\vert S_{1} = 1 \right] & = \dfrac{\mathbb{P}\left[S_{0} = 1, S_{1} = 1 \right]}{\mathbb{P}\left[S_{1} = 1 \right]} \\
	& = \dfrac{\mathbb{P}\left[S_{0} = 1 \right]}{\mathbb{P}\left[S_{1} = 1 \right]} \text{ by Assumption \ref{ASmonotonicity}} \\
	& = \dfrac{\mathbb{P}\left[\left.S = 1 \right\vert D = 0 \right]}{\mathbb{P}\left[\left.S = 1 \right\vert D = 1 \right]} \text{ by Assumption \ref{ASexogeneity}}.
\end{align*}

\subsubsection{Lemma \ref{LEleeuntreated}}

{For ease of notation, we omit from the proof that all probabilities are conditional on covariates $X$.}

Note that
\begin{align*}
	\mathbb{P}\left[\left. Y_{0}^{*} = 0 \right\vert S_{0} = 1, S_{1} = 1 \right] & = \dfrac{\mathbb{P}\left[Y_{0}^{*} = 0, S_{0} = 1, S_{1} = 1 \right]}{\mathbb{P}\left[S_{0} = 1, S_{1} = 1 \right]} \\
	& = \dfrac{\mathbb{P}\left[Y_{0}^{*} = 0, S_{0} = 1 \right]}{\mathbb{P}\left[S_{0} = 1 \right]} \text{ by Assumption \ref{ASmonotonicity}} \\
	& = \dfrac{\mathbb{P}\left[\left. Y = 0, S = 1 \right\vert D = 0 \right]}{\mathbb{P}\left[\left.S = 1 \right\vert D = 0 \right]} \text{ by Assumption \ref{ASexogeneity}} \\
	& = \mathbb{P}\left[\left.Y = 0 \right\vert S = 1, D = 0 \right].
\end{align*}

\subsection{Proof of Proposition \ref{PROPmonotonicityY}}\label{PROOFmonotonicityY}

{For ease of notation, we omit from the proof that all probabilities are conditional on covariates $X$.}

To prove Proposition \ref{PROPmonotonicityY}, we first show that $LB_{1} \leq \theta^{OO}$ and $\theta^{OO} \leq UB_{2}$. Then, we show that $LB_{1}$ and $UB_{2}$ are sharp bounds. For completeness, we state one lemma previously derived in the literature and used in our proofs. We prove i in Appendix \ref{PROOFlemmapoint}.

\begin{lemma}\label{LEpoint}
	\citet{Jun2019}: Under Assumption \ref{ASmonotonicityY}, we have that
	\begin{align*}
		\mathbb{P}\left[\left. Y_{1}^{*} = 1, Y_{0}^{*} = 0  \right\vert S_{0} = 1, S_{1} = 1 \right] = \mathbb{P}\left[\left. Y_{1}^{*} = 1 \right\vert S_{0} = 1, S_{1} = 1 \right] + \mathbb{P}\left[\left. Y_{0}^{*} = 0  \right\vert S_{0} = 1, S_{1} = 1 \right] - 1.
	\end{align*}
\end{lemma}

\subsubsection{Lower Bound: $LB_{1} \leq \theta^{OO}$}

Note that
\begin{align*}
	\theta^{OO} & \coloneqq \mathbb{P}\left[\left. Y_{1}^{*} = 1 \right\vert Y_{0}^{*} = 0, S_{0} = 1, S_{1} = 1 \right] \\
	& = \dfrac{\mathbb{P}\left[\left. Y_{1}^{*} = 1, Y_{0}^{*} = 0 \right\vert S_{0} = 1, S_{1} = 1\right]}{\mathbb{P}\left[\left. Y_{0}^{*} = 0 \right\vert S_{0} = 1, S_{1} = 1\right]} \\
	& = \dfrac{\mathbb{P}\left[\left. Y_{1}^{*} = 1 \right\vert S_{0} = 1, S_{1} = 1 \right] + \mathbb{P}\left[\left. Y_{0}^{*} = 0 \right\vert S_{0} = 1, S_{1} = 1 \right] - 1}{\mathbb{P}\left[\left. Y_{0}^{*} = 0 \right\vert S_{0} = 1, S_{1} = 1\right]} \\
	& \hspace{40pt} \text{ by Lemma \ref{LEpoint}} \\
	& \geq \dfrac{\dfrac{\mathbb{P}\left[\left. Y = 1 \right\vert S = 1, D = 1 \right] - \left(1 - \mathbb{P}\left[\left. S_{0} = 1, S_{1} = 1 \right\vert S_{1} = 1 \right] \right)}{\mathbb{P}\left[\left. S_{0} = 1, S_{1} = 1 \right\vert S_{1} = 1 \right]} + \mathbb{P}\left[\left. Y_{0}^{*} = 0 \right\vert S_{0} = 1, S_{1} = 1 \right] - 1}{\mathbb{P}\left[\left. Y_{0}^{*} = 0 \right\vert S_{0} = 1, S_{1} = 1\right]} \\
	& \hspace{40pt} \text{ by Lemma \ref{LEhorowitz}} \\
	& = \dfrac{\dfrac{\mathbb{P}\left[\left. Y = 1 \right\vert S = 1, D = 1 \right] - \left(1 - \dfrac{\mathbb{P}\left[\left.S = 1 \right\vert D = 0 \right]}{\mathbb{P}\left[\left.S = 1 \right\vert D = 1 \right]} \right)}{\dfrac{\mathbb{P}\left[\left.S = 1 \right\vert D = 0 \right]}{\mathbb{P}\left[\left.S = 1 \right\vert D = 1 \right]}} + \mathbb{P}\left[\left. Y_{0}^{*} = 0 \right\vert S_{0} = 1, S_{1} = 1 \right] - 1}{\mathbb{P}\left[\left. Y_{0}^{*} = 0 \right\vert S_{0} = 1, S_{1} = 1\right]} \\
	& \hspace{40pt} \text{ by Lemma \ref{LEleealpha}} \\
	& = \dfrac{\dfrac{\mathbb{P}\left[\left. Y = 1 \right\vert S = 1, D = 1 \right] - \left(1 - \dfrac{\mathbb{P}\left[\left.S = 1 \right\vert D = 0 \right]}{\mathbb{P}\left[\left.S = 1 \right\vert D = 1 \right]} \right)}{\dfrac{\mathbb{P}\left[\left.S = 1 \right\vert D = 0 \right]}{\mathbb{P}\left[\left.S = 1 \right\vert D = 1 \right]}} + \mathbb{P}\left[\left.Y = 0 \right\vert S = 1, D = 0 \right] - 1}{\mathbb{P}\left[\left.Y = 0 \right\vert S = 1, D = 0 \right]} \\
	& \hspace{40pt} \text{ by Lemma \ref{LEleeuntreated}.}
\end{align*}

Moreover, $\theta^{OO} \geq 0$ by definition.

\subsubsection{Upper Bound: $\theta^{OO} \leq UB_{2}$}

Note that
\begin{align*}
	\theta^{OO} & \coloneqq \mathbb{P}\left[\left. Y_{1}^{*} = 1 \right\vert Y_{0}^{*} = 0, S_{0} = 1, S_{1} = 1 \right] \\
	& = \dfrac{\mathbb{P}\left[\left. Y_{1}^{*} = 1, Y_{0}^{*} = 0 \right\vert S_{0} = 1, S_{1} = 1\right]}{\mathbb{P}\left[\left. Y_{0}^{*} = 0 \right\vert S_{0} = 1, S_{1} = 1\right]} \\
	& = \dfrac{\mathbb{P}\left[\left. Y_{1}^{*} = 1 \right\vert S_{0} = 1, S_{1} = 1 \right] + \mathbb{P}\left[\left. Y_{0}^{*} = 0  \right\vert S_{0} = 1, S_{1} = 1 \right] - 1}{\mathbb{P}\left[\left. Y_{0}^{*} = 0 \right\vert S_{0} = 1, S_{1} = 1\right]} \\
	& \hspace{40pt} \text{ by Lemma \ref{LEpoint}} \\
	& \leq \dfrac{\dfrac{\mathbb{P}\left[\left. Y = 1 \right\vert S = 1, D = 1 \right]}{\mathbb{P}\left[\left. S_{0} = 1, S_{1} = 1 \right\vert S_{1} = 1 \right]} + \mathbb{P}\left[\left. Y_{0}^{*} = 0  \right\vert S_{0} = 1, S_{1} = 1 \right] - 1}{\mathbb{P}\left[\left. Y_{0}^{*} = 0 \right\vert S_{0} = 1, S_{1} = 1\right]} \\
	& \hspace{40pt} \text{ by Lemma \ref{LEhorowitz}} \\
	& = \dfrac{\mathbb{P}\left[\left. Y = 1 \right\vert S = 1, D = 1 \right] \cdot \dfrac{\mathbb{P}\left[\left.S = 1 \right\vert D = 1 \right]}{\mathbb{P}\left[\left.S = 1 \right\vert D = 0 \right]} + \mathbb{P}\left[\left. Y_{0}^{*} = 0  \right\vert S_{0} = 1, S_{1} = 1 \right] - 1}{\mathbb{P}\left[\left. Y_{0}^{*} = 0 \right\vert S_{0} = 1, S_{1} = 1\right]} \\
	& \hspace{40pt} \text{ by Lemma \ref{LEleealpha}} \\
	& = \dfrac{\mathbb{P}\left[\left. Y = 1 \right\vert S = 1, D = 1 \right] \cdot \dfrac{\mathbb{P}\left[\left.S = 1 \right\vert D = 1 \right]}{\mathbb{P}\left[\left.S = 1 \right\vert D = 0 \right]} + \mathbb{P}\left[\left.Y = 0 \right\vert S = 1, D = 0 \right] - 1}{\mathbb{P}\left[\left.Y = 0 \right\vert S = 1, D = 0 \right]} \\
	& \hspace{40pt} \text{ by Lemma \ref{LEleeuntreated}.}
\end{align*}

Moreover, $\theta^{OO} \leq 1$ by definition.

\subsubsection{$LB_{1}$ and $UB_{2}$ are sharp bounds}

The only difference between this proof and the proof in Appendix \ref{PROOFmonotonicity} is the definition of $\mathbb{P}\left[\left. \tilde{Y}^{*}_{0} = y_{0}, \tilde{Y}^{*}_{1} = y_{1} \right\vert \tilde{S}_{0} = s_{0}, \tilde{S}_{1} = s_{1}\right]$ for any $\left(y_{0}, y_{1}, s_{0}, s_{1}\right) \in \left\lbrace 0, 1 \right\rbrace^{4}$. For this reason, we will only construct a conditional distribution $\left. \left(\tilde{Y}^*_0, \tilde{Y}_1^*\right) \right\vert \left(\tilde{S}_0^*, \tilde{S}_1^* \right)$ that is a probability distribution, satisfies Assumption \ref{ASmonotonicityY}, satisfies the data restrictions, and generates a probability of causation $\tilde{\theta}$ respectively equal to:
\begin{enumerate}
	\item[(a)] the lower bound $LB_{1}$;
	\item[(b)] the upper bound $UB_{2}$;
	\item[(c)] any value in the interval $(LB_{1}, UB_2)$.
\end{enumerate}

\

\textbf{(Part a) Constructing a conditional distribution such that $\mathbf{\tilde{\theta} = LB_{1}}$}

Since $\mathbb{P}\left[\tilde{S}_{0} = 1, \tilde{S}_{1} = 0\right] = 0$, we do not need to define $\mathbb{P}\left[\left. \tilde{Y}^{*}_{0} = y_{0}, \tilde{Y}^{*}_{1} = y_{1} \right\vert \tilde{S}_{0} = 1, \tilde{S}_{1} = 0\right]$. We define $\mathbb{P}\left[\left. \tilde{Y}^{*}_{0} = y_{0}, \tilde{Y}^{*}_{1} = y_{1} \right\vert \tilde{S}_{0} = 0, \tilde{S}_{1} = 0\right] = \sfrac{1}{3}$ for any $\left(y_0, y_1\right) \in \{\left(0,0\right), \left(0,1\right), \left(1,1\right)\}^{2}$ and $\mathbb{P}\left[\left. \tilde{Y}^{*}_{0} = 1, \tilde{Y}^{*}_{1} = 0 \right\vert \tilde{S}_{0} = 0, \tilde{S}_{1} = 0\right] = 0$. We also define the constant

\begin{equation*}
	\blacklozenge = \max \left\lbrace \dfrac{\mathbb{P}\left[\left.Y = 1 \right\vert S = 1, D = 1 \right] - \left(1 - \dfrac{\mathbb{P}\left[\left.S = 1 \right\vert D = 0 \right]}{\mathbb{P}\left[\left.S = 1 \right\vert D = 1 \right]} \right)}{\dfrac{\mathbb{P}\left[\left.S = 1 \right\vert D = 0 \right]}{\mathbb{P}\left[\left.S = 1 \right\vert D = 1 \right]}}, 1- \mathbb{P}\left[\left. Y=0 \right\vert S=1, D=0 \right] \right\rbrace,
\end{equation*}
and the conditional probabilities
\begin{align}
	& \mathbb{P}\left[\left. \tilde{Y}^{*}_{0} = 0, \tilde{Y}^{*}_{1} = 1 \right\vert \tilde{S}_{0} = 1, \tilde{S}_{1} = 1\right] = \blacklozenge  + \mathbb{P}\left[\left.Y = 0 \right\vert S = 1, D = 0 \right] - 1 \label{EQlb21} \\
	& \mathbb{P}\left[\left. \tilde{Y}^{*}_{0} = 1, \tilde{Y}^{*}_{1} = 1 \right\vert \tilde{S}_{0} = 1, \tilde{S}_{1} = 1\right]= 1 - \mathbb{P}\left[\left. Y=0 \right\vert S=1, D=0 \right] \label{EQlb22} \\
	& \mathbb{P}\left[\left. \tilde{Y}^{*}_{0} = 0, \tilde{Y}^{*}_{1} = 0 \right\vert \tilde{S}_{0} = 1, \tilde{S}_{1} = 1\right]  \label{EQlb23}\\
	& \hspace{20pt} = \mathbb{P}\left[\left.Y = 0 \right\vert S = 1, D = 0 \right] - \mathbb{P}\left[\left. \tilde{Y}^{*}_{0} = 0, \tilde{Y}^{*}_{1} = 1 \right\vert \tilde{S}_{0} = 1, \tilde{S}_{1} = 1\right], \nonumber \\
	& \mathbb{P}\left[\left. \tilde{Y}^{*}_{0} = 1, \tilde{Y}^{*}_{1} = 0 \right\vert \tilde{S}_{0} = 1, \tilde{S}_{1} = 1\right] = 0 \label{EQlb24} \\
	& \mathbb{P}\left[\left. \tilde{Y}^{*}_{0} = 0, \tilde{Y}^{*}_{1} = 1 \right\vert \tilde{S}_{0} = 0, \tilde{S}_{1} = 1\right] \label{EQlb25}\\
	& \hspace{20pt} = \dfrac{\mathbb{P}\left[\left. Y = 1, \right\vert S = 1, D = 1\right] - \mathbb{P}\left[\left. \tilde{Y}^{*}_{1} = 1 \right\vert \tilde{S}_{0} = 1, \tilde{S}_{1} = 1\right] \cdot \dfrac{\mathbb{P}\left[\left.S = 1 \right\vert D = 0 \right]}{\mathbb{P}\left[\left.S = 1 \right\vert D = 1 \right]}}{1 - \dfrac{\mathbb{P}\left[\left.S = 1 \right\vert D = 0 \right]}{\mathbb{P}\left[\left.S = 1 \right\vert D = 1 \right]}}, \nonumber \\
	& \mathbb{P}\left[\left. \tilde{Y}^{*}_{0} = 1, \tilde{Y}^{*}_{1} = 1 \right\vert \tilde{S}_{0} = 0, \tilde{S}_{1} = 1\right] = 0, \label{EQlb26} \\
	& \mathbb{P}\left[\left. \tilde{Y}^{*}_{0} = 0, \tilde{Y}^{*}_{1} = 0 \right\vert \tilde{S}_{0} = 0, \tilde{S}_{1} = 1\right] = 1 - \mathbb{P}\left[\left. \tilde{Y}^{*}_{0} = 0, \tilde{Y}^{*}_{1} = 1 \right\vert \tilde{S}_{0} = 0, \tilde{S}_{1} = 1\right], \label{EQlb27}\\
	& \mathbb{P}\left[\left. \tilde{Y}^{*}_{0} = 1, \tilde{Y}^{*}_{1} = 0 \right\vert \tilde{S}_{0} = 0, \tilde{S}_{1} = 1\right] = 0. \label{EQlb28}
\end{align}

Note that Equations \eqref{EQlb24} and \eqref{EQlb28} ensure that Assumption \ref{ASmonotonicityY} holds.

\textbf{(Part a.1) The candidate conditional distribution is a probability distribution}

Now, we want to show that the functions described by equations \eqref{EQlb21}-\eqref{EQlb28} are a probability mass function. First, note that:
\begin{align*}
	& \mathbb{P}\left[\left. \tilde{Y}^{*}_{0} = 0, \tilde{Y}^{*}_{1} = 1 \right\vert \tilde{S}_{0} = 1, \tilde{S}_{1} = 1\right] + \mathbb{P}\left[\left. \tilde{Y}^{*}_{0} = 1, \tilde{Y}^{*}_{1} = 1 \right\vert \tilde{S}_{0} = 1, \tilde{S}_{1} = 1\right] \\
	& \hspace{20pt} + \mathbb{P}\left[\left. \tilde{Y}^{*}_{0} = 0, \tilde{Y}^{*}_{1} = 0 \right\vert \tilde{S}_{0} = 1, \tilde{S}_{1} = 1\right] + \mathbb{P}\left[\left. \tilde{Y}^{*}_{0} = 1, \tilde{Y}^{*}_{1} = 0 \right\vert \tilde{S}_{0} = 1, \tilde{S}_{1} = 1\right] = 1
\end{align*}
and
\begin{align*}
	& \mathbb{P}\left[\left. \tilde{Y}^{*}_{0} = 0, \tilde{Y}^{*}_{1} = 1 \right\vert \tilde{S}_{0} = 0, \tilde{S}_{1} = 1\right] + \mathbb{P}\left[\left. \tilde{Y}^{*}_{0} = 1, \tilde{Y}^{*}_{1} = 1 \right\vert \tilde{S}_{0} = 0, \tilde{S}_{1} = 1\right] \\
	& \hspace{20pt} + \mathbb{P}\left[\left. \tilde{Y}^{*}_{0} = 0, \tilde{Y}^{*}_{1} = 0 \right\vert \tilde{S}_{0} = 0, \tilde{S}_{1} = 1\right] + \mathbb{P}\left[\left. \tilde{Y}^{*}_{0} = 1, \tilde{Y}^{*}_{1} = 0 \right\vert \tilde{S}_{0} = 0, \tilde{S}_{1} = 1\right] = 1.
\end{align*}

We must show that all values in \eqref{EQlb21}-\eqref{EQlb28} are in the interval $[0,1]$.

Note that $\blacklozenge \in [0,1]$ for the same reasons explained in Appendix \ref{PROOFmonotonicity}, implying that $\mathbb{P}\left[\left.  \tilde{Y}^*_0=0, \tilde{Y}_1^*=1 \right\vert \tilde{S}_0=1, \tilde{S}_1=1 \right] \in \left[0, 1\right]$. Moreover, observe that Equation \eqref{EQlb23} implies that
\begin{align*}
	\mathbb{P}\left[\left.  \tilde{Y}^*_0=0, \tilde{Y}_1^*=0 \right\vert \tilde{S}_0=1, \tilde{S}_1=1 \right] & = 1 - \blacklozenge \geq 0.
\end{align*}

tIn order to bound, $\mathbb{P}\left[\left. \tilde{Y}^*_0=0, \tilde{Y}_1^*=1 \right\vert  \tilde{S}_0=0, \tilde{S}_1=1 \right]$, note that Equations \eqref{EQlb21} and \eqref{EQlb22} imply that $\mathbb{P}\left[\left. \tilde{Y}_1^*=1 \right\vert \tilde{S}_0=1, \tilde{S}_1=1 \right]=\blacklozenge$. Consequently, Equation \eqref{EQlb25} imply that
\begin{equation*}
	\mathbb{P}\left[\tilde{Y}^*_0=0, \tilde{Y}^*_1=1| \tilde{S}_0=0, \tilde{S}_1=1\right]= \frac{\mathbb{P}\left[ Y=1| S=1, D=1\right] - \blacklozenge \cdot  \frac{\mathbb{P}\left[\left. S=1 \right\vert D=0 \right]}{\mathbb{P}\left[\left. S=1 \right\vert D=1 \right]}}{1-\frac{\mathbb{P}\left[\left. S=1 \right\vert D=0 \right]}{\mathbb{P}\left[\left. S=1 \right\vert D=1 \right]}}.
\end{equation*}

Now, consider two cases:
\begin{itemize}
	\item[Case 1)] $\blacklozenge > 1-\mathbb{P}\left[\left. Y=0 \right\vert S=1, D=0 \right]$.

	In this case, we have that
	\begin{align}
		& \mathbb{P}\left[\left. \tilde{Y}^*_0=0, \tilde{Y}^*_1=1 \right\vert \tilde{S}_0=0, \tilde{S}_1=1\right] \label{Square2} \\
		& \hspace{20pt} = \frac{\mathbb{P}\left[ Y=1| S=1, D=1\right]- \left(\mathbb{P}\left[ Y=1| S=1, D=1\right] - \left( 1- \frac{\mathbb{P}\left[S=1|D=0\right]}{\mathbb{P}\left[S=1|D=1\right]}\right) \right)}{1-\frac{\mathbb{P}\left[S=1|D=0\right]}{\mathbb{P}\left[S=1|D=1\right]}} \nonumber \\
		& \hspace{20pt}=1 \nonumber
	\end{align}

	\item[Case 2)] $\blacklozenge = 1-\mathbb{P}\left[\left. Y=0 \right\vert S=1, D=0 \right]$.

	In this case, we have that
	\begin{align}
		& \mathbb{P}\left[\left. \tilde{Y}^*_0=0, \tilde{Y}^*_1=1 \right\vert \tilde{S}_0=0, \tilde{S}_1=1\right] \label{Square2geq0} \\
		& \hspace{20pt} = \frac{\mathbb{P}\left[ Y=1| S=1, D=1\right] - \left( 1 - \mathbb{P}\left[ Y=0| S=1, D=0\right] \right) \cdot \frac{\mathbb{P}\left[S=1|D=0\right]}{\mathbb{P}\left[S=1|D=1\right]} }{1-\frac{\mathbb{P}\left[S=1|D=0\right]}{\mathbb{P}\left[S=1|D=1\right]}} \nonumber \\
		& \hspace{20pt} = \dfrac{\mathbb{P}\left[\left. Y^*_1=1 \right\vert S_1=1\right] - \mathbb{P}\left[\left. Y^*_0=1 \right\vert S_{0} = 1, S_1=1\right] \cdot \frac{\mathbb{P}\left[S=1|D=0\right]}{\mathbb{P}\left[S=1|D=1\right]} }{1-\frac{\mathbb{P}\left[S=1|D=0\right]}{\mathbb{P}\left[S=1|D=1\right]}} \nonumber \\
		& \hspace{40pt} \text{by Lemma \ref{LEleeuntreated}} \nonumber \\
		& \hspace{20pt} \propto \mathbb{P}\left[\left. Y^*_1=1 \right\vert S_1=1\right] - \mathbb{P}\left[\left. Y^*_0=1 \right\vert S_{0} = 1, S_1=1\right] \cdot \mathbb{P}\left[ \left. S_0=1, S_1=1 \right\vert S_1 = 1\right]  \nonumber \\
		& \hspace{40pt} \text{by Lemma \ref{LEleealpha}} \nonumber \\
		& \hspace{20pt} = \mathbb{P}\left[\left. Y^*_1=1 \right\vert S_{0} = 1, S_1=1\right] \cdot \mathbb{P}\left[ \left. S_0=1, S_1=1 \right\vert S_1 = 1\right] \nonumber \\
		& \hspace{40pt} + \mathbb{P}\left[\left. Y^*_1=1 \right\vert S_{0} = 0, S_1=1\right] \cdot \mathbb{P}\left[ \left. S_0=0, S_1=1 \right\vert S_1 = 1\right] \nonumber \\
		& \hspace{40pt} - \mathbb{P}\left[\left. Y^*_0=1 \right\vert S_{0} = 1, S_1=1\right] \cdot \mathbb{P}\left[ \left. S_0=1, S_1=1 \right\vert S_1 = 1\right]  \nonumber \\
		& \hspace{20pt} = \mathbb{P}\left[\left. Y_0^{*} = 0, Y^*_1=1 \right\vert S_{0} = 1, S_1=1\right] \cdot \mathbb{P}\left[ \left. S_0=1, S_1=1 \right\vert S_1 = 1\right] \nonumber \\
		& \hspace{40pt} + \mathbb{P}\left[\left. Y_0^{*} = 1, Y^*_1=1 \right\vert S_{0} = 1, S_1=1\right] \cdot \mathbb{P}\left[ \left. S_0=1, S_1=1 \right\vert S_1 = 1\right] \nonumber \\
		& \hspace{40pt} + \mathbb{P}\left[\left. Y_0^{*} = 0, Y^*_1=1 \right\vert S_{0} = 0, S_1=1\right] \cdot \mathbb{P}\left[ \left. S_0=0, S_1=1 \right\vert S_1 = 1\right] \nonumber \\
		& \hspace{40pt} + \mathbb{P}\left[\left. Y_0^{*} = 1, Y^*_1=1 \right\vert S_{0} = 0, S_1=1\right] \cdot \mathbb{P}\left[ \left. S_0=0, S_1=1 \right\vert S_1 = 1\right] \nonumber \\
		& \hspace{40pt} - \mathbb{P}\left[\left. Y^*_0=1 \right\vert S_{0} = 1, S_1=1\right] \cdot \mathbb{P}\left[ \left. S_0=1, S_1=1 \right\vert S_1 = 1\right]  \nonumber \\
		& \hspace{20pt} = \mathbb{P}\left[\left. Y_0^{*} = 0, Y^*_1=1 \right\vert S_{0} = 1, S_1=1\right] \cdot \mathbb{P}\left[ \left. S_0=1, S_1=1 \right\vert S_1 = 1\right] \nonumber \\
		& \hspace{40pt} + \mathbb{P}\left[\left. Y_0^{*} = 1, Y^*_1=1 \right\vert S_{0} = 1, S_1=1\right] \cdot \mathbb{P}\left[ \left. S_0=1, S_1=1 \right\vert S_1 = 1\right] \nonumber \\
		& \hspace{40pt} + \mathbb{P}\left[\left. Y_0^{*} = 0, Y^*_1=1 \right\vert S_{0} = 0, S_1=1\right] \cdot \mathbb{P}\left[ \left. S_0=0, S_1=1 \right\vert S_1 = 1\right] \nonumber \\
		& \hspace{40pt} + \mathbb{P}\left[\left. Y_0^{*} = 1, Y^*_1=1 \right\vert S_{0} = 0, S_1=1\right] \cdot \mathbb{P}\left[ \left. S_0=0, S_1=1 \right\vert S_1 = 1\right] \nonumber \\
		& \hspace{40pt} - \mathbb{P}\left[\left. Y^*_0=1, Y_{1}^{*} = 1 \right\vert S_{0} = 1, S_1=1\right] \cdot \mathbb{P}\left[ \left. S_0=1, S_1=1 \right\vert S_1 = 1\right]  \nonumber \\
		& \hspace{40pt} \text{by Assumption \ref{ASmonotonicityY} for the true latent variables} \nonumber \\
		& \hspace{20pt} = \mathbb{P}\left[\left. Y_0^{*} = 0, Y^*_1=1 \right\vert S_{0} = 1, S_1=1\right] \cdot \mathbb{P}\left[ \left. S_0=1, S_1=1 \right\vert S_1 = 1\right] \nonumber \\
		& \hspace{40pt} + \mathbb{P}\left[\left. Y_0^{*} = 0, Y^*_1=1 \right\vert S_{0} = 0, S_1=1\right] \cdot \mathbb{P}\left[ \left. S_0=0, S_1=1 \right\vert S_1 = 1\right] \nonumber \\
		& \hspace{40pt} + \mathbb{P}\left[\left. Y_0^{*} = 1, Y^*_1=1 \right\vert S_{0} = 0, S_1=1\right] \cdot \mathbb{P}\left[ \left. S_0=0, S_1=1 \right\vert S_1 = 1\right] \nonumber \\
		& \hspace{20pt} \geq 0
	\end{align}
	by the definition of a probability.

	Moreover, we have that
	\begin{align}
		& \mathbb{P}\left[\left. \tilde{Y}^*_0=0, \tilde{Y}^*_1=1 \right\vert \tilde{S}_0=0, \tilde{S}_1=1\right] \label{Square2leq1} \\
		& \hspace{20pt} = \frac{\mathbb{P}\left[ Y=1| S=1, D=1\right] - \left( 1 - \mathbb{P}\left[ Y=0| S=1, D=0\right] \right) \cdot \frac{\mathbb{P}\left[S=1|D=0\right]}{\mathbb{P}\left[S=1|D=1\right]} }{1-\frac{\mathbb{P}\left[S=1|D=0\right]}{\mathbb{P}\left[S=1|D=1\right]}} \nonumber \\
		& \hspace{20pt} \leq \frac{\mathbb{P}\left[ Y=1| S=1, D=1\right]- \left(\mathbb{P}\left[ Y=1| S=1, D=1\right] - \left( 1- \frac{\mathbb{P}\left[S=1|D=0\right]}{\mathbb{P}\left[S=1|D=1\right]}\right) \right)}{1-\frac{\mathbb{P}\left[S=1|D=0\right]}{\mathbb{P}\left[S=1|D=1\right]}} \nonumber \\
		& \hspace{40pt} \text{ by the definition of } \blacklozenge \nonumber \\
		& \hspace{20pt}=1. \nonumber
	\end{align}
\end{itemize}

Since $ \mathbb{P}\left[\tilde{Y}^*_0=0, \tilde{Y}^*_1=1| \tilde{S}_0=0, \tilde{S}_1=1\right] \in [0,1]$, Equation \eqref{EQlb27} ensures that
\begin{equation*}
	\mathbb{P}\left[\tilde{Y}^*_0=0, \tilde{Y}^*_1=0| \tilde{S}_0=0, \tilde{S}_1=1\right] \in [0,1].
\end{equation*}

\textbf{(Part 3.a.2) The candidate conditional distribution satisfies its data restrictions}

The data restrictions for $\left. \tilde{Y} \right\vert \tilde{S} = 1, \tilde{D}$ are satisfied because:

\begin{itemize}
	\item $\mathbb{P}\left[\left. \tilde{Y}=1 \right\vert \tilde{S}=1, \tilde{D}=0 \right]=\mathbb{P}\left[ Y=1| S=1, D=0 \right]$;

	To see that,  use Equations \eqref{EQlb22} and \eqref{EQlb24} and the fact that $\mathbb{P}\left[\tilde{S}_0=1,\tilde{S}_1=0\right]=0$ to write:
	\begin{align*}
		& \mathbb{P}\left[\left. \tilde{Y} = 1 \right\vert \tilde{S} = 1, \tilde{D} = 0 \right] \\
		& \hspace{20pt} = \mathbb{P}\left[\left. \tilde{Y}_{0}^{*} = 1 \right\vert \tilde{S}_{0} = 1\right] \\
		& \hspace{20pt} = \mathbb{P}\left[\left. \tilde{Y}_{0}^{*} = 1 \right\vert \tilde{S}_{0} = 1, \tilde{S}_{1} = 1\right] \\
		& \hspace{20pt} = \mathbb{P}\left[\left. \tilde{Y}_{0}^{*} = 1, \tilde{Y}_{1}^{*} = 1 \right\vert \tilde{S}_{0} = 1, \tilde{S}_{1} = 1\right] + \mathbb{P}\left[\left. \tilde{Y}_{0}^{*} = 1, \tilde{Y}_{1}^{*} = 0 \right\vert \tilde{S}_{0} = 1, \tilde{S}_{1} = 1\right] \\
		& \hspace{20pt} = \mathbb{P}\left[\left.Y = 1 \right\vert S = 1, D = 0 \right].
	\end{align*}

	\item $\mathbb{P}\left[\left. \tilde{Y} = 1 \right\vert \tilde{S}=1, \tilde{D}=1 \right]=\mathbb{P}\left[Y = 1|S=1, D=1\right]$.

	To see that, note that we can write:
	\begin{align*}
		& \mathbb{P}\left[\left. \tilde{Y} = 1 \right\vert \tilde{S} = 1, \tilde{D} = 1 \right] \\
		& \hspace{20pt} = \mathbb{P}\left[\left. \tilde{Y}_{1}^{*} = 1 \right\vert \tilde{S}_{1} = 1\right] \\
		& \hspace{20pt} = \mathbb{P}\left[\left. \tilde{Y}_{1}^{*} = 1 \right\vert \tilde{S}_{0} = 1, \tilde{S}_{1} = 1 \right] \cdot \mathbb{P}\left[\left. \tilde{S}_{0} = 1, \tilde{S}_{1} = 1 \right\vert \tilde{S}_{1} = 1 \right] \\
		& \hspace{40pt} + \mathbb{P}\left[\left. \tilde{Y}_{1}^{*} = 1 \right\vert \tilde{S}_{0} = 0, \tilde{S}_{1} = 1 \right] \cdot \left( 1 - \mathbb{P}\left[\left. \tilde{S}_{0} = 1, \tilde{S}_{1} = 1 \right\vert \tilde{S}_{1} = 1 \right] \right) \\
		& \hspace{20pt} = \mathbb{P}\left[\tilde{Y}^*_1=1|\tilde{S}_0=1, \tilde{S}_1=1 \right]\cdot \frac{\mathbb{P}\left[\left. S=1 \right\vert D=0 \right]}{\mathbb{P}\left[\left. S=1 \right\vert D=1 \right]} \\
		& \hspace{40pt} + \frac{\mathbb{P}\left[Y=1| S=1, D=1\right]-\mathbb{P}\left[ \tilde{Y}^*_1| \tilde{S}_0=1, \tilde{S}_1=1\right]\frac{\mathbb{P}\left[\left. S=1 \right\vert D=0 \right]}{\mathbb{P}\left[\left. S=1 \right\vert D=1 \right]}}{1-\frac{\mathbb{P}\left[\left. S=1 \right\vert D=0 \right]}{\mathbb{P}\left[\left. S=1 \right\vert D=1 \right]}} \cdot \left( 1- \frac{\mathbb{P}\left[\left. S=1 \right\vert D=0 \right]}{\mathbb{P}\left[\left. S=1 \right\vert D=1 \right]}    \right) \\
		& \hspace{20pt} = \mathbb{P}\left[\left. Y=1 \right\vert S=1, D=1 \right].
	\end{align*}

\end{itemize}

\textbf{(Part a.3)} \textbf{The probability of causation $\mathbb{\tilde{\theta}}$ reaches the lower bound $\mathbf{LB_{1}}$}

Finally, note that the lower bound $LB_{1}$ is attained because
\begin{align*}
	& \mathbb{P}\left[\left. \tilde{Y}_{1}^{*} = 1 \right\vert \tilde{Y}_{0}^{*} = 0, \tilde{S}_{0} = 1, \tilde{S}_{1} = 1 \right] \\
	& \hspace{5pt} = \dfrac{\mathbb{P}\left[\left. \tilde{Y}_{0}^{*} = 0, \tilde{Y}_{1}^{*} = 1 \right\vert \tilde{S}_{0} = 1, \tilde{S}_{1} = 1 \right]}{\mathbb{P}\left[\left. \tilde{Y}_{0}^{*} = 0 \right\vert \tilde{S}_{0} = 1, \tilde{S}_{1} = 1 \right]} \\
	& \hspace{5pt} = \dfrac{\mathbb{P}\left[\left. \tilde{Y}_{0}^{*} = 0, \tilde{Y}_{1}^{*} = 1 \right\vert \tilde{S}_{0} = 1, \tilde{S}_{1} = 1 \right]}{\mathbb{P}\left[\left. \tilde{Y}_{0}^{*} = 0, \tilde{Y}_{1}^{*} = 1 \right\vert \tilde{S}_{0} = 1, \tilde{S}_{1} = 1 \right] + \mathbb{P}\left[\left. \tilde{Y}_{0}^{*} = 0, \tilde{Y}_{1}^{*} = 0 \right\vert \tilde{S}_{0} = 1, \tilde{S}_{1} = 1 \right]} \\
	& \\
	& \hspace{5pt} = \dfrac{\mathbb{P}\left[\left. \tilde{Y}_{0}^{*} = 0, \tilde{Y}_{1}^{*} = 1 \right\vert \tilde{S}_{0} = 1, \tilde{S}_{1} = 1 \right]}{\mathbb{P}\left[\left. \tilde{Y}_{0}^{*} = 0, \tilde{Y}_{1}^{*} = 1 \right\vert \tilde{S}_{0} = 1, \tilde{S}_{1} = 1 \right] + \mathbb{P}\left[\left.Y = 0 \right\vert S = 1, D = 0 \right] - \mathbb{P}\left[\left. \tilde{Y}^{*}_{0} = 0, \tilde{Y}^{*}_{1} = 1 \right\vert \tilde{S}_{0} = 1, \tilde{S}_{1} = 1\right]} \\
	& \\
	& \hspace{5pt} = \dfrac{\left[\begin{array}{c}
			\max \left\lbrace \dfrac{\mathbb{P}\left[\left.Y = 1 \right\vert S = 1, D = 1 \right] - \left(1 - \dfrac{\mathbb{P}\left[\left.S = 1 \right\vert D = 0 \right]}{\mathbb{P}\left[\left.S = 1 \right\vert D = 1 \right]} \right)}{\dfrac{\mathbb{P}\left[\left.S = 1 \right\vert D = 0 \right]}{\mathbb{P}\left[\left.S = 1 \right\vert D = 1 \right]}}, 1- \mathbb{P}\left[\left. Y=0 \right\vert S=1, D=0 \right] \right\rbrace \\
			\\
			+ \mathbb{P}\left[\left.Y = 0 \right\vert S = 1, D = 0 \right] - 1
		\end{array}\right]}{\mathbb{P}\left[\left.Y = 0 \right\vert S = 1, D = 0 \right]} \\
	& \\
	& \hspace{5pt} = \dfrac{\max \left\lbrace \dfrac{\mathbb{P}\left[\left.Y = 1 \right\vert S = 1, D = 1 \right] - \left(1 - \dfrac{\mathbb{P}\left[\left.S = 1 \right\vert D = 0 \right]}{\mathbb{P}\left[\left.S = 1 \right\vert D = 1 \right]} \right)}{\dfrac{\mathbb{P}\left[\left.S = 1 \right\vert D = 0 \right]}{\mathbb{P}\left[\left.S = 1 \right\vert D = 1 \right]}} + \mathbb{P}\left[\left.Y = 0 \right\vert S = 1, D = 0 \right] - 1, 0 \right\rbrace}{\mathbb{P}\left[\left.Y = 0 \right\vert S = 1, D = 0 \right]} \\
	& \hspace{5pt} = LB_{1}.
\end{align*}

\textbf{(Part b) Constructing a conditional distribution such that} $\mathbf{\tilde{\theta} = UB_{2}}$\\

Since $\mathbb{P}\left[\tilde{S}_{0} = 1, \tilde{S}_{1} = 0\right] = 0$, we do not need to define $\mathbb{P}\left[\left. \tilde{Y}^{*}_{0} = y_{0}, \tilde{Y}^{*}_{1} = y_{1} \right\vert \tilde{S}_{0} = 1, \tilde{S}_{1} = 0\right]$. We define $\mathbb{P}\left[\left. \tilde{Y}^{*}_{0} = y_{0}, \tilde{Y}^{*}_{1} = y_{1} \right\vert \tilde{S}_{0} = 0, \tilde{S}_{1} = 0\right] = \sfrac{1}{3}$ for any $\left(y_0, y_1\right) \in \{\left(0,0\right), \left(0,1\right), \left(1,1\right)\}^{2}$ and $\mathbb{P}\left[\left. \tilde{Y}^{*}_{0} = 1, \tilde{Y}^{*}_{1} = 0 \right\vert \tilde{S}_{0} = 0, \tilde{S}_{1} = 0\right] = 0$. We also define:
\begin{align}
	& \mathbb{P}\left[\left. \tilde{Y}^{*}_{0} = 0, \tilde{Y}^{*}_{1} = 1 \right\vert \tilde{S}_{0} = 1, \tilde{S}_{1} = 1\right] \label{EQub21}\\
	& \hspace{20pt} = \min \left\lbrace \mathbb{P}\left[\left. Y = 1 \right\vert S = 1, D = 1\right] \cdot \dfrac{\mathbb{P}\left[\left. S = 1 \right\vert D = 1\right]}{\mathbb{P}\left[\left. S = 1 \right\vert D = 0\right]}, 1 \right\rbrace  + \mathbb{P}\left[\left. Y = 0 \right\vert S = 1, D = 0\right] - 1, \nonumber \\
	& \mathbb{P}\left[\left. \tilde{Y}^{*}_{0} = 1, \tilde{Y}^{*}_{1} = 1 \right\vert \tilde{S}_{0} = 1, \tilde{S}_{1} = 1\right] = 1 - \mathbb{P}\left[\left. Y = 0 \right\vert S = 1, D = 0\right] \\
	& \mathbb{P}\left[\left. \tilde{Y}^{*}_{0} = 0, \tilde{Y}^{*}_{1} = 0 \right\vert \tilde{S}_{0} = 1, \tilde{S}_{1} = 1\right] \\
	& \hspace{20pt} = \mathbb{P}\left[\left.Y = 0 \right\vert S = 1, D = 0 \right] - \mathbb{P}\left[\left. \tilde{Y}^{*}_{0} = 0, \tilde{Y}^{*}_{1} = 1 \right\vert \tilde{S}_{0} = 1, \tilde{S}_{1} = 1\right], \nonumber \\
	& \mathbb{P}\left[\left. \tilde{Y}^{*}_{0} = 1, \tilde{Y}^{*}_{1} = 0 \right\vert \tilde{S}_{0} = 1, \tilde{S}_{1} = 1\right] = 0 \label{EQub24} \\
	& \mathbb{P}\left[\left. \tilde{Y}^{*}_{0} = 0, \tilde{Y}^{*}_{1} = 1 \right\vert \tilde{S}_{0} = 0, \tilde{S}_{1} = 1\right] \\
	& \hspace{20pt} = \max\left\lbrace \dfrac{\mathbb{P}\left[\left. Y = 1 \right\vert S = 1, D = 1\right] - \dfrac{\mathbb{P}\left[\left. S = 1 \right\vert D = 0\right]}{\mathbb{P}\left[\left. S = 1 \right\vert D = 1\right]}}{1 - \dfrac{\mathbb{P}\left[\left. S = 1 \right\vert D = 0\right]}{\mathbb{P}\left[\left. S = 1 \right\vert D = 1\right]}} , 0 \right\rbrace, \nonumber \\
	& \mathbb{P}\left[\left. \tilde{Y}^{*}_{0} = 1, \tilde{Y}^{*}_{1} = 1 \right\vert \tilde{S}_{0} = 0, \tilde{S}_{1} = 1\right] = 0, \\
	& \mathbb{P}\left[\left. \tilde{Y}^{*}_{0} = 0, \tilde{Y}^{*}_{1} = 0 \right\vert \tilde{S}_{0} = 0, \tilde{S}_{1} = 1\right] = 1 - \mathbb{P}\left[\left. \tilde{Y}^{*}_{0} = 0, \tilde{Y}^{*}_{1} = 1 \right\vert \tilde{S}_{0} = 0, \tilde{S}_{1} = 1\right], \\
	& \mathbb{P}\left[\left. \tilde{Y}^{*}_{0} = 1, \tilde{Y}^{*}_{1} = 0 \right\vert \tilde{S}_{0} = 0, \tilde{S}_{1} = 1\right] = 0. \label{EQub28}
\end{align}

Note that Equations \eqref{EQub24} and \eqref{EQub28} ensure that Assumption \ref{ASmonotonicityY} hold.

Moreover, observe that
\begin{align*}
	& \mathbb{P}\left[\left. \tilde{Y}^{*}_{0} = 0, \tilde{Y}^{*}_{1} = 1 \right\vert \tilde{S}_{0} = 1, \tilde{S}_{1} = 1\right] \\
	& \hspace{20pt} = \min \left\lbrace \mathbb{P}\left[\left. Y = 1 \right\vert S = 1, D = 1\right] \cdot \dfrac{\mathbb{P}\left[\left. S = 1 \right\vert D = 1\right]}{\mathbb{P}\left[\left. S = 1 \right\vert D = 0\right]}, 1 \right\rbrace  + \mathbb{P}\left[\left. Y = 0 \right\vert S = 1, D = 0\right] - 1 \\
	& \hspace{20pt} \geq \mathbb{P}\left[\left. Y_{1}^{*} = 1 \right\vert S_{0} = 1, S_{1} = 1 \right]   + \mathbb{P}\left[\left. Y = 0 \right\vert S = 1, D = 0\right] - 1 \\
	& \hspace{40pt} \text{ by Lemmas \ref{LEhorowitz} and \ref{LEleealpha}} \\
	& \hspace{20pt} = \mathbb{P}\left[\left. Y_{1}^{*} = 1 \right\vert S_{0} = 1, S_{1} = 1 \right]   + \mathbb{P}\left[\left. Y_{0}^{*} = 0  \right\vert S_{0} = 1, S_{1} = 1 \right] - 1 \\
	& \hspace{40pt} \text{ by Lemma \ref{LEleeuntreated}} \\
	& \hspace{20pt} = \mathbb{P}\left[\left. Y_{1}^{*} = 1, Y_{0}^{*} = 0  \right\vert S_{0} = 1, S_{1} = 1 \right] \\
	& \hspace{40pt} \text{ by Lemma \ref{LEpoint}} \\
	& \hspace{20pt} \geq 0,
\end{align*}
and
\begin{align*}
	& \mathbb{P}\left[\left. \tilde{Y}^{*}_{0} = 0, \tilde{Y}^{*}_{1} = 0 \right\vert \tilde{S}_{0} = 1, \tilde{S}_{1} = 1\right] \\
	& \hspace{20pt} \geq \mathbb{P}\left[\left.Y = 0 \right\vert S = 1, D = 0 \right] - \mathbb{P}\left[\left.Y = 0 \right\vert S = 1, D = 0 \right] \\
	& \hspace{20pt} \geq 0,
\end{align*}
and
\begin{align*}
	& \mathbb{P}\left[\left. \tilde{Y}^{*}_{0} = 0, \tilde{Y}^{*}_{1} = 1 \right\vert \tilde{S}_{0} = 1, \tilde{S}_{1} = 1\right] + \mathbb{P}\left[\left. \tilde{Y}^{*}_{0} = 1, \tilde{Y}^{*}_{1} = 1 \right\vert \tilde{S}_{0} = 1, \tilde{S}_{1} = 1\right] \\
	& \hspace{20pt} + \mathbb{P}\left[\left. \tilde{Y}^{*}_{0} = 0, \tilde{Y}^{*}_{1} = 0 \right\vert \tilde{S}_{0} = 1, \tilde{S}_{1} = 1\right] + \mathbb{P}\left[\left. \tilde{Y}^{*}_{0} = 1, \tilde{Y}^{*}_{1} = 0 \right\vert \tilde{S}_{0} = 1, \tilde{S}_{1} = 1\right] = 1.
\end{align*}
Moreover, note that $\mathbb{P}\left[\left. \tilde{Y}^{*}_{0} = 0, \tilde{Y}^{*}_{1} = 1 \right\vert \tilde{S}_{0} = 0, \tilde{S}_{1} = 1\right] \in \left[0, 1 \right]$ by construction.

Notice also that the data restrictions are satisfied because
\begin{align*}
	& \mathbb{P}\left[\left. \tilde{Y} = 1 \right\vert \tilde{S} = 1, \tilde{D} = 0 \right] \\
	& \hspace{20pt} = \mathbb{P}\left[\left. \tilde{Y}_{0}^{*} = 1 \right\vert \tilde{S}_{0} = 1\right] \\
	& \hspace{20pt} = \mathbb{P}\left[\left. \tilde{Y}_{0}^{*} = 1 \right\vert \tilde{S}_{0} = 1, S_{1} = 1\right] \\
	& \hspace{20pt} = \mathbb{P}\left[\left. \tilde{Y}_{0}^{*} = 1, \tilde{Y}_{1}^{*} = 1 \right\vert \tilde{S}_{0} = 1, S_{1} = 1\right] + \mathbb{P}\left[\left. \tilde{Y}_{0}^{*} = 1, \tilde{Y}_{1}^{*} = 0 \right\vert \tilde{S}_{0} = 1, S_{1} = 1\right] \\
	& \hspace{20pt} = \mathbb{P}\left[\left.Y = 1 \right\vert S = 1, D = 0 \right]
\end{align*}
and
\begin{align*}
	& \mathbb{P}\left[\left. \tilde{Y} = 1 \right\vert \tilde{S} = 1, \tilde{D} = 1 \right] \\
	& \hspace{20pt} = \mathbb{P}\left[\left. \tilde{Y}_{1}^{*} = 1 \right\vert \tilde{S}_{1} = 1\right] \\
	& \hspace{20pt} = \mathbb{P}\left[\left. \tilde{Y}_{1}^{*} = 1 \right\vert \tilde{S}_{0} = 1, \tilde{S}_{1} = 1 \right] \cdot \mathbb{P}\left[\left. \tilde{S}_{0} = 1, \tilde{S}_{1} = 1 \right\vert \tilde{S}_{1} = 1 \right] \\
	& \hspace{40pt} + \mathbb{P}\left[\left. \tilde{Y}_{1}^{*} = 1 \right\vert \tilde{S}_{0} = 0, \tilde{S}_{1} = 1 \right] \cdot \left( 1 - \mathbb{P}\left[\left. \tilde{S}_{0} = 1, \tilde{S}_{1} = 1 \right\vert \tilde{S}_{1} = 1 \right] \right), \\
	& \hspace{20pt} = \left(\mathbb{P}\left[\left. \tilde{Y}_{0}^{*} = 0, \tilde{Y}_{1}^{*} = 1 \right\vert \tilde{S}_{0} = 1, \tilde{S}_{1} = 1 \right] + \mathbb{P}\left[\left. \tilde{Y}_{0}^{*} = 1, \tilde{Y}_{1}^{*} = 1 \right\vert \tilde{S}_{0} = 1, \tilde{S}_{1} = 1 \right]\right) \\
	& \hspace{60pt} \cdot \mathbb{P}\left[\left. \tilde{S}_{0} = 1, \tilde{S}_{1} = 1 \right\vert \tilde{S}_{1} = 1 \right] \\
	& \hspace{40pt} + \left(\mathbb{P}\left[\left. \tilde{Y}_{0}^{*} = 0, \tilde{Y}_{1}^{*} = 1 \right\vert \tilde{S}_{0} = 0, \tilde{S}_{1} = 1 \right] + \mathbb{P}\left[\left. \tilde{Y}_{0}^{*} = 1, \tilde{Y}_{1}^{*} = 1 \right\vert \tilde{S}_{0} = 0, \tilde{S}_{1} = 1 \right]\right) \\
	& \hspace{60pt} \cdot \left( 1 - \mathbb{P}\left[\left. \tilde{S}_{0} = 1, \tilde{S}_{1} = 1 \right\vert \tilde{S}_{1} = 1 \right] \right) \\
	& \hspace{20pt} = \min\left\lbrace \mathbb{P}\left[\left. Y = 1 \right\vert S = 1, D = 1\right] \cdot \dfrac{\mathbb{P}\left[\left. S = 1 \right\vert D = 1\right]}{\mathbb{P}\left[\left. S = 1 \right\vert D = 0\right]} , 1 \right\rbrace \cdot \dfrac{\mathbb{P}\left[\left. S = 1 \right\vert D = 0\right]}{\mathbb{P}\left[\left. S = 1 \right\vert D = 1\right]} \\
	& \hspace{40pt} + \max\left\lbrace \dfrac{\mathbb{P}\left[\left. Y = 1 \right\vert S = 1, D = 1\right] - \dfrac{\mathbb{P}\left[\left. S = 1 \right\vert D = 0\right]}{\mathbb{P}\left[\left. S = 1 \right\vert D = 1\right]}}{1 - \dfrac{\mathbb{P}\left[\left. S = 1 \right\vert D = 0\right]}{\mathbb{P}\left[\left. S = 1 \right\vert D = 1\right]}} , 0 \right\rbrace \cdot \left(1 - \dfrac{\mathbb{P}\left[\left. S = 1 \right\vert D = 0\right]}{\mathbb{P}\left[\left. S = 1 \right\vert D = 1\right]}\right) \\
	& \hspace{20pt} = \min\left\lbrace \mathbb{P}\left[\left. Y = 1 \right\vert S = 1, D = 1\right], \dfrac{\mathbb{P}\left[\left. S = 1 \right\vert D = 0\right]}{\mathbb{P}\left[\left. S = 1 \right\vert D = 1\right]} \right\rbrace \\
	& \hspace{40pt} + \max\left\lbrace \mathbb{P}\left[\left. Y = 1 \right\vert S = 1, D = 1\right] - \dfrac{\mathbb{P}\left[\left. S = 1 \right\vert D = 0\right]}{\mathbb{P}\left[\left. S = 1 \right\vert D = 1\right]} , 0 \right\rbrace \\
	& \hspace{20pt} = \mathbb{P}\left[\left. Y = 1 \right\vert S = 1, D = 1\right].
\end{align*}

Finally, note that
\begin{align*}
	& \mathbb{P}\left[\left. \tilde{Y}_{1}^{*} = 1 \right\vert \tilde{Y}_{0}^{*} = 0, \tilde{S}_{0} = 1, \tilde{S}_{1} = 1 \right] \\
	& \hspace{5pt} = \dfrac{\mathbb{P}\left[\left. \tilde{Y}_{0}^{*} = 0, \tilde{Y}_{1}^{*} = 1 \right\vert \tilde{S}_{0} = 1, \tilde{S}_{1} = 1 \right]}{\mathbb{P}\left[\left. \tilde{Y}_{0}^{*} = 0 \right\vert \tilde{S}_{0} = 1, \tilde{S}_{1} = 1 \right]} \\
	& \hspace{5pt} = \dfrac{\mathbb{P}\left[\left. \tilde{Y}_{0}^{*} = 0, \tilde{Y}_{1}^{*} = 1 \right\vert \tilde{S}_{0} = 1, \tilde{S}_{1} = 1 \right]}{\mathbb{P}\left[\left. \tilde{Y}_{0}^{*} = 0, \tilde{Y}_{1}^{*} = 1 \right\vert \tilde{S}_{0} = 1, \tilde{S}_{1} = 1 \right] + \mathbb{P}\left[\left. \tilde{Y}_{0}^{*} = 0, \tilde{Y}_{1}^{*} = 0 \right\vert \tilde{S}_{0} = 1, \tilde{S}_{1} = 1 \right]} \\
	& \hspace{5pt} = \dfrac{\mathbb{P}\left[\left. \tilde{Y}_{0}^{*} = 0, \tilde{Y}_{1}^{*} = 1 \right\vert \tilde{S}_{0} = 1, \tilde{S}_{1} = 1 \right]}{\mathbb{P}\left[\left. \tilde{Y}_{0}^{*} = 0, \tilde{Y}_{1}^{*} = 1 \right\vert \tilde{S}_{0} = 1, \tilde{S}_{1} = 1 \right] + \mathbb{P}\left[\left.Y = 0 \right\vert S = 1, D = 0 \right] - \mathbb{P}\left[\left. \tilde{Y}^{*}_{0} = 0, \tilde{Y}^{*}_{1} = 1 \right\vert \tilde{S}_{0} = 1, \tilde{S}_{1} = 1\right]} \\
	& \hspace{5pt} = \dfrac{\min \left\lbrace \mathbb{P}\left[\left. Y = 1 \right\vert S = 1, D = 1\right] \cdot \dfrac{\mathbb{P}\left[\left. S = 1 \right\vert D = 1\right]}{\mathbb{P}\left[\left. S = 1 \right\vert D = 0\right]}, 1 \right\rbrace  + \mathbb{P}\left[\left. Y = 0 \right\vert S = 1, D = 0\right] - 1}{\mathbb{P}\left[\left.Y = 0 \right\vert S = 1, D = 0 \right]} \\
	& \hspace{5pt} = UB_{2}.
\end{align*}

\textbf{(Part c) Constructing a conditional distribution that attains any $\mathbf{\tilde{\theta} \in \left(LB_{1}, UB_{2}\right)}$}

This part of the proof is identical to the proof explained in Appendix \ref{PROOFmonotonicity}.

\subsection{Proof of Lemma \ref{LEpoint}}\label{PROOFlemmapoint}

{For ease of notation, we omit from the proof that all probabilities are conditional on covariates $X$.}

Observe that
\begin{align*}
	& \mathbb{P}\left[\left. Y_{1}^{*} = 1, Y_{0}^{*} = 0  \right\vert S_{0} = 1, S_{1} = 1 \right] \\
	& \hspace{20pt} = \mathbb{P}\left[\left. Y_{1}^{*} = 1, Y_{0}^{*} = 1  \right\vert S_{0} = 1, S_{1} = 1 \right] + \mathbb{P}\left[\left. Y_{1}^{*} = 1, Y_{0}^{*} = 0  \right\vert S_{0} = 1, S_{1} = 1 \right] \\
	& \hspace{40pt} - \mathbb{P}\left[\left. Y_{1}^{*} = 1, Y_{0}^{*} = 1  \right\vert S_{0} = 1, S_{1} = 1 \right]  \\
	& \hspace{20pt} = \mathbb{P}\left[\left. Y_{1}^{*} = 1 \right\vert S_{0} = 1, S_{1} = 1 \right] - \mathbb{P}\left[\left. Y_{1}^{*} = 1, Y_{0}^{*} = 1  \right\vert S_{0} = 1, S_{1} = 1 \right] \\
	& \hspace{20pt} = \mathbb{P}\left[\left. Y_{1}^{*} = 1 \right\vert S_{0} = 1, S_{1} = 1 \right] - \mathbb{P}\left[\left. Y_{0}^{*} = 1  \right\vert S_{0} = 1, S_{1} = 1 \right] \\
	& \hspace{40pt} \text{by Assumption \ref{ASmonotonicityY}} \\
	& \hspace{20pt} = \mathbb{P}\left[\left. Y_{1}^{*} = 1 \right\vert S_{0} = 1, S_{1} = 1 \right] + \mathbb{P}\left[\left. Y_{0}^{*} = 0  \right\vert S_{0} = 1, S_{1} = 1 \right] - 1.
\end{align*}

\subsection{Proof of Proposition \ref{PROPstochastic}}\label{PROOFstochastic}

{For ease of notation, we omit from the proof that all probabilities are conditional on covariates $X$.}

To prove Proposition \ref{PROPstochastic}, we first show that $LB_{3} \leq \theta^{OO}$ and $\theta^{OO} \leq UB_{2}$. Then, we show that $LB_{3}$ and $UB_{2}$ are sharp bounds. For completeness, we state one lemma previously derived in the literature and is used in our proofs. We prove it in Appendix \ref{PROOFlemmastochastic}.

\begin{lemma}\label{LEstochastic}
	\citet{Chen2015}: Under Assumptions \ref{ASexogeneity}, \ref{ASpositive} and \ref{ASstochastic}, we have that
	\begin{align*}
		\mathbb{P}\left[\left. Y_{1}^{*} = 1 \right\vert S_{0} = 1, S_{1} = 1 \right] \geq \mathbb{P}\left[\left. Y = 1 \right\vert S = 1, D = 1 \right].
	\end{align*}
\end{lemma}

\subsubsection{Lower Bound: $LB_{3} \leq \theta^{OO}$}

Note that
\begin{align*}
	\theta^{OO} & \coloneqq \mathbb{P}\left[\left. Y_{1}^{*} = 1 \right\vert Y_{0}^{*} = 0, S_{0} = 1, S_{1} = 1 \right] \\
	& = \dfrac{\mathbb{P}\left[\left. Y_{1}^{*} = 1, Y_{0}^{*} = 0 \right\vert S_{0} = 1, S_{1} = 1\right]}{\mathbb{P}\left[\left. Y_{0}^{*} = 0 \right\vert S_{0} = 1, S_{1} = 1\right]} \\
	& = \dfrac{\mathbb{P}\left[\left. Y_{1}^{*} = 1 \right\vert S_{0} = 1, S_{1} = 1 \right] + \mathbb{P}\left[\left. Y_{0}^{*} = 0 \right\vert S_{0} = 1, S_{1} = 1 \right] - 1}{\mathbb{P}\left[\left. Y_{0}^{*} = 0 \right\vert S_{0} = 1, S_{1} = 1\right]} \\
	& \hspace{40pt} \text{ by Lemma \ref{LEpoint}} \\
	& \geq \dfrac{\mathbb{P}\left[\left. Y = 1 \right\vert S = 1, D = 1 \right] + \mathbb{P}\left[\left. Y_{0}^{*} = 0 \right\vert S_{0} = 1, S_{1} = 1 \right] - 1}{\mathbb{P}\left[\left. Y_{0}^{*} = 0 \right\vert S_{0} = 1, S_{1} = 1\right]} \\
	& \hspace{40pt} \text{ by Lemma \ref{LEstochastic}} \\
	& = \dfrac{\mathbb{P}\left[\left. Y = 1 \right\vert S = 1, D = 1 \right] + \mathbb{P}\left[\left.Y = 0 \right\vert S = 1, D = 0 \right] - 1}{\mathbb{P}\left[\left.Y = 0 \right\vert S = 1, D = 0 \right]} \\
	& \hspace{40pt} \text{ by Lemma \ref{LEleeuntreated}.}
\end{align*}

Moreover, $\theta^{OO} \geq 0$ by definition.

\subsubsection{Upper Bound: $\theta^{OO} \leq UB_{2}$}

The proof is identical to the proof explained in Appendix \ref{PROOFmonotonicityY}.

\subsubsection{$LB_{1}$ and $UB_{2}$ are sharp bounds}

The only difference between this proof and the proof in Appendix \ref{PROOFmonotonicity} is the definition of $\mathbb{P}\left[\left. \tilde{Y}^{*}_{0} = y_{0}, \tilde{Y}^{*}_{1} = y_{1} \right\vert \tilde{S}_{0} = s_{0}, \tilde{S}_{1} = s_{1}\right]$ for any $\left(y_{0}, y_{1}, s_{0}, s_{1}\right) \in \left\lbrace 0, 1 \right\rbrace^{4}$. For this reason, we will only construct a conditional distribution $\left. \left(\tilde{Y}^*_0, \tilde{Y}_1^*\right) \right\vert \left(\tilde{S}_0^*, \tilde{S}_1^* \right)$ that is a probability distribution, satisfies Assumption \ref{ASstochastic}, satisfies the data restrictions, and generates a probability of causation $\tilde{\theta}$ respectively equal to:
\begin{enumerate}
	\item[(a)] the lower bound $LB_{3}$;
	\item[(b)] the upper bound $UB_{2}$;
	\item[(c)] any value in the interval $(LB_3, UB_{2})$.
\end{enumerate}

\

\textbf{(Part a) Constructing a conditional distribution such that $\mathbf{\tilde{\theta} = LB_{3}}$}

Since $\mathbb{P}\left[\tilde{S}_{0} = 1, \tilde{S}_{1} = 0\right] = 0$, we do not need to define $\mathbb{P}\left[\left. \tilde{Y}^{*}_{0} = y_{0}, \tilde{Y}^{*}_{1} = y_{1} \right\vert \tilde{S}_{0} = 1, \tilde{S}_{1} = 0\right]$. We define $\mathbb{P}\left[\left. \tilde{Y}^{*}_{0} = y_{0}, \tilde{Y}^{*}_{1} = y_{1} \right\vert \tilde{S}_{0} = 0, \tilde{S}_{1} = 0\right] = \sfrac{1}{3}$ for any $\left(y_0, y_1\right) \in \{\left(0,0\right), \left(0,1\right), \left(1,1\right)\}^{2}$ and $\mathbb{P}\left[\left. \tilde{Y}^{*}_{0} = 1, \tilde{Y}^{*}_{1} = 0 \right\vert \tilde{S}_{0} = 0, \tilde{S}_{1} = 0\right] = 0$. We also define the conditional probabilities
\begin{align}
	& \mathbb{P}\left[\left. \tilde{Y}^{*}_{0} = 0, \tilde{Y}^{*}_{1} = 1 \right\vert \tilde{S}_{0} = 1, \tilde{S}_{1} = 1\right] \label{EQlb31} \\
	& \hspace{20pt} = \max \left\lbrace \mathbb{P}\left[\left.Y = 1 \right\vert S = 1, D = 1 \right] + \mathbb{P}\left[\left. Y=0 \right\vert S=1, D=0 \right] - 1, 0  \right\rbrace \nonumber  \\
	& \mathbb{P}\left[\left. \tilde{Y}^{*}_{0} = 1, \tilde{Y}^{*}_{1} = 1 \right\vert \tilde{S}_{0} = 1, \tilde{S}_{1} = 1\right]= 1 - \mathbb{P}\left[\left. Y=0 \right\vert S=1, D=0 \right] \label{EQlb32} \\
	& \mathbb{P}\left[\left. \tilde{Y}^{*}_{0} = 0, \tilde{Y}^{*}_{1} = 0 \right\vert \tilde{S}_{0} = 1, \tilde{S}_{1} = 1\right]  \label{EQlb33}\\
	& \hspace{20pt} = \min \left\lbrace 1 - \mathbb{P}\left[\left.Y = 1 \right\vert S = 1, D = 1 \right], \mathbb{P}\left[\left. Y=0 \right\vert S=1, D=0 \right]  \right\rbrace, \nonumber \\
	& \mathbb{P}\left[\left. \tilde{Y}^{*}_{0} = 1, \tilde{Y}^{*}_{1} = 0 \right\vert \tilde{S}_{0} = 1, \tilde{S}_{1} = 1\right] = 0 \label{EQlb34} \\
	& \mathbb{P}\left[\left. \tilde{Y}^{*}_{0} = 0, \tilde{Y}^{*}_{1} = 1 \right\vert \tilde{S}_{0} = 0, \tilde{S}_{1} = 1\right] \label{EQlb35}\\
	& \hspace{20pt} = \dfrac{\mathbb{P}\left[\left. Y = 1, \right\vert S = 1, D = 1\right] - \mathbb{P}\left[\left. \tilde{Y}^{*}_{1} = 1 \right\vert \tilde{S}_{0} = 1, \tilde{S}_{1} = 1\right] \cdot \dfrac{\mathbb{P}\left[\left.S = 1 \right\vert D = 0 \right]}{\mathbb{P}\left[\left.S = 1 \right\vert D = 1 \right]}}{1 - \dfrac{\mathbb{P}\left[\left.S = 1 \right\vert D = 0 \right]}{\mathbb{P}\left[\left.S = 1 \right\vert D = 1 \right]}}, \nonumber \\
	& \mathbb{P}\left[\left. \tilde{Y}^{*}_{0} = 1, \tilde{Y}^{*}_{1} = 1 \right\vert \tilde{S}_{0} = 0, \tilde{S}_{1} = 1\right] = 0, \label{EQlb36} \\
	& \mathbb{P}\left[\left. \tilde{Y}^{*}_{0} = 0, \tilde{Y}^{*}_{1} = 0 \right\vert \tilde{S}_{0} = 0, \tilde{S}_{1} = 1\right] = 1 - \mathbb{P}\left[\left. \tilde{Y}^{*}_{0} = 0, \tilde{Y}^{*}_{1} = 1 \right\vert \tilde{S}_{0} = 0, \tilde{S}_{1} = 1\right], \label{EQlb37}\\
	& \mathbb{P}\left[\left. \tilde{Y}^{*}_{0} = 1, \tilde{Y}^{*}_{1} = 0 \right\vert \tilde{S}_{0} = 0, \tilde{S}_{1} = 1\right] = 0. \label{EQlb38}
\end{align}

To check that Assumption \ref{ASstochastic} holds, we have to analyze two cases.

\begin{itemize}
	\item[Case 1)] $\mathbb{P}\left[\left. \tilde{Y}^{*}_{0} = 0, \tilde{Y}^{*}_{1} = 1 \right\vert \tilde{S}_{0} = 1, \tilde{S}_{1} = 1\right] > 0$

	In this case, we have that
	\begin{align*}
		& \mathbb{P}\left[\left. \tilde{Y}^{*}_{1} = 1 \right\vert \tilde{S}_{0} = 1, \tilde{S}_{1} = 1\right] \\
		& \hspace{20pt} = \mathbb{P}\left[\left. \tilde{Y}^{*}_{0} = 0, \tilde{Y}^{*}_{1} = 1 \right\vert \tilde{S}_{0} = 1, \tilde{S}_{1} = 1\right] + \mathbb{P}\left[\left. \tilde{Y}^{*}_{0} = 1, \tilde{Y}^{*}_{1} = 1 \right\vert \tilde{S}_{0} = 1, \tilde{S}_{1} = 1\right] \\
		& \hspace{20pt} = \mathbb{P}\left[\left. Y = 1 \right\vert S = 1, D = 1\right] \\
		& \hspace{20pt} = \mathbb{P}\left[\left. \tilde{Y}^{*}_{0} = 0, \tilde{Y}^{*}_{1} = 1 \right\vert \tilde{S}_{0} = 0, \tilde{S}_{1} = 1\right] + \mathbb{P}\left[\left. \tilde{Y}^{*}_{0} = 1, \tilde{Y}^{*}_{1} = 1 \right\vert \tilde{S}_{0} = 0, \tilde{S}_{1} = 1\right] \\
		& \hspace{20pt} =  \mathbb{P}\left[\left. \tilde{Y}^{*}_{1} = 1 \right\vert \tilde{S}_{0} = 0, \tilde{S}_{1} = 1\right].
	\end{align*}

	\item[Case 2)] $\mathbb{P}\left[\left. \tilde{Y}^{*}_{0} = 0, \tilde{Y}^{*}_{1} = 1 \right\vert \tilde{S}_{0} = 1, \tilde{S}_{1} = 1\right] = 0$

	In this case, we have that
	\begin{align*}
		& \mathbb{P}\left[\left. \tilde{Y}^{*}_{1} = 1 \right\vert \tilde{S}_{0} = 1, \tilde{S}_{1} = 1\right] \\
		& \hspace{20pt} = \mathbb{P}\left[\left. \tilde{Y}^{*}_{0} = 0, \tilde{Y}^{*}_{1} = 1 \right\vert \tilde{S}_{0} = 1, \tilde{S}_{1} = 1\right] + \mathbb{P}\left[\left. \tilde{Y}^{*}_{0} = 1, \tilde{Y}^{*}_{1} = 1 \right\vert \tilde{S}_{0} = 1, \tilde{S}_{1} = 1\right] \\
		& \hspace{20pt} = \mathbb{P}\left[\left. Y = 1 \right\vert S = 1, D = 0\right]
	\end{align*}
	and
	\begin{align*}
		& \mathbb{P}\left[\left. \tilde{Y}^{*}_{1} = 1 \right\vert \tilde{S}_{0} = 0, \tilde{S}_{1} = 1\right] \\
		& \hspace{20pt} = \mathbb{P}\left[\left. \tilde{Y}^{*}_{0} = 0, \tilde{Y}^{*}_{1} = 1 \right\vert \tilde{S}_{0} = 0, \tilde{S}_{1} = 1\right] + \mathbb{P}\left[\left. \tilde{Y}^{*}_{0} = 1, \tilde{Y}^{*}_{1} = 1 \right\vert \tilde{S}_{0} = 0, \tilde{S}_{1} = 1\right] \\
		& \hspace{20pt} = \dfrac{\mathbb{P}\left[\left. Y = 1, \right\vert S = 1, D = 1\right] - \mathbb{P}\left[\left. Y = 1 \right\vert S = 1, D = 0\right] \cdot \dfrac{\mathbb{P}\left[\left.S = 1 \right\vert D = 0 \right]}{\mathbb{P}\left[\left.S = 1 \right\vert D = 1 \right]}}{1 - \dfrac{\mathbb{P}\left[\left.S = 1 \right\vert D = 0 \right]}{\mathbb{P}\left[\left.S = 1 \right\vert D = 1 \right]}} \\
		& \hspace{40pt} \text{by Equation \eqref{EQlb35} and the last result},
	\end{align*}
	implying that
	\begin{align*}
		& \mathbb{P}\left[\left. \tilde{Y}^{*}_{1} = 1 \right\vert \tilde{S}_{0} = 1, \tilde{S}_{1} = 1\right] - \mathbb{P}\left[\left. \tilde{Y}^{*}_{1} = 1 \right\vert \tilde{S}_{0} = 0, \tilde{S}_{1} = 1\right] \\
		& \hspace{20pt} = \dfrac{\left(1 - \dfrac{\mathbb{P}\left[\left.S = 1 \right\vert D = 0 \right]}{\mathbb{P}\left[\left.S = 1 \right\vert D = 1 \right]}\right) \cdot \mathbb{P}\left[\left. Y = 1 \right\vert S = 1, D = 0\right]}{1 - \dfrac{\mathbb{P}\left[\left.S = 1 \right\vert D = 0 \right]}{\mathbb{P}\left[\left.S = 1 \right\vert D = 1 \right]}} \\
		& \hspace{40pt} - \dfrac{ \mathbb{P}\left[\left. Y = 1, \right\vert S = 1, D = 1\right] - \mathbb{P}\left[\left. Y = 1 \right\vert S = 1, D = 0\right] \cdot \dfrac{\mathbb{P}\left[\left.S = 1 \right\vert D = 0 \right]}{\mathbb{P}\left[\left.S = 1 \right\vert D = 1 \right]}}{1 - \dfrac{\mathbb{P}\left[\left.S = 1 \right\vert D = 0 \right]}{\mathbb{P}\left[\left.S = 1 \right\vert D = 1 \right]}} \\
		& \hspace{20pt} = \dfrac{\mathbb{P}\left[\left. Y = 1 \right\vert S = 1, D = 0\right] - \mathbb{P}\left[\left. Y = 1, \right\vert S = 1, D = 1\right]}{1 - \dfrac{\mathbb{P}\left[\left.S = 1 \right\vert D = 0 \right]}{\mathbb{P}\left[\left.S = 1 \right\vert D = 1 \right]}} \\
		& \hspace{20pt} \geq 0
	\end{align*}
	by Equation \eqref{EQlb31} and the assumption that $\mathbb{P}\left[\left. \tilde{Y}^{*}_{0} = 0, \tilde{Y}^{*}_{1} = 1 \right\vert \tilde{S}_{0} = 1, \tilde{S}_{1} = 1\right] = 0$.
\end{itemize}

\textbf{(Part a.1) The candidate conditional distribution is a probability distribution}

Now, we only have to show that $\mathbb{P}\left[\left. \tilde{Y}^{*}_{0} = 0, \tilde{Y}^{*}_{1} = 1 \right\vert \tilde{S}_{0} = 0, \tilde{S}_{1} = 1\right] \in \left[0, 1\right]$. We have to analyze two cases.

\begin{itemize}

	\item[Case 1)] $\mathbb{P}\left[\left. \tilde{Y}^{*}_{0} = 0, \tilde{Y}^{*}_{1} = 1 \right\vert \tilde{S}_{0} = 1, \tilde{S}_{1} = 1\right] > 0$

	In this case, we have that $$\mathbb{P}\left[\left. \tilde{Y}^{*}_{0} = 0,  \tilde{Y}^{*}_{1} = 1 \right\vert \tilde{S}_{0} = 0, \tilde{S}_{1} = 1\right] = \mathbb{P}\left[\left. Y = 1 \right\vert S = 1, D = 1\right] \in \left[0, 1\right]$$ according to Equations \eqref{EQlb31}, \eqref{EQlb32} and \eqref{EQlb35}.

	\item[Case 2)] $\mathbb{P}\left[\left. \tilde{Y}^{*}_{0} = 0, \tilde{Y}^{*}_{1} = 1 \right\vert \tilde{S}_{0} = 1, \tilde{S}_{1} = 1\right] = 0$

	In this case, we have that
	\begin{align*}
		& \mathbb{P}\left[\left. \tilde{Y}^{*}_{0} = 0, \tilde{Y}^{*}_{1} = 1 \right\vert \tilde{S}_{0} = 0, \tilde{S}_{1} = 1\right] \\
		& \hspace{20pt} = \dfrac{\mathbb{P}\left[\left. Y = 1 \right\vert S = 1, D = 1\right] - \mathbb{P}\left[\left. Y = 1 \right\vert S = 1, D = 0\right] \cdot \dfrac{\mathbb{P}\left[\left.S = 1 \right\vert D = 0 \right]}{\mathbb{P}\left[\left.S = 1 \right\vert D = 1 \right]}}{1 - \dfrac{\mathbb{P}\left[\left.S = 1 \right\vert D = 0 \right]}{\mathbb{P}\left[\left.S = 1 \right\vert D = 1 \right]}} \\
		& \hspace{20pt} \propto \mathbb{P}\left[\left. Y = 1 \right\vert S = 1, D = 1\right] - \mathbb{P}\left[\left. Y = 1 \right\vert S = 1, D = 0\right] \cdot \dfrac{\mathbb{P}\left[\left.S = 1 \right\vert D = 0 \right]}{\mathbb{P}\left[\left.S = 1 \right\vert D = 1 \right]} \\
		& \hspace{40pt} \text{ by Lemma \ref{LEleealpha}} \\
		& \hspace{20pt} \propto \mathbb{P}\left[\left. Y = 1 \right\vert S = 1, D = 1\right] \cdot \mathbb{P}\left[\left.S = 1 \right\vert D = 1 \right] - \mathbb{P}\left[\left. Y = 1 \right\vert S = 1, D = 0\right] \cdot \mathbb{P}\left[\left.S = 1 \right\vert D = 0 \right] \\
		& \hspace{20pt} = \mathbb{P}\left[Y_{1}^{*} = 1, S_{1} = 1 \right] - \mathbb{P}\left[Y_{0}^{*} = 1, S_{0} = 1 \right] \\
		& \hspace{40pt} \text{by Assumption \ref{ASexogeneity}} \\
		& \hspace{20pt} = \mathbb{P}\left[Y_{1}^{*} = 0, Y_{1}^{*} = 1, S_{0} = 0, S_{1} = 1 \right] + \mathbb{P}\left[Y_{1}^{*} = 1, Y_{1}^{*} = 1, S_{0} = 0, S_{1} = 1 \right] \\
		& \hspace{40pt} + \mathbb{P}\left[Y_{1}^{*} = 0, Y_{1}^{*} = 1, S_{0} = 1, S_{1} = 1 \right] + \mathbb{P}\left[Y_{1}^{*} = 1, Y_{1}^{*} = 1, S_{0} = 1, S_{1} = 1 \right] \\
		& \hspace{40pt} - \mathbb{P}\left[Y_{0}^{*} = 1, Y_{1}^{*} = 1, S_{0} = 1, S_{1} = 1 \right] \\
		& \hspace{40pt} \text{by Assumptions \ref{ASmonotonicity} and \ref{ASmonotonicityY}} \\
		& \hspace{20pt} = \mathbb{P}\left[Y_{1}^{*} = 0, Y_{1}^{*} = 1, S_{0} = 0, S_{1} = 1 \right] + \mathbb{P}\left[Y_{1}^{*} = 1, Y_{1}^{*} = 1, S_{0} = 0, S_{1} = 1 \right] \\
		& \hspace{40pt} + \mathbb{P}\left[Y_{1}^{*} = 0, Y_{1}^{*} = 1, S_{0} = 1, S_{1} = 1 \right] \\
		& \hspace{20pt} \geq 0.
	\end{align*}

	We also have that
	\begin{align*}
		& \mathbb{P}\left[\left. \tilde{Y}^{*}_{0} = 0, \tilde{Y}^{*}_{1} = 1 \right\vert \tilde{S}_{0} = 0, \tilde{S}_{1} = 1\right] \\
		& \hspace{20pt} = \dfrac{\mathbb{P}\left[\left. Y = 1 \right\vert S = 1, D = 1\right] - \mathbb{P}\left[\left. Y = 1 \right\vert S = 1, D = 0\right] \cdot \dfrac{\mathbb{P}\left[\left.S = 1 \right\vert D = 0 \right]}{\mathbb{P}\left[\left.S = 1 \right\vert D = 1 \right]}}{1 - \dfrac{\mathbb{P}\left[\left.S = 1 \right\vert D = 0 \right]}{\mathbb{P}\left[\left.S = 1 \right\vert D = 1 \right]}} \\
		& \hspace{20pt} \leq \dfrac{\mathbb{P}\left[\left. Y = 1 \right\vert S = 1, D = 1\right] - \mathbb{P}\left[\left. Y = 1 \right\vert S = 1, D = 1\right] \cdot \dfrac{\mathbb{P}\left[\left.S = 1 \right\vert D = 0 \right]}{\mathbb{P}\left[\left.S = 1 \right\vert D = 1 \right]}}{1 - \dfrac{\mathbb{P}\left[\left.S = 1 \right\vert D = 0 \right]}{\mathbb{P}\left[\left.S = 1 \right\vert D = 1 \right]}} \\
		& \hspace{20pt} = \mathbb{P}\left[\left. Y = 1, \right\vert S = 1, D = 1\right] \\
		& \hspace{20pt} \leq 1
	\end{align*}
	by Equation \eqref{EQlb31} and the assumption that $\mathbb{P}\left[\left. \tilde{Y}^{*}_{0} = 0, \tilde{Y}^{*}_{1} = 1 \right\vert \tilde{S}_{0} = 1, \tilde{S}_{1} = 1\right] = 0$.
\end{itemize}

\textbf{(Part a.2) The candidate conditional distribution satisfies its data restrictions}

This part of the proof follows the same steps of the proof explained in Appendix \ref{PROOFmonotonicityY}.

\

\textbf{(Part a.3)} \textbf{The probability of causation $\mathbb{\tilde{\theta}}$ reaches the lower bound $\mathbf{LB_3}$}

Note that the lower bound $LB_{3}$ is attained because
\begin{align*}
	& \mathbb{P}\left[\left. \tilde{Y}_{1}^{*} = 1 \right\vert \tilde{Y}_{0}^{*} = 0, \tilde{S}_{0} = 1, \tilde{S}_{1} = 1 \right] \\
	& \hspace{5pt} = \dfrac{\mathbb{P}\left[\left. \tilde{Y}_{0}^{*} = 0, \tilde{Y}_{1}^{*} = 1 \right\vert \tilde{S}_{0} = 1, \tilde{S}_{1} = 1 \right]}{\mathbb{P}\left[\left. \tilde{Y}_{0}^{*} = 0 \right\vert \tilde{S}_{0} = 1, \tilde{S}_{1} = 1 \right]} \\
	& \hspace{5pt} = \dfrac{\mathbb{P}\left[\left. \tilde{Y}_{0}^{*} = 0, \tilde{Y}_{1}^{*} = 1 \right\vert \tilde{S}_{0} = 1, \tilde{S}_{1} = 1 \right]}{\mathbb{P}\left[\left. \tilde{Y}_{0}^{*} = 0, \tilde{Y}_{1}^{*} = 1 \right\vert \tilde{S}_{0} = 1, \tilde{S}_{1} = 1 \right] + \mathbb{P}\left[\left. \tilde{Y}_{0}^{*} = 0, \tilde{Y}_{1}^{*} = 0 \right\vert \tilde{S}_{0} = 1, \tilde{S}_{1} = 1 \right]} \\
	& \hspace{5pt} = \dfrac{\max \left\lbrace \mathbb{P}\left[\left.Y = 1 \right\vert S = 1, D = 1 \right] + \mathbb{P}\left[\left. Y=0 \right\vert S=1, D=0 \right] - 1, 0  \right\rbrace}{\left[\begin{array}{c} \max \left\lbrace \mathbb{P}\left[\left.Y = 1 \right\vert S = 1, D = 1 \right] + \mathbb{P}\left[\left. Y=0 \right\vert S=1, D=0 \right] - 1, 0  \right\rbrace \\ + \min \left\lbrace 1- \mathbb{P}\left[\left.Y = 1 \right\vert S = 1, D = 1 \right], \mathbb{P}\left[\left. Y=0 \right\vert S=1, D=0 \right] \right\rbrace \end{array}\right]} \\
	& \\
	& \hspace{5pt} = \dfrac{\max \left\lbrace \mathbb{P}\left[\left.Y = 1 \right\vert S = 1, D = 1 \right] + \mathbb{P}\left[\left. Y=0 \right\vert S=1, D=0 \right] - 1, 0  \right\rbrace}{\mathbb{P}\left[\left. Y=0 \right\vert S=1, D=0 \right]} \\
	& \hspace{5pt} = LB_{3}.
\end{align*}

\

\textbf{(Part b) Constructing a conditional distribution such that} $\mathbf{\tilde{\theta} = UB_{2}}$

Here, we use the same distribution that attains the upper bound $UB_{2}$ in Appendix \ref{PROOFmonotonicityY}. For this reason, we only have to show that the distribution in  Appendix \ref{PROOFmonotonicityY} also satisfies Assumption \ref{ASstochastic}. Note that Equations \eqref{EQub21}-\eqref{EQub28} imply that
\begin{align*}
	& \mathbb{P}\left[\left. \tilde{Y}^{*}_{1} = 1 \right\vert \tilde{S}_{0} = 1, \tilde{S}_{1} = 1\right] \\
	& \hspace{20pt} = \mathbb{P}\left[\left. \tilde{Y}^{*}_{0} = 0, \tilde{Y}^{*}_{1} = 1 \right\vert \tilde{S}_{0} = 1, \tilde{S}_{1} = 1\right] + \mathbb{P}\left[\left. \tilde{Y}^{*}_{0} = 1, \tilde{Y}^{*}_{1} = 1 \right\vert \tilde{S}_{0} = 1, \tilde{S}_{1} = 1\right] \\
	& \hspace{20pt} = \min \left\lbrace \mathbb{P}\left[\left. Y = 1 \right\vert S = 1, D = 1\right] \cdot \dfrac{\mathbb{P}\left[\left. S = 1 \right\vert D = 1\right]}{\mathbb{P}\left[\left. S = 1 \right\vert D = 0\right]}, 1 \right\rbrace
\end{align*}
and
\begin{align*}
	& \mathbb{P}\left[\left. \tilde{Y}^{*}_{1} = 1 \right\vert \tilde{S}_{0} = 0, \tilde{S}_{1} = 1\right] \\
	& \hspace{20pt} = \mathbb{P}\left[\left. \tilde{Y}^{*}_{0} = 0, \tilde{Y}^{*}_{1} = 1 \right\vert \tilde{S}_{0} = 0, \tilde{S}_{1} = 1\right] + \mathbb{P}\left[\left. \tilde{Y}^{*}_{0} = 1, \tilde{Y}^{*}_{1} = 1 \right\vert \tilde{S}_{0} = 0, \tilde{S}_{1} = 1\right] \\
	& \hspace{20pt} = \max\left\lbrace \dfrac{\mathbb{P}\left[\left. Y = 1 \right\vert S = 1, D = 1\right] - \dfrac{\mathbb{P}\left[\left. S = 1 \right\vert D = 0\right]}{\mathbb{P}\left[\left. S = 1 \right\vert D = 1\right]}}{1 - \dfrac{\mathbb{P}\left[\left. S = 1 \right\vert D = 0\right]}{\mathbb{P}\left[\left. S = 1 \right\vert D = 1\right]}} , 0 \right\rbrace.
\end{align*}
Consequently, we have to analyze two cases. If $\mathbb{P}\left[\left. \tilde{Y}^{*}_{1} = 1 \right\vert \tilde{S}_{0} = 1, \tilde{S}_{1} = 1\right] < 1$, then $\mathbb{P}\left[\left. \tilde{Y}^{*}_{1} = 1 \right\vert \tilde{S}_{0} = 0, \tilde{S}_{1} = 1\right] = 0$ and Assumption \ref{ASstochastic} holds. If $\mathbb{P}\left[\left. \tilde{Y}^{*}_{1} = 1 \right\vert \tilde{S}_{0} = 1, \tilde{S}_{1} = 1\right] = 1$, then $\mathbb{P}\left[\left. \tilde{Y}^{*}_{1} = 1 \right\vert \tilde{S}_{0} = 0, \tilde{S}_{1} = 1\right] = \mathbb{P}\left[\left. \tilde{Y}^{*}_{0} = 0, \tilde{Y}^{*}_{1} = 1 \right\vert \tilde{S}_{0} = 0, \tilde{S}_{1} = 1\right] \leq 1$ according to Appendix \ref{PROOFmonotonicityY}, implying that Assumption \ref{ASstochastic} holds.

\

\textbf{(Part c) Constructing a conditional distribution that attains any $\mathbf{\tilde{\theta} \in \left(LB_{3}, UB_{2}\right)}$}

This part of the proof is identical to the proof explained in Appendix \ref{PROOFmonotonicity}.

\subsection{Proof of Lemma \ref{LEstochastic}}\label{PROOFlemmastochastic}

{For ease of notation, we omit from the proof that all probabilities are conditional on covariates $X$.}

Observe that
\begin{align*}
	& \mathbb{P}\left[\left. Y = 1 \right\vert S = 1, D = 1 \right] \\
	& \hspace{20pt} = \mathbb{P}\left[\left. Y_{1}^{*} = 1 \right\vert S_{1} = 1 \right] \\
	& \hspace{40pt} \text{by Assumption \ref{ASexogeneity}} \\
	& \hspace{20pt} = \mathbb{P}\left[\left. Y_{1}^{*} = 1 \right\vert S_{0} = 1, S_{1} = 1 \right] \cdot \mathbb{P}\left[\left. S_{0} = 1, S_{1} = 1 \right\vert S_{1} = 1 \right] \\
	& \hspace{40pt} + \mathbb{P}\left[\left. Y_{1}^{*} = 1 \right\vert S_{0} = 0, S_{1} = 1 \right] \cdot \left(1 - \mathbb{P}\left[\left. S_{0} = 1, S_{1} = 1 \right\vert S_{1} = 1 \right]\right) \\
	& \hspace{20pt} \leq \mathbb{P}\left[\left. Y_{1}^{*} = 1 \right\vert S_{0} = 1, S_{1} = 1 \right] \cdot \mathbb{P}\left[\left. S_{0} = 1, S_{1} = 1 \right\vert S_{1} = 1 \right] \\
	& \hspace{40pt} + \mathbb{P}\left[\left. Y_{1}^{*} = 1 \right\vert S_{0} = 1, S_{1} = 1 \right] \cdot \left(1 - \mathbb{P}\left[\left. S_{0} = 1, S_{1} = 1 \right\vert S_{1} = 1 \right]\right) \\
	& \hspace{40pt} \text{by Assumption \ref{ASstochastic}} \\
	& \hspace{20pt} = \mathbb{P}\left[\left. Y_{1}^{*} = 1 \right\vert S_{0} = 1, S_{1} = 1 \right].
\end{align*}

\subsection{Proof of Lemma \ref{LEcovariates}}\label{PROOFcovariates}

Fix $x \in \mathcal{X}$ arbitrarily. Observe that
\begin{align*}
\omega\left(x\right) & = \mathbb{P}\left[\left. X \right\vert Y_{0}^{*} = 0, S_{0} = 1, S_{1} = 1\right] \\
& = \dfrac{\mathbb{P}\left[ Y_{0}^{*} = 0, S_{0} = 1, S_{1} = 1, X = x\right]}{\mathbb{P}\left[Y_{0}^{*} = 0, S_{0} = 1, S_{1} = 1\right]} \\
& \hspace{20pt} \text{by the definition of a conditional probability} \\
& = \dfrac{\mathbb{P}\left[ Y_{0}^{*} = 0, S_{0} = 1, S_{1} = 1, X = x\right]}{\sum_{x \in \mathcal{X}} \mathbb{P}\left[Y_{0}^{*} = 0, S_{0} = 1, S_{1} = 1, X = x^{\prime}\right]} \\
& \hspace{20pt} \text{by the the law of total probability} \\
& = \dfrac{\mathbb{P}\left[\left. Y_{0}^{*} = 0, S_{0} = 1, S_{1} = 1 \right\vert X = x\right] \cdot \mathbb{P}\left[X = x\right]}{\sum_{x \in \mathcal{X}} \mathbb{P}\left[\left. Y_{0}^{*} = 0, S_{0} = 1, S_{1} = 1 \right\vert X = x^{\prime}\right] \cdot \mathbb{P}\left[ X = x^{\prime}\right]} \\
& \hspace{20pt} \text{by the definition of a conditional probability} \\
& = \dfrac{\mathbb{P}\left[\left. Y_{0}^{*} = 0, S_{0} = 1\right\vert X = x\right] \cdot \mathbb{P}\left[X = x\right]}{\sum_{x \in \mathcal{X}} \mathbb{P}\left[\left. Y_{0}^{*} = 0, S_{0} = 1\right\vert X = x^{\prime}\right] \cdot \mathbb{P}\left[ X = x^{\prime}\right]} \\
& \hspace{20pt} \text{by Assumption \ref{ASmonotonicity}} \\
& = \dfrac{\mathbb{P}\left[\left. Y = 0, S = 1 \right\vert D = 0, X = x \right] \cdot \mathbb{P}\left[X = x\right]}{\sum_{x^{\prime} \in \mathcal{X}} \mathbb{P}\left[\left. Y = 0, S = 1 \right\vert D = 0, X = x^{\prime} \right] \cdot \mathbb{P}\left[X = x^{\prime}\right]} \\
& \hspace{20pt} \text{by Assumption \ref{ASexogeneity}}.
\end{align*}

{

\subsection{Proof of Corollary \ref{CorIDprop2}}\label{PROOFidProp2}

For ease of notation, we omit from the proof that all probabilities are conditional on covariates $X$.

To prove this result, we have to show that
\begin{equation*}
\mathbb{P}\left[Y_{0}^{*} = 0, S_{0} = 1\right] > \mathbb{P}\left[Y_{1}^{*} = 0, S_{1} = 1 \right]
\end{equation*}
implies that $LB_{1} > 0$ and that
\begin{equation*}
\mathbb{P}\left[Y_{0}^{*} = 0, S_{0} = 1\right] > \mathbb{P}\left[Y_{1}^{*} = 1, S_{1} = 1 \right]
\end{equation*}
implies that $UB_{1} < 1$.

First, note that
\begin{align*}
& \mathbb{P}\left[Y_{0}^{*} = 0, S_{0} = 1\right] > \mathbb{P}\left[Y_{1}^{*} = 0, S_{1} = 1 \right] \\
& \hspace{5pt} \Rightarrow \mathbb{P}\left[\left. Y_{0}^{*} = 0 \right\vert S_{0} = 1\right] \cdot \mathbb{P}\left[ S_{0} = 1\right] > \mathbb{P}\left[\left. Y_{1}^{*} = 0 \right\vert S_{1} = 1 \right] \cdot \mathbb{P}\left[S_{1} = 1 \right] \\
& \hspace{20pt} \text{by the definition of a conditional probability} \\
& \hspace{5pt} \Rightarrow \mathbb{P}\left[\left. Y_{0}^{*} = 0 \right\vert S_{0} = 1\right] \cdot \dfrac{\mathbb{P}\left[ S_{0} = 1\right]}{\mathbb{P}\left[S_{1} = 1 \right]} > \mathbb{P}\left[\left. Y_{1}^{*} = 0 \right\vert S_{1} = 1 \right] \\
& \hspace{20pt} \text{because $\mathbb{P}\left[S_{1} = 1 \right] > 0$ by Assumption \ref{ASpositive}} \\
& \hspace{5pt} \Rightarrow C \cdot A > 1 - B \text{ by Assumption \ref{ASexogeneity}} \\
& \hspace{5pt} \Rightarrow B - 1 + C \cdot A > 0 \\
& \hspace{5pt} \Rightarrow \dfrac{B}{A} - \dfrac{1}{A} + C > 0 \text{ because $A > 0$ by  Assumptions \ref{ASexogeneity} and \ref{ASpositive}} \\
& \hspace{5pt} \Rightarrow \dfrac{B}{A} - \dfrac{1}{A} + 1 + C - 1 > 0 \\
& \hspace{5pt} \Rightarrow \dfrac{\left[B - \left(1 - A\right)\right] \cdot A^{-1} + C - 1}{C} > 0 \text{ because $C > 0$  by  Assumptions \ref{ASexogeneity} and \ref{ASpositive}} \\
& \hspace{5pt} \Rightarrow LB_{1} > 0.
\end{align*}

Second, observe that
\begin{align*}
& \mathbb{P}\left[Y_{0}^{*} = 0, S_{0} = 1\right] > \mathbb{P}\left[Y_{1}^{*} = 1, S_{1} = 1 \right] \\
& \hspace{5pt} \Rightarrow \mathbb{P}\left[\left. Y_{0}^{*} = 0 \right\vert S_{0} = 1\right] \cdot \mathbb{P}\left[ S_{0} = 1\right] > \mathbb{P}\left[\left. Y_{1}^{*} = 1 \right\vert S_{1} = 1 \right] \cdot \mathbb{P}\left[S_{1} = 1 \right] \\
& \hspace{20pt} \text{by the definition of a conditional probability} \\
& \hspace{5pt} \Rightarrow \mathbb{P}\left[\left. Y_{0}^{*} = 0 \right\vert S_{0} = 1\right] \cdot \dfrac{\mathbb{P}\left[ S_{0} = 1\right]}{\mathbb{P}\left[S_{1} = 1 \right]} > \mathbb{P}\left[\left. Y_{1}^{*} = 1 \right\vert S_{1} = 1 \right] \\
& \hspace{20pt} \text{because $\mathbb{P}\left[S_{1} = 1 \right] > 0$ by Assumption \ref{ASpositive}} \\
& \hspace{5pt} \Rightarrow C \cdot A > B \text{ by Assumption \ref{ASexogeneity}} \\
& \hspace{5pt} \Rightarrow \dfrac{B \cdot A^{-1}}{C} < 1 \text{ because $A > 0$ and $C > 0$  by  Assumptions \ref{ASexogeneity} and \ref{ASpositive}} \\
& \hspace{5pt} \Rightarrow UB_{1} < 1.
\end{align*}

\subsection{Proof of Corollary \ref{CorIDprop3}}\label{PROOFidProp3}

For ease of notation, we omit from the proof that all probabilities are conditional on covariates $X$.

To prove this result, it suffices to show that
\begin{equation*}
\mathbb{P}\left[\left. Y_{0}^{*} = 1, Y_{1}^{*} = 1 \right\vert S_{0} = 1, S_{1} = 1\right] > 0
\end{equation*}
implies that $UB_{2} < UB_{1}$.

Notice that
\begin{align*}
& \mathbb{P}\left[\left. Y_{0}^{*} = 1, Y_{1}^{*} = 1 \right\vert S_{0} = 1, S_{1} = 1\right] > 0 \\
& \hspace{5pt} \Rightarrow \mathbb{P}\left[\left. Y_{0}^{*} = 1 \right\vert S_{0} = 1, S_{1} = 1\right] > 0 \text{ by Assumption \ref{ASmonotonicityY}} \\
& \hspace{5pt} \Rightarrow \mathbb{P}\left[\left. Y_{0}^{*} = 0 \right\vert S_{0} = 1, S_{1} = 1\right] < 1 \\
& \hspace{5pt} \Rightarrow \mathbb{P}\left[\left. Y_{0}^{*} = 0 \right\vert S_{0} = 1\right] < 1 \text{ by Assumption \ref{ASmonotonicity}} \\
& \hspace{5pt} \Rightarrow C < 1 \text{ by Assumption \ref{ASexogeneity},}
\end{align*}
implying that $UB_{2} = \dfrac{B \cdot A^{-1} + C - 1}{C} < \dfrac{B \cdot A^{-1}}{C} = UB_{1}$.

\subsection{Proof of Corollary \ref{CorIDprop4}}\label{PROOFidProp4}

For ease of notation, we omit from the proof that all probabilities are conditional on covariates $X$.
To prove this result, it suffices to show that $$\mathbb{P}\left[S_{0} = 0, S_{1} = 1 \right] > 0$$ and $$\mathbb{P}\left[\left. Y_{0}^{*} = 0, Y_{1}^{*} = 0 \right\vert S_{1} = 1\right] > 0$$ implies that $LB_{3} > LB_{1}$.

First, observe that
\begin{align}
& \mathbb{P}\left[S_{0} = 0, S_{1} = 1 \right] > 0 \nonumber \\
& \hspace{5pt} \Rightarrow \mathbb{P}\left[S_{0} = 0, S_{1} = 1 \right] + \mathbb{P}\left[S_{0} = 1, S_{1} = 1 \right] > \mathbb{P}\left[S_{0} = 1, S_{1} = 1 \right] \nonumber \\
& \hspace{5pt} \Rightarrow \mathbb{P}\left[S_{1} = 1 \right] > \mathbb{P}\left[S_{0} = 1\right] \text{ by Assumption \ref{ASmonotonicity}} \nonumber \\
& \hspace{5pt} \Rightarrow \dfrac{\mathbb{P}\left[S_{0} = 1\right]}{\mathbb{P}\left[S_{1} = 1 \right]} < 1 \text{ because $\mathbb{P}\left[S_{1} = 1 \right] > 0$ by Assumption \ref{ASpositive}}\nonumber \\
& \hspace{5pt} \Rightarrow A < 1 \text{ by Assumption \ref{ASexogeneity}}.\label{EqAppIneq}
\end{align}

Finally, note that
\begin{align*}
& \mathbb{P}\left[\left. Y_{0}^{*} = 0, Y_{1}^{*} = 0 \right\vert S_{1} = 1\right] > 0 \\
& \hspace{5pt} \Rightarrow \mathbb{P}\left[\left. Y_{1}^{*} = 0 \right\vert S_{1} = 1\right] > 0 \text{ by the Law of Total Probability} \\
& \hspace{5pt} \Rightarrow \mathbb{P}\left[\left. Y_{1}^{*} = 1 \right\vert S_{1} = 1\right] < 1 \\
& \hspace{5pt} \Rightarrow B < 1 \text{ by Assumption \ref{ASexogeneity}} \\
& \hspace{5pt} \Rightarrow B \cdot \left(1 - A\right) < 1 - A \text{ by Inequality \eqref{EqAppIneq}} \\
& \hspace{5pt} \Rightarrow B \cdot \left(1 - A\right) \cdot A^{-1} < \left(1 - A\right)  \cdot A^{-1} \\
& \hspace{20pt} \text{ because $A > 0$ by  Assumptions \ref{ASexogeneity} and \ref{ASpositive}} \\
& \hspace{5pt} \Rightarrow B \cdot A^{-1} - B < \left(1 - A\right)  \cdot A^{-1} \\
& \hspace{5pt} \Rightarrow B \cdot A^{-1} - \left(1 - A\right)  \cdot A^{-1} < B \\
& \hspace{5pt} \Rightarrow \left[B - \left(1 - A\right) \right] \cdot A^{-1} < B \\
& \hspace{5pt} \Rightarrow \left[B - \left(1 - A\right) \right] \cdot A^{-1} + C - 1 < B + C - 1 \\
& \hspace{5pt} \Rightarrow \dfrac{\left[B - \left(1 - A\right) \right] \cdot A^{-1} + C - 1}{C} < \dfrac{B + C - 1}{C} \\
& \hspace{20pt} \text{because $C > 0$  by  Assumptions \ref{ASexogeneity} and \ref{ASpositive}} \\
& \hspace{5pt} \Rightarrow LB_{3} > LB_{1}.
\end{align*}

}

\newpage

\section{Numerical Example}\label{Sexample}
\setcounter{table}{0}
\renewcommand\thetable{B.\arabic{table}}

\setcounter{figure}{0}
\renewcommand\thefigure{B.\arabic{figure}}

\setcounter{equation}{0}
\renewcommand\theequation{B.\arabic{equation}}

\setcounter{theorem}{0}
\renewcommand\thetheorem{B.\arabic{theorem}}

\setcounter{lemma}{0}
\renewcommand\thelemma{B.\arabic{lemma}}

\setcounter{proposition}{0}
\renewcommand\theproposition{B.\arabic{proposition}}

\setcounter{corollary}{0}
\renewcommand\thecorollary{B.\arabic{corollary}}

\setcounter{assumption}{0}
\renewcommand\theassumption{B.\arabic{assumption}}

In this appendix, we use a numerical example to intuitively explain our partial identification results from Section \ref{Sident}. We focus on understanding the factors that determine the length of our bounds in each proposition and the reason why each additional assumption tightens our bounds.

{
Let our data-generating process be given by $\mathbb{P}\left[D = 1\right] = \sfrac{1}{2}$ and the conditional probability mass function described in Table \ref{TabNumerical}.

\begin{table}[!hp]
	\centering
	\caption{$\mathbb{P}\left[\left. Y_{0}^{*} = \cdot, Y_{1}^{*} = \cdot, S_{0} = \cdot, S_{1} = \cdot \right\vert D = d\right]$ for any $d \in \left\lbrace 0, 1 \right\rbrace$} \label{TabNumerical}
	\begin{lrbox}{\tablebox}
		\begin{tabular}{ccccccccccccccccccc}
			\hline
			\hline
			\multicolumn{4}{c}{Panel A:} &  & \multicolumn{4}{c}{Panel B:} &  & \multicolumn{4}{c}{Panel C:} &  & \multicolumn{4}{c}{Panel D:} \\
			\multicolumn{4}{c}{$S_{0} = 1, S_{1} = 1$} &  & \multicolumn{4}{c}{$S_{0} = 0, S_{1} = 1$} &  & \multicolumn{4}{c}{$S_{0} = 1, S_{1} = 0$} &  & \multicolumn{4}{c}{$S_{0} = 0, S_{1} = 0$} \\ \cline{1-4} \cline{6-9} \cline{11-14} \cline{16-19}
			&  & \multicolumn{2}{c}{$Y_{0}^{*} = $} &  &  &  & \multicolumn{2}{c}{$Y_{0}^{*} = $} &  &  &  & \multicolumn{2}{c}{$Y_{0}^{*} = $} &  &  &  & \multicolumn{2}{c}{$Y_{0}^{*} = $} \\ \cline{3-4} \cline{8-9} \cline{13-14} \cline{18-19}
			& \multicolumn{1}{c|}{} & \multicolumn{1}{c|}{0} & \multicolumn{1}{c|}{1} &  &  & \multicolumn{1}{c|}{} & \multicolumn{1}{c|}{0} & \multicolumn{1}{c|}{1} &  &  & \multicolumn{1}{c|}{} & \multicolumn{1}{c|}{0} & \multicolumn{1}{c|}{1} &  &  & \multicolumn{1}{c|}{} & \multicolumn{1}{c|}{0} & \multicolumn{1}{c|}{1} \\ \cline{2-4} \cline{7-9} \cline{12-14} \cline{17-19}
			\multicolumn{1}{c|}{\multirow{2}{*}{$Y_{1}^{*} = $}} & \multicolumn{1}{c|}{0} & \multicolumn{1}{c|}{$\sfrac{3}{16}$} & \multicolumn{1}{c|}{$0$} &  & \multicolumn{1}{c|}{\multirow{2}{*}{$Y_{1}^{*} = $}} & \multicolumn{1}{c|}{0} & \multicolumn{1}{c|}{$\sfrac{2}{16}$} & \multicolumn{1}{c|}{$0$} &  & \multicolumn{1}{c|}{\multirow{2}{*}{$Y_{1}^{*} = $}} & \multicolumn{1}{c|}{0} & \multicolumn{1}{c|}{$0$} & \multicolumn{1}{c|}{$0$} &  & \multicolumn{1}{c|}{\multirow{2}{*}{$Y_{1}^{*} = $}} & \multicolumn{1}{c|}{0} & \multicolumn{1}{c|}{$\sfrac{1}{16}$} & \multicolumn{1}{c|}{$0$} \\ \cline{2-4} \cline{7-9} \cline{12-14} \cline{17-19}
			\multicolumn{1}{c|}{} & \multicolumn{1}{c|}{1} & \multicolumn{1}{c|}{$\sfrac{4}{16}$} & \multicolumn{1}{c|}{$\sfrac{2}{16}$} &  & \multicolumn{1}{c|}{} & \multicolumn{1}{c|}{1} & \multicolumn{1}{c|}{$\sfrac{1}{16}$} & \multicolumn{1}{c|}{$\sfrac{1}{16}$} &  & \multicolumn{1}{c|}{} & \multicolumn{1}{c|}{1} & \multicolumn{1}{c|}{$0$} & \multicolumn{1}{c|}{$0$} &  & \multicolumn{1}{c|}{} & \multicolumn{1}{c|}{1} & \multicolumn{1}{c|}{$\sfrac{1}{16}$} & \multicolumn{1}{c|}{$\sfrac{1}{16}$} \\ \cline{2-4} \cline{7-9} \cline{12-14} \cline{17-19}
			&  &  &  &  &  &  &  &  &  &  &  &  &  &  &  &  &  &  \\ \hline
		\end{tabular}
	\end{lrbox}
	\usebox{\tablebox}\\
	\settowidth{\tableboxwidth}{\usebox{\tablebox}} \parbox{\tableboxwidth}{\footnotesize{\textit{Notes}: Each cell reports $\mathbb{P}\left[\left. Y_{0}^{*} = y_{0}, Y_{1}^{*} = y_{1}, S_{0} = s_{0}, S_{1} = s_{1} \right\vert D = d\right]$ for the values $s_{0}$ and $s_{1}$ described in the panels, the value $y_{0}$ described in the columns and the value of $y_{1}$ described in the rows.}}
\end{table}

}

Note that this data-generating process satisfies Assumptions \ref{ASexogeneity}-\ref{ASmonotonicityY} by construction. Observe also that $\mathbb{P}\left[\left. Y_{1}^{*} = 1 \right\vert S_{0} = 1, S_{1} = 1\right] = \sfrac{2}{3}$ and $\mathbb{P}\left[\left. Y_{1}^{*} = 1 \right\vert S_{0} = 0, S_{1} = 1\right] = \sfrac{1}{2}$, implying that Assumption \ref{ASstochastic} is valid too.

Finally, notice that our target parameter --- the probability of causation for the always-employed --- is given by $$\theta^{OO} = \mathbb{P}\left[\left. Y_{1}^{*} = 1 \right\vert Y_{0}^{*} = 0, S_{0} = 1, S_{1} = 1\right] \approx 0.571.$$

Now, we carefully derive our bounds to understand the factors determining the length of our bounds in each proposition and why each additional assumption tightens our bounds.

To understand the intuition behind Proposition \ref{PROPmonotonicity}, note that
\begin{align*}
\theta^{OO} & = \mathbb{P}\left[\left. Y_{1}^{*} = 1 \right\vert Y_{0}^{*} = 0, S_{0} = 1, S_{1} = 1\right] \\
& = \dfrac{\mathbb{P}\left[\left. Y_{0}^{*} = 0, Y_{1}^{*} = 1 \right\vert S_{0} = 1, S_{1} = 1\right]}{\mathbb{P}\left[\left. Y_{0}^{*} = 0 \right\vert, S_{0} = 1, S_{1} = 1\right]}.
\end{align*}
Since the denominator is point-identified by $\mathbb{P}\left[\left.Y = 0 \right\vert S = 1, D = 0 \right]$ (Lemma \ref{LEleeuntreated}), we have that
\begin{equation}\label{EQex1}
\theta^{OO} = \dfrac{\mathbb{P}\left[\left. Y_{0}^{*} = 0, Y_{1}^{*} = 1 \right\vert S_{0} = 1, S_{1} = 1\right]}{\mathbb{P}\left[\left.Y = 0 \right\vert S = 1, D = 0 \right]}.
\end{equation}

We want to bound the numerator in Equation \eqref{EQex1} using information from the marginal distributions of $\left. Y_{0}^{*} \right\vert \left( S_{0} = 1, S_{1} = 1 \right)$ and $\left. Y_{1}^{*} \right\vert \left( S_{0} = 1, S_{1} = 1 \right)$. To do so, we use the Boole-Frechet inequalities (Lemma \ref{LEboole}) and find that
\begin{equation*}
\theta^{OO} \leq \dfrac{\min \left\lbrace \mathbb{P}\left[\left. Y_{1}^{*} = 1 \right\vert S_{0} = 1, S_{1} = 1\right], \mathbb{P}\left[\left. Y_{0}^{*} = 0\right\vert S_{0} = 1, S_{1} = 1\right] \right\rbrace}{\mathbb{P}\left[\left.Y = 0 \right\vert S = 1, D = 0 \right]}
\end{equation*}
and that
\begin{equation}\label{EQex2}
\theta^{OO} \geq \dfrac{\mathbb{P}\left[\left. Y_{1}^{*} = 1 \right\vert S_{0} = 1, S_{1} = 1\right] + \mathbb{P}\left[\left. Y_{0}^{*} = 0\right\vert S_{0} = 1, S_{1} = 1\right] - 1}{\mathbb{P}\left[\left.Y = 0 \right\vert S = 1, D = 0 \right]}.
\end{equation}

Note, once more, that $\mathbb{P}\left[\left. Y_{0}^{*} = 0\right\vert S_{0} = 1, S_{1} = 1\right]$ is point-identified by $\mathbb{P}\left[\left.Y = 0 \right\vert S = 1, D = 0 \right]$ (Lemma \ref{LEleeuntreated}), implying that
\begin{equation*}
\theta^{OO} \leq \min \left\lbrace \dfrac{  \mathbb{P}\left[\left. Y_{1}^{*} = 1 \right\vert S_{0} = 1, S_{1} = 1\right] }{\mathbb{P}\left[\left.Y = 0 \right\vert S = 1, D = 0 \right]}, 1 \right\rbrace
\end{equation*}
and that
\begin{equation}\label{EQex3}
\theta^{OO} \geq \dfrac{\mathbb{P}\left[\left. Y_{1}^{*} = 1 \right\vert S_{0} = 1, S_{1} = 1\right] + \mathbb{P}\left[\left.Y = 0 \right\vert S = 1, D = 0 \right] - 1}{\mathbb{P}\left[\left.Y = 0 \right\vert S = 1, D = 0 \right]}.
\end{equation}

Now, we address the sample selection issue in the term $\mathbb{P}\left[\left. Y_{1}^{*} = 1 \right\vert S_{0} = 1, S_{1} = 1\right]$. To do so, we use the trimming bounds proposed by \cite{Horowitz1995} and \cite{Lee2009} (Lemma \ref{LEhorowitz}) and find that
\begin{equation*}
\theta^{OO} \leq \min \left\lbrace \dfrac{  \dfrac{\mathbb{P}\left[\left.Y = 1 \right\vert S = 1, D = 1 \right]}{\mathbb{P}\left[\left. S_{0} = 1, S_{1} = 1 \right\vert S_{1} = 1\right]} }{\mathbb{P}\left[\left.Y = 0 \right\vert S = 1, D = 0 \right]}, 1 \right\rbrace
\end{equation*}
and that
\begin{equation}\label{EQex4}
\theta^{OO} \geq \dfrac{\dfrac{\mathbb{P}\left[\left.Y = 1 \right\vert S = 1, D = 1 \right]  - \left(1 - \mathbb{P}\left[\left. S_{0} = 1, S_{1} = 1 \right\vert S_{1} = 1\right]\right)}{\mathbb{P}\left[\left. S_{0} = 1, S_{1} = 1 \right\vert S_{1} = 1\right]} + \mathbb{P}\left[\left.Y = 0 \right\vert S = 1, D = 0 \right] - 1}{\mathbb{P}\left[\left.Y = 0 \right\vert S = 1, D = 0 \right]}.
\end{equation}

The last two inequalities illustrate the first factor that intuitively explains the length of our bounds. Observe that the upper bound is smaller and the lower bound is greater if the share of the always-employed among the ones who are employed when treated $\left(\mathbb{P}\left[\left. S_{0} = 1, S_{1} = 1 \right\vert S_{1} = 1\right]\right)$ is large.

Finally, to derive the last expression of the bounds in Proposition \ref{PROPmonotonicity}, we use Assumption \ref{ASmonotonicity} to pointy identify $\mathbb{P}\left[\left. S_{0} = 1, S_{1} = 1 \right\vert S_{1} = 1\right]$ (Lemma \ref{LEleealpha}). Applying the analytic expressions from Proposition \ref{PROPmonotonicity}, our data-generating process implies that $LB_{1} \approx 0.286$ and $UB_{1} = 1$.

Now, we focus on the bounds in Proposition \ref{PROPmonotonicityY}. Since $UB_{2} \leq UB_{1}$, we want to understand why Assumption \ref{ASmonotonicityY} can reduce the upper bound around the target parameter. Using the Monotone Treatment Response Assumption, the joint probability $\mathbb{P}\left[\left. Y_{0}^{*} = 0, Y_{1}^{*} = 1 \right\vert S_{0} = 1, S_{1} = 1\right]$ is equal to $\mathbb{P}\left[\left. Y_{1}^{*} = 1 \right\vert S_{0} = 1, S_{1} = 1\right] + \mathbb{P}\left[\left. Y_{0}^{*} = 0\right\vert S_{0} = 1, S_{1} = 1\right] - 1$ (Lemma \ref{LEpoint}). Combining this result with Equation \eqref{EQex1}, we find that
\begin{equation}\label{EQex5}
\theta^{OO} = \dfrac{\mathbb{P}\left[\left. Y_{1}^{*} = 1 \right\vert S_{0} = 1, S_{1} = 1\right] + \mathbb{P}\left[\left. Y_{0}^{*} = 0\right\vert S_{0} = 1, S_{1} = 1\right] - 1}{\mathbb{P}\left[\left.Y = 0 \right\vert S = 1, D = 0 \right]}.
\end{equation}
Since the right-hand side term in Equation \eqref{EQex5} is equal to the lower bound in Inequality \eqref{EQex2}, we can conclude that the upper bound in Proposition \ref{PROPmonotonicityY} is less than or equal to the upper bound in Proposition \ref{PROPmonotonicity}. This result intuitively explains the identifying power of Assumption \ref{ASmonotonicityY}.

Now, to derive the last expression of the bounds in Proposition \ref{PROPmonotonicityY}, we follow the same steps used to derive the bounds in Proposition \ref{PROPmonotonicity}. Finally, applying the analytic expressions from Proposition \ref{PROPmonotonicityY}, our data-generating process implies that $LB_{1} \approx 0.286$ and $UB_{2} \approx 0.857$, numerically illustrating that Assumption \ref{ASmonotonicityY} reduces the upper bound substantially.

To conclude this section, we focus on the bounds in Proposition \ref{PROPstochastic}. Since $LB_{3} \geq LB_{1}$, we want to understand why Assumption \ref{ASstochastic} can increase the lower bound around the target parameter. To do so, we return to Inequality \eqref{EQex3}. Since $\mathbb{P}\left[\left. Y_{1}^{*} = 1 \right\vert S_{0} = 1, S_{1} = 1 \right] \geq \mathbb{P}\left[\left. Y = 1 \right\vert S = 1, D = 1 \right]$ due to the stochastic dominance assumption (Lemma \ref{LEstochastic}), there is no need to use the trimming bounds in Inequality \eqref{EQex4}. Consequently, we have that $$\theta^{OO} \geq \dfrac{\mathbb{P}\left[\left.Y = 1 \right\vert S = 1, D = 1 \right] + \mathbb{P}\left[\left.Y = 0 \right\vert S = 1, D = 0 \right] - 1}{\mathbb{P}\left[\left.Y = 0 \right\vert S = 1, D = 0 \right]},$$ which is greater than the expression in Inequality \eqref{EQex4} and the lower bound in Proposition \ref{PROPmonotonicityY}. This result intuitively explains the identifying power of Assumption \ref{ASstochastic}.

Finally, applying the analytic expressions from Proposition \ref{PROPstochastic}, our data-generating process implies that $LB_{3} \approx 0.505$ and $UB_{2} \approx 0.857$, numerically illustrating that Assumption \ref{ASstochastic} increases the lower bound substantially. Importantly, our shortest identified interval contains the target parameter and is not wide.

We can also compare our identified bounds against an estimand that would identify the probability of causation if Assumptions \ref{ASexogeneity}-\ref{ASmonotonicityY} were valid and all agents were observed $\left(\mathbb{P}\left[S_{0} = 1, S_{1} = 1\right] = 1\right)$. In this case, the probability of causation would be point-identified by the lower bound $LB_{3}$ in Proposition \ref{PROPstochastic}. If we ignored sample selection and used this estimand, we would underestimate the true probability of causation for the always-employed in this numerical example.

\newpage

\section{Detailed Discussion on the Testable Restrictions}\label{Stestable}
\setcounter{table}{0}
\renewcommand\thetable{C.\arabic{table}}

\setcounter{figure}{0}
\renewcommand\thefigure{C.\arabic{figure}}

\setcounter{equation}{0}
\renewcommand\theequation{C.\arabic{equation}}

\setcounter{theorem}{0}
\renewcommand\thetheorem{C.\arabic{theorem}}

\setcounter{lemma}{0}
\renewcommand\thelemma{C.\arabic{lemma}}

\setcounter{proposition}{0}
\renewcommand\theproposition{C.\arabic{proposition}}

\setcounter{corollary}{0}
\renewcommand\thecorollary{C.\arabic{corollary}}

\setcounter{assumption}{0}
\renewcommand\theassumption{C.\arabic{assumption}}

{

In this appendix, we discuss the relationship between the testable restrictions in Subsection \ref{Srestriction} and the bounds in Propositions \ref{PROPmonotonicity} and \ref{PROPmonotonicityY}. In this discussion, we omit that all probabilities are conditional on covariates $X$ for ease of notation, and we impose that Assumptions \ref{ASexogeneity} and \ref{ASpositive} hold.

We start by showing two results. First, Inequality \eqref{EQrestriction1} is sufficient (but not necessary) for the property that the bounds in Proposition \ref{PROPmonotonicity} do not cross, i.e., $LB_{1} \leq UB_{1}$. Second, Inequalities \eqref{EQrestriction1} and \eqref{EQrestriction2} are necessary and sufficient for the property that the bounds in Proposition \ref{PROPmonotonicityY} do not cross, i.e., $LB_{1} \leq UB_{2}$.

At the end, we discuss the implications of these two results with respect to testing our identifying assumptions.

\subsection{Relationship between Inequality \eqref{EQrestriction1} and Proposition \ref{PROPmonotonicity}}\label{AppTestProp2}

\subsubsection{Inequality \eqref{EQrestriction1} implies $LB_{1} \leq UB_{1}$.}\label{AppR1sufficient}

We assume that Inequality \eqref{EQrestriction1} holds, i.e., $\mathbb{P}\left[\left. S = 1 \right\vert D = 1\right] - \mathbb{P}\left[\left. S = 1 \right\vert D = 0\right] \geq 0$. We want to show that $LB_{1} \leq UB_{1}$. To do so, we need to check three inequalities.
\begin{enumerate}
	\item $\dfrac{\left[B - \left(1 - A\right)\right] \cdot A^{-1} + C - 1}{C} \leq 1$

	Note that
	\begin{align*}
	& B \leq 1 \text{ because $B$ is a probability} \\
	& \hspace{5pt} \Rightarrow \dfrac{B - 1}{A} \leq 0 \text{ because $A > 0$ by  Assumptions \ref{ASexogeneity} and \ref{ASpositive}} \\
	& \hspace{5pt} \Rightarrow \dfrac{B - 1}{A} + 1 \leq 1 \\
	& \hspace{5pt} \Rightarrow \left[B - \left(1 - A\right)\right] \cdot A^{-1} \leq 1 \\
	& \hspace{5pt} \Rightarrow \left[B - \left(1 - A\right)\right] \cdot A^{-1} + C - 1 \leq C \\
	& \hspace{5pt} \Rightarrow \dfrac{\left[B - \left(1 - A\right)\right] \cdot A^{-1} + C - 1}{C} \leq 1 \text{ because $C > 0$  by Assumptions \ref{ASexogeneity} and \ref{ASpositive}.}
	\end{align*}

	\item $\dfrac{B \cdot A^{-1}}{C} \geq 0$

	Observe that the above inequality holds because all objects on the left-hand side are probabilities.

	\item $\dfrac{\left[B - \left(1 - A\right)\right] \cdot A^{-1} + C - 1}{C} \leq \dfrac{B \cdot A^{-1}}{C} $

	Notice that
	\begin{align*}
	\dfrac{\left[B - \left(1 - A\right)\right] \cdot A^{-1} + C - 1}{C} & \leq \dfrac{\left[B - \left(1 - A\right)\right] \cdot A^{-1}}{C} \\
	& \hspace{20pt} \text{because $C \leq 1$ since $C$ is a probability} \\
	& \leq \dfrac{B \cdot A^{-1}}{C} \\
	& \hspace{20pt} \text{because $A \leq 1$ since Inequality \eqref{EQrestriction1} holds.}
	\end{align*}
\end{enumerate}

\subsubsection{Inequality \eqref{EQrestriction1} is not implied by $LB_{1} \leq UB_{1}$.}

To show that Inequality \eqref{EQrestriction1} is not implied by $LB_{1} \leq UB_{1}$, we need a data-generating process that implies $LB_{1} \leq UB_{1}$ and $\mathbb{P}\left[\left. S = 1 \right\vert D = 1\right] - \mathbb{P}\left[\left. S = 1 \right\vert D = 0\right] < 0$.

Let our data-generating process be given by $\mathbb{P}\left[D = 1\right] = \sfrac{1}{2}$ and the conditional probability mass function described in Table \ref{TabNumericalInvalid}.

\begin{table}[!htbp]
	\centering
	\caption{$\mathbb{P}\left[\left. Y_{0}^{*} = \cdot, Y_{1}^{*} = \cdot, S_{0} = \cdot, S_{1} = \cdot \right\vert D = d\right]$ for any $d \in \left\lbrace 0, 1 \right\rbrace$} \label{TabNumericalInvalid}
	\begin{lrbox}{\tablebox}
		\begin{tabular}{ccccccccccccccccccc}
			\hline
			\hline
			\multicolumn{4}{c}{Panel A:} &  & \multicolumn{4}{c}{Panel B:} &  & \multicolumn{4}{c}{Panel C:} &  & \multicolumn{4}{c}{Panel D:} \\
			\multicolumn{4}{c}{$S_{0} = 1, S_{1} = 1$} &  & \multicolumn{4}{c}{$S_{0} = 0, S_{1} = 1$} &  & \multicolumn{4}{c}{$S_{0} = 1, S_{1} = 0$} &  & \multicolumn{4}{c}{$S_{0} = 0, S_{1} = 0$} \\ \cline{1-4} \cline{6-9} \cline{11-14} \cline{16-19}
			&  & \multicolumn{2}{c}{$Y_{0}^{*} = $} &  &  &  & \multicolumn{2}{c}{$Y_{0}^{*} = $} &  &  &  & \multicolumn{2}{c}{$Y_{0}^{*} = $} &  &  &  & \multicolumn{2}{c}{$Y_{0}^{*} = $} \\ \cline{3-4} \cline{8-9} \cline{13-14} \cline{18-19}
			& \multicolumn{1}{c|}{} & \multicolumn{1}{c|}{0} & \multicolumn{1}{c|}{1} &  &  & \multicolumn{1}{c|}{} & \multicolumn{1}{c|}{0} & \multicolumn{1}{c|}{1} &  &  & \multicolumn{1}{c|}{} & \multicolumn{1}{c|}{0} & \multicolumn{1}{c|}{1} &  &  & \multicolumn{1}{c|}{} & \multicolumn{1}{c|}{0} & \multicolumn{1}{c|}{1} \\ \cline{2-4} \cline{7-9} \cline{12-14} \cline{17-19}
			\multicolumn{1}{c|}{\multirow{2}{*}{$Y_{1}^{*} = $}} & \multicolumn{1}{c|}{0} & \multicolumn{1}{c|}{$\sfrac{3}{16}$} & \multicolumn{1}{c|}{$0$} &  & \multicolumn{1}{c|}{\multirow{2}{*}{$Y_{1}^{*} = $}} & \multicolumn{1}{c|}{0} & \multicolumn{1}{c|}{$\sfrac{1}{16}$} & \multicolumn{1}{c|}{$0$} &  & \multicolumn{1}{c|}{\multirow{2}{*}{$Y_{1}^{*} = $}} & \multicolumn{1}{c|}{0} & \multicolumn{1}{c|}{0} & \multicolumn{1}{c|}{0} &  & \multicolumn{1}{c|}{\multirow{2}{*}{$Y_{1}^{*} = $}} & \multicolumn{1}{c|}{0} & \multicolumn{1}{c|}{0} & \multicolumn{1}{c|}{0} \\ \cline{2-4} \cline{7-9} \cline{12-14} \cline{17-19}
			\multicolumn{1}{c|}{} & \multicolumn{1}{c|}{1} & \multicolumn{1}{c|}{$\sfrac{4}{16}$} & \multicolumn{1}{c|}{$\sfrac{2}{16}$} &  & \multicolumn{1}{c|}{} & \multicolumn{1}{c|}{1} & \multicolumn{1}{c|}{$\sfrac{1}{16}$} & \multicolumn{1}{c|}{$\sfrac{1}{16}$} &  & \multicolumn{1}{c|}{} & \multicolumn{1}{c|}{1} & \multicolumn{1}{c|}{0} & \multicolumn{1}{c|}{$\sfrac{4}{16}$} &  & \multicolumn{1}{c|}{} & \multicolumn{1}{c|}{1} & \multicolumn{1}{c|}{0} & \multicolumn{1}{c|}{0} \\ \cline{2-4} \cline{7-9} \cline{12-14} \cline{17-19}
			&  &  &  &  &  &  &  &  &  &  &  &  &  &  &  &  &  &  \\ \hline
		\end{tabular}
	\end{lrbox}
	\usebox{\tablebox}\\
	\settowidth{\tableboxwidth}{\usebox{\tablebox}} \parbox{\tableboxwidth}{\footnotesize{\textit{Notes}: Each cell reports $\mathbb{P}\left[\left. Y_{0}^{*} = y_{0}, Y_{1}^{*} = y_{1}, S_{0} = s_{0}, S_{1} = s_{1} \right\vert D = d\right]$ for the values $s_{0}$ and $s_{1}$ described in the panels, the value $y_{0}$ described in the columns and the value of $y_{1}$ described in the rows.}}
\end{table}

Note that this data-generating process satisfies Assumptions \ref{ASexogeneity}, \ref{ASpositive}, \ref{ASmonotonicityY} and \ref{ASstochastic} by construction. More importantly, we have that $LB_{1} \approx .43 \leq 1 = UB_{1}$. However, we also have that $\mathbb{P}\left[\left. S = 1 \right\vert D = 1\right] - \mathbb{P}\left[\left. S = 1 \right\vert D = 0\right] = .75 - .8125 = - .0625 < 0$.

\subsection{Relationship between Inequalities \eqref{EQrestriction1} and \eqref{EQrestriction2} and Proposition \ref{PROPmonotonicityY}}\label{AppTestProp3}

\subsubsection{Inequalities \eqref{EQrestriction1} and \eqref{EQrestriction2} imply $LB_{1} \leq UB_{2}$.}\label{AppR2sufficient}

We assume that Inequalities \eqref{EQrestriction1} and \eqref{EQrestriction2} hold, i.e., $\mathbb{P}\left[\left. S = 1 \right\vert D = 1\right] - \mathbb{P}\left[\left. S = 1 \right\vert D = 0\right] \geq 0$ and $\mathbb{P}\left[\left. Y = 1 \right\vert D = 1\right] - \mathbb{P}\left[\left. Y = 1 \right\vert D = 0\right] \geq 0$. We want to show that $LB_{1} \leq UB_{2}$. To do so, we need to check three inequalities.
\begin{enumerate}
	\item $\dfrac{\left[B - \left(1 - A\right)\right] \cdot A^{-1} + C - 1}{C} \leq 1$

	This inequality holds as shown in Appendix \ref{AppR1sufficient}.

	\item\label{StepIneq3} $\dfrac{B \cdot A^{-1} + C - 1}{C} \geq 0$

	Note that
	\begin{align*}
	& \mathbb{P}\left[\left. Y = 1 \right\vert D = 1\right] \geq \mathbb{P}\left[\left. Y = 1 \right\vert D = 0\right] \text{ because Inequality \eqref{EQrestriction2} holds} \\
	& \hspace{5pt} \Leftrightarrow \mathbb{P}\left[\left. Y = 1, S = 1 \right\vert D = 1\right] \geq \mathbb{P}\left[\left. Y = 1, S = 1 \right\vert D = 0\right] \text{ by Equation \eqref{EQoutcomes} } \\
	& \hspace{5pt} \Leftrightarrow \mathbb{P}\left[\left. Y = 1 \right\vert S = 1, D = 1\right] \cdot \mathbb{P}\left[\left. S = 1 \right\vert D = 1\right] \geq \mathbb{P}\left[\left. Y = 1 \right\vert S = 1, D = 0\right] \cdot \mathbb{P}\left[\left. S = 1 \right\vert D = 0\right] \\
	& \hspace{15pt} \text{by the definition of conditional probability} \\
	& \hspace{5pt} \Leftrightarrow \dfrac{\mathbb{P}\left[\left. Y = 1 \right\vert S = 1, D = 1\right] \cdot \mathbb{P}\left[\left. S = 1 \right\vert D = 1\right]}{\mathbb{P}\left[\left. S = 1 \right\vert D = 0\right]} \geq \mathbb{P}\left[\left. Y = 1 \right\vert S = 1, D = 0\right] \\
        & \hspace{15pt} \text{because }  \mathbb{P}\left[\left. S = 1 \right\vert D = 0\right] > 0 \text{ by Assumption \ref{ASpositive}} \\
	& \hspace{5pt} \Leftrightarrow B \cdot A^{-1} \geq 1 - C \\
	& \hspace{5pt} \Leftrightarrow \dfrac{B \cdot A^{-1} + C - 1}{C} \geq 0 \text{ because $C > 0$  by  Assumptions \ref{ASexogeneity} and \ref{ASpositive}.}
	\end{align*}

	\item\label{StepIneq4} $\dfrac{\left[B - \left(1 - A\right)\right] \cdot A^{-1} + C - 1}{C} \leq \dfrac{B \cdot A^{-1} + C - 1}{C} $

	Observe that that
	\begin{align*}
	& A \leq 1 \text{ because Inequality \eqref{EQrestriction1} holds} \\
	& \hspace{5pt} \Leftrightarrow B - \left(1 - A\right) \leq B \\
	& \hspace{5pt} \Leftrightarrow \left[B - \left(1 - A\right)\right] \cdot A^{-1} \leq B \cdot A^{-1} \text{ because $A > 0$ by  Assumptions \ref{ASexogeneity} and \ref{ASpositive}} \\
	& \hspace{5pt} \Leftrightarrow  \dfrac{\left[B - \left(1 - A\right)\right] \cdot A^{-1} + C - 1}{C} \leq \dfrac{B \cdot A^{-1} + C - 1}{C} \\
	& \hspace{15pt} \text{ because $C > 0$  by Assumptions \ref{ASexogeneity} and \ref{ASpositive}.}
	\end{align*}
\end{enumerate}

\subsubsection{Inequalities \eqref{EQrestriction1} and \eqref{EQrestriction2} are implied by $LB_{1} \leq UB_{2}$.}

We assume that $LB_{1} \leq UB_{2}$. We want to show that Inequalities \eqref{EQrestriction1} and \eqref{EQrestriction2} hold. Note that the proof of this result is located in Steps \ref{StepIneq3} and \ref{StepIneq4} in Appendix \ref{AppR2sufficient}.

\subsection{Implications for Testing our Identifying Assumptions}

In this appendix, we discuss the implications of Appendices \ref{AppTestProp2} and \ref{AppTestProp3} for testing our identifying assumptions.

Appendix \ref{AppTestProp2} shows that the testable restriction in Lemma \ref{LErestriction1} is more stringent than testing that the bounds in Proposition \ref{PROPmonotonicity} do not cross. In other words, there are data-generating processes that violate the testable restriction in Lemma \ref{LErestriction1} but produce well-behaved bounds ($LB_{1} \leq UB_{1}$). Consequently, testing Inequality \eqref{EQrestriction1} seems more likely to detect violations of Assumption \ref{ASmonotonicity} than testing that the bounds in Proposition \ref{PROPmonotonicity} do not cross.\footnote{A formal proof of this claim is beyond the scope of this paper.} For this reason, we recommend testing Inequality \eqref{EQrestriction1} directly when implementing the methods proposed in this paper.

Appendix \ref{AppTestProp3} shows that the testable restrictions in Proposition \ref{PROPrestriction2} are equivalent to testing that the bounds in Proposition \ref{PROPmonotonicityY} do not cross. However, when implementing the methods proposed in this paper, we recommend testing Inequalities \eqref{EQrestriction1} and \eqref{EQrestriction2} directly instead of testing that $LB_{1} \leq UB_{2}$. In particular, Inequalities \eqref{EQrestriction1} and \eqref{EQrestriction2} can be tested using standard regression methods (Section \ref{Sestimation}) while testing that $LB_{1} \leq UB_{2}$ requires more complicated inferential methods.

}

\newpage

\section{Comparing the probability of causation parameter against other treatment effect parameters}\label{Spoc}
\setcounter{table}{0}
\renewcommand\thetable{D.\arabic{table}}

\setcounter{figure}{0}
\renewcommand\thefigure{D.\arabic{figure}}

\setcounter{equation}{0}
\renewcommand\theequation{D.\arabic{equation}}

\setcounter{theorem}{0}
\renewcommand\thetheorem{D.\arabic{theorem}}

\setcounter{lemma}{0}
\renewcommand\thelemma{D.\arabic{lemma}}

\setcounter{proposition}{0}
\renewcommand\theproposition{D.\arabic{proposition}}

\setcounter{corollary}{0}
\renewcommand\thecorollary{D.\arabic{corollary}}

\setcounter{assumption}{0}
\renewcommand\theassumption{D.\arabic{assumption}}

{

In this appendix, we compare the probability of causation parameter against other treatment effect parameters. For brevity, we omit covariates. To have a focused discussion, we also assume that there is no sample selection problem because the previous literature has not discussed this parameter in the presence of sample selection. In this case, our target parameter is simply the probability of causation, i.e., $$\theta \coloneqq \mathbb{P}\left[\left. Y_{1}^{*} = 1 \right\vert Y_{0}^{*} = 0\right].$$

In the Econometrics literature, four treatment effect parameters are related to the probability of causation parameter. The first is the persuasion effect \citep{Jun2019}. The second and third ones are the distribution of gains at selected base state values and the probability of ``employed with treatment, not employed without treatment'' \citep{Heckman1997}. The fourth one is the average treatment effect.

First, the persuasion effect and the probability of causation parameter are identical. \cite{Jun2019} prefer to use the expression ``persuasion effect'' because their empirical application focuses on an informational treatment whose goal is to persuade an individual to modify their political opinions, beliefs or behaviors. \cite{Pearl1999} and \cite{Tian2000} prefer to use the expression ``probability of causation'' because they emphasize that this parameter captures the probability that a positive outcome is caused by the treatment, i.e., the probability of a positive outcome when treated given a negative outcome when untreated.

Second, \cite{Heckman1997} analyze the distribution of gains at selected base state values. Adapting their parameter to our notation and focusing on a binary outcome, the distribution of gains at selected base state values is formally defined as $$\tau\left(\Delta\right) \coloneqq \mathbb{P}\left[\left. Y_{1}^{*} - Y_{0}^{*} = \Delta \right\vert D = 1, Y_{0}^{*} = y_{0}\right],$$ where $\Delta \in \left\lbrace -1, 0, 1 \right\rbrace$ and $y_{0} \in \left\lbrace 0, 1 \right\rbrace$. When $y_{0} = 0$ and $\Delta = 1$, the distribution of gains at selected base state values equals the probability of causation for the treated individuals. Therefore, the main difference between $\theta$ and $\tau$ is whether the researcher conditions on receiving the treatment, i.e., $D = 1$.

Third, \cite{Heckman1997} discuss the probability of ``employed with treatment, not employed without treatment''. Since employment is the main outcome of interest in their empirical application, this parameter is formally defined as $$P_{0,1} \coloneqq \mathbb{P}\left[ Y_{0}^{*} = 0, Y_{1}^{*} = 1\right].$$ Note that $\theta = \sfrac{P_{0,1}}{\mathbb{P}\left[ Y_{0}^{*} = 0\right]}.$ Therefore, the main difference between $\theta$ and $P_{0,1}$ is whether the researcher conditions on having a negative untreated outcome, i.e., $Y_{0}^{*} = 0$.

Finally, the average treatment effect is defined as $$ATE \coloneqq \mathbb{E}\left[Y_{1}^{*} - Y_{0}^{*}\right].$$ When the monotone treatment response assumption is valid, we have that $ATE = P_{0,1}$. This equality clarifies when a researcher should focus on $P_{0,1}$ or $\theta$ to evaluate a policy. When the policy maker is equally concerned with every individual, focusing on the average treatment effect $\left(ATE = P_{0,1}\right)$ is natural. However, when a negative outcome is particularly severe (i.e., $Y^{*} = 0$ denotes that the individual died, was famished or was in extreme poverty), the policymaker may be particularly concerned with individuals who would have a negative outcome if untreated. In this case, focusing on the probability of causation parameter is justified.

}

\newpage

\section{Details on the Estimation and Inference Procedures}\label{AppDetailEstimationInference}
\setcounter{table}{0}
\renewcommand\thetable{E.\arabic{table}}

\setcounter{figure}{0}
\renewcommand\thefigure{E.\arabic{figure}}

\setcounter{equation}{0}
\renewcommand\theequation{E.\arabic{equation}}

\setcounter{theorem}{0}
\renewcommand\thetheorem{E.\arabic{theorem}}

\setcounter{lemma}{0}
\renewcommand\thelemma{E.\arabic{lemma}}

\setcounter{proposition}{0}
\renewcommand\theproposition{E.\arabic{proposition}}

\setcounter{corollary}{0}
\renewcommand\thecorollary{E.\arabic{corollary}}

\setcounter{assumption}{0}
\renewcommand\theassumption{E.\arabic{assumption}}

\subsection{Details on the Estimation Procedure}\label{AppDetailEstimation}

In this section, we present the details of our estimators for the bounds described in Propositions \ref{PROPmonotonicity}-\ref{PROPstochastic} and Corollary \ref{CorTarget}, and the weights in Lemma \ref{LEcovariates}.

We estimate these objects parametrically using maximum likelihood estimators. Let $\lambda\left(\cdot\right)$ be a link function, such as the logistic link function or the normal link function. Our parametric regression models are given by:
\begin{enumerate}
	\item $\mathbb{P}\left[\left. S = 1\right\vert D = d, X = x \right] = \lambda\left(\alpha_{0} + \alpha_{1} \cdot d + \alpha_{x}\right)$,

	\item $\mathbb{P}\left[\left. Y = 1\right\vert S = 1, D = d, X = x \right] = \lambda\left(\beta_{0} + \beta_{1} \cdot d + \beta_{x}\right)$, where we only use the employed subsample to estimate $\beta_{0}$, $\beta_{1}$ and $\beta_{x}$, and

	\item $\mathbb{P}\left[\left. W = 1 \right\vert D = d, X = x \right] = \lambda\left(\gamma_{0} + \gamma_{1} \cdot d + \gamma_{x}\right)$, where $W \coloneqq \mathbf{1}\left\lbrace Y = 0, S = 1 \right\rbrace$.
\end{enumerate}

Denoting our coefficients' estimators with the hat notation, we define:
\begin{enumerate}
	\item $\hat{A}\left(x\right) = \dfrac{\lambda\left(\hat{\alpha}_{0} + \hat{\alpha}_{x}\right)}{\lambda\left(\hat{\alpha}_{0} + \hat{\alpha}_{1} + \hat{\alpha}_{x}\right)}$,

	\item $\hat{B}\left(x\right) = \lambda\left(\hat{\beta}_{0} + \hat{\beta}_{1} + \hat{\beta}_{x}\right)$, and

	\item $\hat{C}\left(x\right) = 1 - \lambda\left(\hat{\beta}_{0} + \hat{\beta}_{x}\right)$
\end{enumerate}
for any $x \in \mathcal{X}$.

Consequently, the bounds in Propositions \ref{PROPmonotonicity}-\ref{PROPstochastic} can be estimated using the following objects:
\begin{align*}
\widehat{LB}_{1}\left(x\right) & \coloneqq \max\left\lbrace \dfrac{ \left[\hat{B}\left(x\right) - \left(1 - \hat{A}\left(x\right) \right)\right] \cdot \left[\hat{A}\left(x\right)\right]^{-1} + \hat{C}\left(x\right) - 1}{\hat{C}\left(x\right)} , 0 \right\rbrace, \\
\widehat{UB}_{1}\left(x\right) & \coloneqq \min \left\lbrace \dfrac{\hat{B}\left(x\right) \cdot \left[\hat{A}\left(x\right)\right]^{-1}}{\hat{C}\left(x\right)} , 1 \right\rbrace, \\
\widehat{UB}_{2}\left(x\right) & \coloneqq \min \left\lbrace \dfrac{\hat{B}\left(x\right) \cdot \left[\hat{A}\left(x\right)\right]^{-1} + \hat{C}\left(x\right) - 1}{\hat{C}\left(x\right)} , 1 \right\rbrace, \text{ and} \\
\widehat{LB}_{3}\left(x\right) & \coloneqq \max\left\lbrace \dfrac{\hat{B}\left(x\right) + \hat{C}\left(x\right) - 1}{\hat{C}\left(x\right)} , 0 \right\rbrace
\end{align*}
for any $x \in \mathcal{X}$.

Furthermore, the weights in Lemma \ref{LEcovariates} can be estimated by $$\hat{\omega}\left(x\right) = \dfrac{\lambda\left(\hat{\gamma}_{0} + \hat{\gamma}_{x}\right) \cdot \sum_{i = 1}^{N} \mathbf{1}\left\lbrace X_{i} = x \right\rbrace}{\sum_{x^{\prime} \in \mathcal{X}} \lambda\left(\hat{\gamma}_{0} + \hat{\gamma}_{x^{\prime}}\right) \cdot \sum_{i = 1}^{N} \mathbf{1}\left\lbrace X_{i} = x^{\prime} \right\rbrace}.$$

Finally, the bounds in Corollary \ref{CorTarget} can be estimated using the following objects:
\begin{align*}
\hat{\theta}^{OO}_{LB,1} & \coloneqq \sum_{x \in \mathcal{X}} \widehat{LB}_{1}\left(x\right) \cdot \hat{\omega}\left(x\right), \\
\hat{\theta}^{OO}_{UB,1} & \coloneqq \sum_{x \in \mathcal{X}} \widehat{UB}_{1}\left(x\right) \cdot \hat{\omega}\left(x\right), \\
\hat{\theta}^{OO}_{UB,2} & \coloneqq \sum_{x \in \mathcal{X}} \widehat{UB}_{2}\left(x\right) \cdot \hat{\omega}\left(x\right), \text{ and}\\
\hat{\theta}^{OO}_{LB,3} & \coloneqq \sum_{x \in \mathcal{X}} \widehat{LB}_{3}\left(x\right) \cdot \hat{\omega}\left(x\right).
\end{align*}

\subsection{Details on the Inference Procedure}\label{AppDetailInference}

This section is divided into three parts.

In the first part, we show that the unfeasible random set $R_{N}$ is a confidence region around the identified set. As a caveat, we highlight that we do not show that the feasible set $\widehat{R}_{N}$ is a valid $p$-confidence region around the identified set. We hope that proving that the unfeasible set $R_{N}$ is a confidence region may work as an intuitive argument for future work that rigorously address feasible inference around the identified set $\left[\sum_{x \in \mathcal{X}} LB_{3}\left(x\right) \cdot \omega\left(x\right), \sum_{x \in \mathcal{X}} UB_{2}\left(x\right) \cdot \omega\left(x\right) \right]$.

In the second part, we explain how to implement the precision-corrected estimators proposed by \cite{Chernozhukov2013}.

In the third part, we describe a Monte Carlo Simulation that illustrates the finite sample properties of the feasible inference procedure proposed in Section \ref{Sinference}. Although we have not formally proved that the feasible set $\widehat{R}_{N}$ is a valid $p$-confidence region, we find that, in our simulated data-generating process, $\widehat{R}_{N}$ covers the identified set more frequently than its nominal confidence level of 90\%. This result suggests that using the feasible set $\widehat{R}_{N}$ in place of the unfeasible set ${R}_{N}$ may work appropriately despite the absence of a formal proof.

\subsubsection{The unfeasible set $R_{N}$ is a confidence region.}
In this part, we show that the unfeasible set $R_{N}$ proposed in Equation \eqref{EqConfRegionUnfeasible} satisfies Equation \eqref{EQplimR} with $p = 90\%$ if $p_{Q} = 99.96\%$ when we replace $\widehat{R}_{N}$ by $R_{N}$.

First, we formally show that Equation \eqref{EQplimQ} holds. Fix $x \in \mathcal{X}$ and $p_{Q} \in \left(\sfrac{1}{2},1\right)$ arbitrarily. Note that
\begin{align*}
& \mathbb{P}\left[ \left[ LB_{3}\left(x\right), UB_{2}\left(x\right) \right] \subseteq Q_{N}\left(x\right) \right] \\
& \hspace{20pt} = \mathbb{P}\left[ \left[ LB_{3}\left(x\right), UB_{2}\left(x\right) \right] \subseteq \left[ \widehat{LB}_{3,N}^{CLR}\left(x,\sfrac{\left(1 + p_{Q}\right)}{2}\right), \widehat{UB}^{CLR}_{2,N}\left(x,\sfrac{\left(1 + p_{Q}\right)}{2}\right) \right] \right] \\
& \hspace{40pt} \text{according to the definition of } Q_{N}\left(x\right) \\
& \hspace{20pt} = \mathbb{P}\left[\left\lbrace \widehat{LB}_{3,N}^{CLR}\left(x,\sfrac{\left(1 + p_{Q}\right)}{2}\right) \leq LB_{3}\left(x\right) \right\rbrace \bigcap \left\lbrace UB_{2}\left(x\right) \leq \widehat{UB}^{CLR}_{2,N}\left(x,\sfrac{\left(1 + p_{Q}\right)}{2}\right) \right\rbrace \right] \\
& \hspace{20pt} = \mathbb{P}\left[\left\lbrace \widehat{LB}_{3,N}^{CLR}\left(x,\sfrac{\left(1 + p_{Q}\right)}{2}\right) \leq LB_{3}\left(x\right) \right\rbrace \right] + \mathbb{P}\left[\left\lbrace UB_{2}\left(x\right) \leq \widehat{UB}^{CLR}_{2,N}\left(x,\sfrac{\left(1 + p_{Q}\right)}{2}\right) \right\rbrace \right] \\
& \hspace{40pt} - \mathbb{P}\left[\left\lbrace \widehat{LB}_{3,N}^{CLR}\left(x,\sfrac{\left(1 + p_{Q}\right)}{2}\right) \leq LB_{3}\left(x\right) \right\rbrace \bigcup \left\lbrace UB_{2}\left(x\right) \leq \widehat{UB}^{CLR}_{2,N}\left(x,\sfrac{\left(1 + p_{Q}\right)}{2}\right) \right\rbrace \right] \\
& \hspace{60pt} \text{by the Addition Rule for Probabilities} \\
& \hspace{20pt} \geq \mathbb{P}\left[\left\lbrace \widehat{LB}_{3,N}^{CLR}\left(x,\sfrac{\left(1 + p_{Q}\right)}{2}\right) \leq LB_{3}\left(x\right) \right\rbrace \right] + \mathbb{P}\left[\left\lbrace UB_{2}\left(x\right) \leq \widehat{UB}^{CLR}_{2,N}\left(x,\sfrac{\left(1 + p_{Q}\right)}{2}\right) \right\rbrace \right] - 1 \\
& \hspace{40pt} \text{because any probability is less than 1} \\
& \hspace{20pt} \geq \dfrac{1 + p_{Q}}{2} - o\left(1\right) + \dfrac{1 + p_{Q}}{2} - o\left(1\right) - 1 \\
& \hspace{40pt} \text{according to \citet[Theorem 1]{Chernozhukov2013}} \\
& \hspace{20pt} \geq p_{Q} - o\left(1\right),
\end{align*}
implying that Equation \eqref{EQplimQ} holds.

Second, we show that Equation \eqref{EQplimR} holds for $R_{N}$ in place of $\widehat{R}_{N}$ and $p = 1 - K \cdot \left(1 - p_{Q}\right)$, where $K$ is the number of strata in our empirical application, i.e., $K \coloneqq \left\vert \mathcal{X} \right\vert$, and $\mathcal{X} = \left\lbrace 1, 2, \ldots, K \right\rbrace$. Observe that
\begin{align*}
& \mathbb{P}\left[ \left[ \sum_{x \in \mathcal{X}} LB_{3}\left(x\right) \cdot \omega\left(x\right), \sum_{x \in \mathcal{X}} UB_{2}\left(x\right) \cdot \omega\left(x\right) \right] \subseteq R_{N} \right] \\
& \hspace{20pt} = \mathbb{P}\left[ \begin{matrix*}[l]
\left[ \sum_{x \in \mathcal{X}} LB_{3}\left(x\right) \cdot \omega\left(x\right), \sum_{x \in \mathcal{X}} UB_{2}\left(x\right) \cdot \omega\left(x\right) \right] \\
\hspace{20pt} \subseteq \left[ \sum_{x \in \mathcal{X}} \widehat{LB}_{3,N}^{CLR}\left(x,\sfrac{\left(1 + p_{Q}\right)}{2}\right) \cdot {\omega}\left(x\right), \sum_{x \in \mathcal{X}} \widehat{UB}^{CLR}_{2,N}\left(x,\sfrac{\left(1 + p_{Q}\right)}{2}\right) \cdot {\omega}\left(x\right) \right]
\end{matrix*}
\right] \\
& \hspace{40pt} \text{according to Equation \eqref{EqConfRegionUnfeasible}} \\
& \hspace{20pt} \geq \mathbb{P}\left[\bigcap_{x \in \mathcal{X}} \left\lbrace \left[ LB_{3}\left(x\right), UB_{2}\left(x\right) \right] \subseteq \left[ \widehat{LB}_{3,N}^{CLR}\left(x,\sfrac{\left(1 + p_{Q}\right)}{2}\right), \widehat{UB}^{CLR}_{2,N}\left(x,\sfrac{\left(1 + p_{Q}\right)}{2}\right) \right] \right\rbrace \right] \\
& \hspace{40pt} \text{because } \left[ LB_{3}\left(x\right), UB_{2}\left(x\right) \right] \subseteq \left[ \widehat{LB}_{3,N}^{CLR}\left(x,\sfrac{\left(1 + p_{Q}\right)}{2}\right), \widehat{UB}^{CLR}_{2,N}\left(x,\sfrac{\left(1 + p_{Q}\right)}{2}\right) \right] \\
& \hspace{40pt} \text{for every } x \in \mathcal{X} \text{ implies} \\
& \hspace{40pt} \begin{matrix*}[l]
\left[ \sum_{x \in \mathcal{X}} LB_{3}\left(x\right) \cdot \omega\left(x\right), \sum_{x \in \mathcal{X}} UB_{2}\left(x\right) \cdot \omega\left(x\right) \right] \\
\hspace{20pt} \subseteq \left[ \sum_{x \in \mathcal{X}} \widehat{LB}_{3,N}^{CLR}\left(x,\sfrac{\left(1 + p_{Q}\right)}{2}\right) \cdot {\omega}\left(x\right), \sum_{x \in \mathcal{X}} \widehat{UB}^{CLR}_{2,N}\left(x,\sfrac{\left(1 + p_{Q}\right)}{2}\right) \cdot {\omega}\left(x\right) \right]
\end{matrix*} \\
& \hspace{20pt} = \mathbb{P}\left[\bigcap_{x \in \mathcal{X}} \left\lbrace \left[ LB_{3}\left(x\right), UB_{2}\left(x\right) \right] \subseteq Q_{N}\left(x\right) \right\rbrace \right] \\
& \hspace{40pt} \text{according to the definition of } Q_{N}\left(x\right) \\
& \hspace{20pt} = \mathbb{P}\left[\bigcap_{k = 1}^{K} \left\lbrace \left[ LB_{3}\left(k\right), UB_{2}\left(k\right) \right] \subseteq Q_{N}\left(k\right) \right\rbrace \right] \\
& \hspace{40pt} \text{because } \mathcal{X} = \left\lbrace 1, 2, \ldots, K \right\rbrace \\
& \hspace{20pt} = \mathbb{P}\left[\left\lbrace \left[ LB_{3}\left(1\right), UB_{2}\left(1\right) \right] \subseteq Q_{N}\left(1\right) \right\rbrace \bigcap \left\lbrace \bigcap_{k = 2}^{K} \left\lbrace \left[ LB_{3}\left(k\right), UB_{2}\left(k\right) \right] \subseteq Q_{N}\left(k\right) \right\rbrace \right\rbrace \right] \\
& \hspace{20pt} = \mathbb{P}\left[\left[ LB_{3}\left(1\right), UB_{2}\left(1\right) \right] \subseteq Q_{N}\left(1\right) \right] + \mathbb{P}\left[ \bigcap_{k = 2}^{K} \left\lbrace \left[ LB_{3}\left(k\right), UB_{2}\left(k\right) \right] \subseteq Q_{N}\left(k\right) \right\rbrace \right] \\
& \hspace{40pt} - \mathbb{P}\left[\left\lbrace \left[ LB_{3}\left(1\right), UB_{2}\left(1\right) \right] \subseteq Q_{N}\left(1\right) \right\rbrace \bigcup \left\lbrace \bigcap_{k = 2}^{K} \left\lbrace \left[ LB_{3}\left(k\right), UB_{2}\left(k\right) \right] \subseteq Q_{N}\left(k\right) \right\rbrace \right\rbrace \right] \\
& \hspace{60pt} \text{by the Addition Rule for Probabilities} \\
& \hspace{20pt} \geq \mathbb{P}\left[\left[ LB_{3}\left(1\right), UB_{2}\left(1\right) \right] \subseteq Q_{N}\left(1\right) \right] + \mathbb{P}\left[ \bigcap_{k = 2}^{K} \left\lbrace \left[ LB_{3}\left(k\right), UB_{2}\left(k\right) \right] \subseteq Q_{N}\left(k\right) \right\rbrace \right] - 1 \\
& \hspace{40pt} \text{because any probability is less than 1} \\
& \hspace{20pt} = \mathbb{P}\left[\left[ LB_{3}\left(1\right), UB_{2}\left(1\right) \right] \subseteq Q_{N}\left(1\right) \right] - 1 \\
& \hspace{40pt} + \mathbb{P}\left[\left\lbrace \left[ LB_{3}\left(2\right), UB_{2}\left(2\right) \right] \subseteq Q_{N}\left(2\right) \right\rbrace \bigcap \left\lbrace \bigcap_{k = 3}^{K} \left\lbrace \left[ LB_{3}\left(k\right), UB_{2}\left(k\right) \right] \subseteq Q_{N}\left(k\right) \right\rbrace \right\rbrace \right] \\
& \hspace{20pt} = \mathbb{P}\left[\left[ LB_{3}\left(1\right), UB_{2}\left(1\right) \right] \subseteq Q_{N}\left(1\right) \right] - 1 \\
& \hspace{40pt} + \mathbb{P}\left[\left[ LB_{3}\left(2\right), UB_{2}\left(2\right) \right] \subseteq Q_{N}\left(2\right) \right] + \mathbb{P}\left[ \bigcap_{k = 3}^{K} \left\lbrace \left[ LB_{3}\left(k\right), UB_{2}\left(k\right) \right] \subseteq Q_{N}\left(k\right) \right\rbrace \right] \\
& \hspace{40pt} - \mathbb{P}\left[\left\lbrace \left[ LB_{3}\left(2\right), UB_{2}\left(2\right) \right] \subseteq Q_{N}\left(2\right) \right\rbrace \bigcup \left\lbrace \bigcap_{k = 2}^{K} \left\lbrace \left[ LB_{3}\left(k\right), UB_{2}\left(k\right) \right] \subseteq Q_{N}\left(k\right) \right\rbrace \right\rbrace \right] \\
& \hspace{60pt} \text{by the Addition Rule for Probabilities} \\
& \hspace{20pt} \geq \left\lbrace \sum_{k = 1}^{2} \mathbb{P}\left[\left[ LB_{3}\left(k\right), UB_{2}\left(k\right) \right] \subseteq Q_{N}\left(k\right) \right] \right\rbrace - 2 + \mathbb{P}\left[ \bigcap_{k = 3}^{K} \left\lbrace \left[ LB_{3}\left(k\right), UB_{2}\left(k\right) \right] \subseteq Q_{N}\left(k\right) \right\rbrace \right] \\
& \hspace{40pt} \text{because any probability is less than 1} \\
& \hspace{20pt} \vdotswithin{=} \\
& \hspace{20pt} \geq \left\lbrace \sum_{k = 1}^{K} \mathbb{P}\left[\left[ LB_{3}\left(k\right), UB_{2}\left(k\right) \right] \subseteq Q_{N}\left(k\right) \right] \right\rbrace - \left(K - 1\right) \\
& \hspace{20pt} \geq \left\lbrace \sum_{k = 1}^{K} p_{Q} \right\rbrace - \left(K - 1\right) - o\left(1\right) \\
& \hspace{40pt} \text{according to Equation \eqref{EQplimQ}} \\
& \hspace{20pt} = 1 - K \cdot \left(1 - p_{Q}\right) - o\left(1\right),
\end{align*}
implying that Equation \eqref{EQplimR} holds for $p = 1 - K \cdot \left(1 - p_{Q}\right)$.

Finally, notice that $K = 246$ strata (as in our empirical application) and $p_{Q} = 99.96\%$ implies that $p = 90\%$ in the last equation. Consequently, when we replace $\widehat{R}_{N}$ by $R_{N}$, the random set $R_{N}$ proposed in Equation \eqref{EqConfRegion} satisfies Equation \eqref{EQplimR} with $p = 90\%$ if $p_{Q} = 99.96\%$. Observe also that, if our goal was to derive half-median unbiased estimators, we could use $p_{Q} = 99.8\%$.

\subsubsection{Implementing the precision-corrected estimators proposed by \cite{Chernozhukov2013}}

In this part, we explain how to implement the precision-corrected estimators $\widehat{LB}_{3,N}^{CLR}\left(x,\sfrac{\left(1 + p_{Q}\right)}{2}\right)$ and $\widehat{UB}^{CLR}_{2,N}\left(x,\sfrac{\left(1 + p_{Q}\right)}{2}\right)$ for each $x \in \mathcal{X}$. This part relies heavily on the work done by \cite{Flores2013}, who intuitively explain the method proposed by \cite{Chernozhukov2013}.

Fix $x \in \mathcal{X}$ arbitrarily. For brevity, we write our estimators in Appendix \ref{AppDetailEstimation} as $$\widehat{UB}_{2}\left(x\right) = \min \left\lbrace \hat{f}_{U}\left(x\right) , 1 \right\rbrace \text{ and } \widehat{LB}_{3}\left(x\right) = \max\left\lbrace \hat{f}_{L}\left(x\right) , 0 \right\rbrace,$$ where $$\hat{f}_{U}\left(x\right) \coloneqq \dfrac{\hat{B}\left(x\right) \cdot \left[\hat{A}\left(x\right)\right]^{-1} + \hat{C}\left(x\right) - 1}{\hat{C}\left(x\right)} \text{ and } \hat{f}_{L}\left(x\right) \coloneqq \dfrac{\hat{B}\left(x\right) + \hat{C}\left(x\right) - 1}{\hat{C}\left(x\right)},$$ and define $q \coloneqq \dfrac{1 + p_{Q}}{2}$.

To compute $\widehat{UB}^{CLR}_{2,N}\left(x,q\right)$, we follow 5 steps.
\begin{enumerate}
	\item Using the weighted bootstrap, obtain a consistent estimate $\hat{s}_{U}\left(x\right)$ of the standard error of $\hat{f}_{U}\left(x\right)$.\footnote{In our empirical application, we specifically use a cluster weighted bootstrap where we cluster our standard error at the stratum level. To do so, in each bootstrap iteration, we draw standard exponential weights for each stratum and re-run the regressions described in Appendix \ref{AppDetailEstimation} using weighted maximum likelihood estimators where each observation is weighted according to its stratum's weight. We use 5,000 bootstrap iterations.}

	\item Simulate $R$ draws from a standard normal distribution and denote them by $Z_{1}^{*}, \ldots,Z_{R}^{*}$.

	\item Let $Q_{z}\left(Z\right)$ denote the $z$-th quantile of a random variable Z and $c_{N} = 1 - \left(\dfrac{0.1}{\ln N}\right)$. Compute $$\kappa_{N}^{U}\left(c_{N}\right) \coloneqq Q_{c_{N}}\left(\max \left\lbrace Z_{r}^{*}, 0 \right\rbrace, r = 1, \ldots, R\right).$$

	\item Check if $\hat{f}_{U}\left(x\right) + \kappa_{N}^{U}\left(c_{N}\right) \cdot \hat{s}_{U}\left(x\right) < 1$.
	\begin{enumerate}
		\item If $\hat{f}_{U}\left(x\right) + \kappa_{N}^{U}\left(c_{N}\right) \cdot \hat{s}_{U}\left(x\right) < 1$, compute $$\hat{\kappa}_{N}^{U}\left(x, q\right) \coloneqq Q_{q}\left(Z_{r}^{*}, r = 1, \ldots, R\right).$$

		\item If $\hat{f}_{U}\left(x\right) + \kappa_{N}^{U}\left(c_{N}\right) \cdot \hat{s}_{U}\left(x\right) \geq 1$, compute $$\hat{\kappa}_{N}^{U}\left(x, q\right) \coloneqq Q_{q}\left(\max \left\lbrace Z_{r}^{*}, 0 \right\rbrace, r = 1, \ldots, R\right).$$
	\end{enumerate}

	\item Compute $\widehat{UB}^{CLR}_{2,N}\left(x,q\right) \coloneqq \min \left\lbrace \hat{f}_{U}\left(x\right) + \hat{\kappa}_{N}^{U}\left(x, q\right) \cdot \hat{s}_{u}\left(x\right) , 1 \right\rbrace$.
\end{enumerate}

To compute $\widehat{LB}^{CLR}_{3,N}\left(x,q\right)$, we follow 5 steps.
\begin{enumerate}
	\item Using the weighted bootstrap, obtain a consistent estimate $\hat{s}_{L}\left(x\right)$ of the standard error of $\hat{f}_{L}\left(x\right)$.

	\item Simulate $R$ draws from a standard normal distribution and denote them by $Z_{1}^{*}, \ldots,Z_{R}^{*}$.

	\item Let $Q_{z}\left(Z\right)$ denote the $z$-th quantile of a random variable Z and $c_{N} = 1 - \left(\dfrac{0.1}{\ln N}\right)$. Compute $$\kappa_{N}^{L}\left(c_{N}\right) \coloneqq Q_{c_{N}}\left(\max \left\lbrace Z_{r}^{*}, 0 \right\rbrace, r = 1, \ldots, R\right).$$

	\item Check if $\hat{f}_{L}\left(x\right) - \kappa_{N}^{L}\left(c_{N}\right) \cdot \hat{s}_{L}\left(x\right) > 0$.
	\begin{enumerate}
		\item If $\hat{f}_{L}\left(x\right) - \kappa_{N}^{L}\left(c_{N}\right) \cdot \hat{s}_{L}\left(x\right) > 0$, compute $$\hat{\kappa}_{N}^{L}\left(x, q\right) \coloneqq Q_{q}\left(Z_{r}^{*}, r = 1, \ldots, R\right).$$

		\item If $\hat{f}_{L}\left(x\right) - \kappa_{N}^{L}\left(c_{N}\right) \cdot \hat{s}_{L}\left(x\right) \leq 0$, compute $$\hat{\kappa}_{N}^{L}\left(x, q\right) \coloneqq Q_{q}\left(\max \left\lbrace Z_{r}^{*}, 0 \right\rbrace, r = 1, \ldots, R\right).$$
	\end{enumerate}

	\item Compute $\widehat{LB}^{CLR}_{3,N}\left(x,q\right) \coloneqq \min \left\lbrace \hat{f}_{L}\left(x\right) - \hat{\kappa}_{N}^{L}\left(x, q\right) \cdot \hat{s}_{L}\left(x\right) , 0 \right\rbrace$.
\end{enumerate}

\subsubsection{Monte Carlo Simulation: Using the feasible confidence region $\widehat{R}_{N}$}\label{AppMC}

In this part, we implement a Monte Carlo simulation to illustrate the finite sample properties of the feasible inference procedure proposed in Section \ref{Sinference}. Although we have not formally proved that the feasible set $\widehat{R}_{N}$ is a valid $p$-confidence region, we find that, in our simulated data-generating process, $\widehat{R}_{N}$ covers the identified set more frequently than its nominal confidence level of 90\%.

Our Monte Carlo Simulation creates 1,000 samples with 10,000 observations. Our data-generating process assigns treatment independently from all other variables with probability $\sfrac{1}{2}$. It also assigns a covariate $X$ independently from all other variables, where $\mathbb{P}\left[X = 1\right] = \sfrac{1}{2}$ and $\mathbb{P}\left[X = 2\right] = \sfrac{1}{2}$.

When $X = 1$, the distribution of the latent variables $\left(Y_{0}^{*}, Y_{1}^{*}, S_{0}, S_{1}\right)$ follows the same distribution proposed in our numerical example in Appendix \ref{Sexample}. We copy it here for convenience.

\begin{table}[!hbtp]
	\centering
	\caption{$\mathbb{P}\left[\left. Y_{0}^{*} = \cdot, Y_{1}^{*} = \cdot, S_{0} = \cdot, S_{1} = \cdot \right\vert D = d, X = 1\right]$ for any $d \in \left\lbrace 0, 1 \right\rbrace$} \label{TabDGP1}
	\begin{lrbox}{\tablebox}
		\begin{tabular}{ccccccccccccccccccc}
			\hline
			\hline
			\multicolumn{4}{c}{Panel A:} &  & \multicolumn{4}{c}{Panel B:} &  & \multicolumn{4}{c}{Panel C:} &  & \multicolumn{4}{c}{Panel D:} \\
			\multicolumn{4}{c}{$S_{0} = 1, S_{1} = 1$} &  & \multicolumn{4}{c}{$S_{0} = 0, S_{1} = 1$} &  & \multicolumn{4}{c}{$S_{0} = 1, S_{1} = 0$} &  & \multicolumn{4}{c}{$S_{0} = 0, S_{1} = 0$} \\ \cline{1-4} \cline{6-9} \cline{11-14} \cline{16-19}
			&  & \multicolumn{2}{c}{$Y_{0}^{*} = $} &  &  &  & \multicolumn{2}{c}{$Y_{0}^{*} = $} &  &  &  & \multicolumn{2}{c}{$Y_{0}^{*} = $} &  &  &  & \multicolumn{2}{c}{$Y_{0}^{*} = $} \\ \cline{3-4} \cline{8-9} \cline{13-14} \cline{18-19}
			& \multicolumn{1}{c|}{} & \multicolumn{1}{c|}{0} & \multicolumn{1}{c|}{1} &  &  & \multicolumn{1}{c|}{} & \multicolumn{1}{c|}{0} & \multicolumn{1}{c|}{1} &  &  & \multicolumn{1}{c|}{} & \multicolumn{1}{c|}{0} & \multicolumn{1}{c|}{1} &  &  & \multicolumn{1}{c|}{} & \multicolumn{1}{c|}{0} & \multicolumn{1}{c|}{1} \\ \cline{2-4} \cline{7-9} \cline{12-14} \cline{17-19}
			\multicolumn{1}{c|}{\multirow{2}{*}{$Y_{1}^{*} = $}} & \multicolumn{1}{c|}{0} & \multicolumn{1}{c|}{$\sfrac{3}{16}$} & \multicolumn{1}{c|}{$0$} &  & \multicolumn{1}{c|}{\multirow{2}{*}{$Y_{1}^{*} = $}} & \multicolumn{1}{c|}{0} & \multicolumn{1}{c|}{$\sfrac{2}{16}$} & \multicolumn{1}{c|}{$0$} &  & \multicolumn{1}{c|}{\multirow{2}{*}{$Y_{1}^{*} = $}} & \multicolumn{1}{c|}{0} & \multicolumn{1}{c|}{$0$} & \multicolumn{1}{c|}{$0$} &  & \multicolumn{1}{c|}{\multirow{2}{*}{$Y_{1}^{*} = $}} & \multicolumn{1}{c|}{0} & \multicolumn{1}{c|}{$\sfrac{1}{16}$} & \multicolumn{1}{c|}{$0$} \\ \cline{2-4} \cline{7-9} \cline{12-14} \cline{17-19}
			\multicolumn{1}{c|}{} & \multicolumn{1}{c|}{1} & \multicolumn{1}{c|}{$\sfrac{4}{16}$} & \multicolumn{1}{c|}{$\sfrac{2}{16}$} &  & \multicolumn{1}{c|}{} & \multicolumn{1}{c|}{1} & \multicolumn{1}{c|}{$\sfrac{1}{16}$} & \multicolumn{1}{c|}{$\sfrac{1}{16}$} &  & \multicolumn{1}{c|}{} & \multicolumn{1}{c|}{1} & \multicolumn{1}{c|}{$0$} & \multicolumn{1}{c|}{$0$} &  & \multicolumn{1}{c|}{} & \multicolumn{1}{c|}{1} & \multicolumn{1}{c|}{$\sfrac{1}{16}$} & \multicolumn{1}{c|}{$\sfrac{1}{16}$} \\ \cline{2-4} \cline{7-9} \cline{12-14} \cline{17-19}
			&  &  &  &  &  &  &  &  &  &  &  &  &  &  &  &  &  &  \\ \hline
		\end{tabular}
	\end{lrbox}
	\usebox{\tablebox}\\
	\settowidth{\tableboxwidth}{\usebox{\tablebox}} \parbox{\tableboxwidth}{\footnotesize{\textit{Notes}: Each cell reports $\mathbb{P}\left[\left. Y_{0}^{*} = y_{0}, Y_{1}^{*} = y_{1}, S_{0} = s_{0}, S_{1} = s_{1} \right\vert D = d\right]$ for the values $s_{0}$ and $s_{1}$ described in the panels, the value $y_{0}$ described in the columns and the value of $y_{1}$ described in the rows.}}
\end{table}

Note that, for individuals with $X = 1$, the conditional probability of causation within the population that is always observed is given by $$\theta^{OO}\left(1\right) = \mathbb{P}\left[\left. Y_{1}^{*} = 1 \right\vert Y_{0}^{*} = 0, S_{0} = 1, S_{1} = 1, X = 1\right] \approx 0.571$$ and the conditional identified region is given by $$\left[LB_{3}\left(1\right), UB_{2}\left(1\right)\right] \approx \left[0.505, 0.857\right]$$ under Assumptions \ref{ASexogeneity}-\ref{ASstochastic}.

When $X = 2$, the distribution of the latent variables $\left(Y_{0}^{*}, Y_{1}^{*}, S_{0}, S_{1}\right)$ follows the distribution described in Table \ref{TabDGP2}.

\begin{table}[!hbtp]
	\centering
	\caption{$\mathbb{P}\left[\left. Y_{0}^{*} = \cdot, Y_{1}^{*} = \cdot, S_{0} = \cdot, S_{1} = \cdot \right\vert D = d, X = 2\right]$ for any $d \in \left\lbrace 0, 1 \right\rbrace$} \label{TabDGP2}
	\begin{lrbox}{\tablebox}
		\begin{tabular}{ccccccccccccccccccc}
			\hline
			\hline
			\multicolumn{4}{c}{Panel A:} &  & \multicolumn{4}{c}{Panel B:} &  & \multicolumn{4}{c}{Panel C:} &  & \multicolumn{4}{c}{Panel D:} \\
			\multicolumn{4}{c}{$S_{0} = 1, S_{1} = 1$} &  & \multicolumn{4}{c}{$S_{0} = 0, S_{1} = 1$} &  & \multicolumn{4}{c}{$S_{0} = 1, S_{1} = 0$} &  & \multicolumn{4}{c}{$S_{0} = 0, S_{1} = 0$} \\ \cline{1-4} \cline{6-9} \cline{11-14} \cline{16-19}
			&  & \multicolumn{2}{c}{$Y_{0}^{*} = $} &  &  &  & \multicolumn{2}{c}{$Y_{0}^{*} = $} &  &  &  & \multicolumn{2}{c}{$Y_{0}^{*} = $} &  &  &  & \multicolumn{2}{c}{$Y_{0}^{*} = $} \\ \cline{3-4} \cline{8-9} \cline{13-14} \cline{18-19}
			& \multicolumn{1}{c|}{} & \multicolumn{1}{c|}{0} & \multicolumn{1}{c|}{1} &  &  & \multicolumn{1}{c|}{} & \multicolumn{1}{c|}{0} & \multicolumn{1}{c|}{1} &  &  & \multicolumn{1}{c|}{} & \multicolumn{1}{c|}{0} & \multicolumn{1}{c|}{1} &  &  & \multicolumn{1}{c|}{} & \multicolumn{1}{c|}{0} & \multicolumn{1}{c|}{1} \\ \cline{2-4} \cline{7-9} \cline{12-14} \cline{17-19}
			\multicolumn{1}{c|}{\multirow{2}{*}{$Y_{1}^{*} = $}} & \multicolumn{1}{c|}{0} & \multicolumn{1}{c|}{$\sfrac{4}{16}$} & \multicolumn{1}{c|}{$0$} &  & \multicolumn{1}{c|}{\multirow{2}{*}{$Y_{1}^{*} = $}} & \multicolumn{1}{c|}{0} & \multicolumn{1}{c|}{$\sfrac{2}{16}$} & \multicolumn{1}{c|}{$0$} &  & \multicolumn{1}{c|}{\multirow{2}{*}{$Y_{1}^{*} = $}} & \multicolumn{1}{c|}{0} & \multicolumn{1}{c|}{$0$} & \multicolumn{1}{c|}{$0$} &  & \multicolumn{1}{c|}{\multirow{2}{*}{$Y_{1}^{*} = $}} & \multicolumn{1}{c|}{0} & \multicolumn{1}{c|}{$0$} & \multicolumn{1}{c|}{$0$} \\ \cline{2-4} \cline{7-9} \cline{12-14} \cline{17-19}
			\multicolumn{1}{c|}{} & \multicolumn{1}{c|}{1} & \multicolumn{1}{c|}{$\sfrac{2}{16}$} & \multicolumn{1}{c|}{$\sfrac{3}{16}$} &  & \multicolumn{1}{c|}{} & \multicolumn{1}{c|}{1} & \multicolumn{1}{c|}{$\sfrac{2}{16}$} & \multicolumn{1}{c|}{$0$} &  & \multicolumn{1}{c|}{} & \multicolumn{1}{c|}{1} & \multicolumn{1}{c|}{$0$} & \multicolumn{1}{c|}{$0$} &  & \multicolumn{1}{c|}{} & \multicolumn{1}{c|}{1} & \multicolumn{1}{c|}{$\sfrac{1}{16}$} & \multicolumn{1}{c|}{$\sfrac{2}{16}$} \\ \cline{2-4} \cline{7-9} \cline{12-14} \cline{17-19}
			&  &  &  &  &  &  &  &  &  &  &  &  &  &  &  &  &  &  \\ \hline
		\end{tabular}
	\end{lrbox}
	\usebox{\tablebox}\\
	\settowidth{\tableboxwidth}{\usebox{\tablebox}} \parbox{\tableboxwidth}{\footnotesize{\textit{Notes}: Each cell reports $\mathbb{P}\left[\left. Y_{0}^{*} = y_{0}, Y_{1}^{*} = y_{1}, S_{0} = s_{0}, S_{1} = s_{1} \right\vert D = d\right]$ for the values $s_{0}$ and $s_{1}$ described in the panels, the value $y_{0}$ described in the columns and the value of $y_{1}$ described in the rows.}}
\end{table}

Note that, for individuals with $X = 2$, the conditional probability of causation within the population that is always observed is given by $$\theta^{OO}\left(2\right) = \mathbb{P}\left[\left. Y_{1}^{*} = 1 \right\vert Y_{0}^{*} = 0, S_{0} = 1, S_{1} = 1, X = 2\right] = \dfrac{1}{3}$$ and the conditional identified region is given by $$\left[LB_{3}\left(2\right), UB_{2}\left(2\right)\right] \approx \left[0.308, 0.667\right]$$ under Assumptions \ref{ASexogeneity}-\ref{ASstochastic}.

Moreover, our data-generating process imposes that $$\omega\left(x\right) = \mathbb{P}\left[\left. X = x \right\vert Y_{0}^{*} = 0, S_{0} = 1, S_{1} = 1\right] = \dfrac{1}{2}$$ for any $x \in \left\lbrace 1, 2 \right\rbrace$. As a consequence, we have that unconditional probability of causation within the population that is always observed is given by $$\theta^{OO} = \mathbb{P}\left[\left. Y_{1}^{*} = 1 \right\vert Y_{0}^{*} = 0, S_{0} = 1, S_{1} = 1\right] \approx 0.452$$ and the unconditional identified region is given by $$\left[LB_{3}, UB_{2}\right] \approx \left[0.407, 0.762\right]$$ under Assumptions \ref{ASexogeneity}-\ref{ASstochastic}, where $LB_{3} \coloneqq LB_{3}\left(1\right) \cdot \sfrac{1}{2} + LB_{3}\left(2\right) \cdot \sfrac{1}{2}$ and  $UB_{2} \coloneqq UB_{2}\left(1\right) \cdot \sfrac{1}{2} + UB_{2}\left(2\right) \cdot \sfrac{1}{2}$.

In each Monte Carlo iteration, we estimate the bounds $LB_{3}\left(1\right)$, $LB_{3}\left(2\right)$, $LB_{3}$, $UB_{2}\left(1\right)$, $UB_{2}\left(2\right)$, $UB_{2}$ using the estimators proposed in Section \ref{Sestimator} with the Probit and Logit models as the link function $\lambda\left(\cdot\right)$. To conduct inference in each Monte Carlo iteration, we also estimate the feasible sets $\widehat{Q}_{N}\left(1\right)$, $\widehat{Q}_{N}\left(2\right)$ and $\widehat{R}_{N}$ proposed in Section \ref{Sinference} using 399 bootstrap iterations. Based on this procedure, we aim to estimate 95\%-confidence regions around the conditional identified regions $\left(\left[LB_{3}\left(x\right), UB_{2}\left(x\right)\right]\right)$ and to estimate 90\%-confidence regions around the unconditional identified region $\left[LB_{3}, UB_{2}\right]$.

Table \ref{TabMC} reports the coverage rate across Monte Carlo iterations of our estimated confidence regions.

\begin{table}[!hbtp]
	\centering
	\caption{Coverage Rates of $\widehat{Q}_{N}\left(1\right)$, $\widehat{Q}_{N}\left(2\right)$ and $\widehat{R}_{N}$} \label{TabMC}
	\begin{lrbox}{\tablebox}
		\begin{tabular}{lccc}
			\hline \hline
			& Probit  & Logit  & Nominal \\
			& Model & Model & Confidence Level \\
			& (1) & (2) & (3) \\ \hline
			$\mathbb{P}\left[ \left[ LB_{3}\left(1\right), UB_{2}\left(1\right) \right] \subseteq \widehat{Q}_{N}\left(1\right) \right]$ & 0.72 & 0.73 & 0.95 \\
			$\mathbb{P}\left[ \left[ LB_{3}\left(2\right), UB_{2}\left(2\right) \right] \subseteq \widehat{Q}_{N}\left(2\right) \right]$ & 0.59 & 0.58 & 0.95 \\
			$\mathbb{P}\left[ \left[ LB_{3}, UB_{2} \right] \subseteq \widehat{R}_{N} \right]$ & 1.0 & 1.0 & 0.90 \\ \hline
		\end{tabular}
	\end{lrbox}
	\usebox{\tablebox}\\
	\settowidth{\tableboxwidth}{\usebox{\tablebox}} \parbox{\tableboxwidth}{\footnotesize{\textit{Notes}: Each cell reports the coverage rates of the feasible sets $\widehat{Q}_{N}\left(1\right)$, $\widehat{Q}_{N}\left(2\right)$ and $\widehat{R}_{N}$ proposed in Section \ref{Sinference} using 399 bootstrap iterations. We estimate the bounds $LB_{3}\left(1\right)$, $LB_{3}\left(2\right)$, $LB_{3}$, $UB_{2}\left(1\right)$, $UB_{2}\left(2\right)$, $UB_{2}$ using the estimators proposed in Section \ref{Sestimator} with the Probit Model in Column (1) and with the Logit Model in Column (2).}}
\end{table}

We focus on the results associated with the Probit Estimator (Column (1)) because the results with the Logit Estimator (Column (2)) are similar.

First, we note that the feasible confidence regions $\widehat{Q}_{N}\left(1\right)$ and $\widehat{Q}_{N}\left(2\right)$ cover the conditional identified regions $\left[ LB_{3}\left(1\right), UB_{2}\left(1\right) \right]$ and $\left[ LB_{3}\left(2\right), UB_{2}\left(2\right) \right]$ with a probability strictly less than their nominal confidence levels. This finding is not surprising because the estimating models in Section \ref{Sestimator} do not interact the treatment variable with the covariate variable, implying that they misspecified.

Second, we highlight that the feasible confidence region $\widehat{R}_{N}$ covers the unconditional identified region $\left[ LB_{3}, UB_{2} \right]$ with a probability strictly greater than its nominal confidence level. Consequently, the conservative Bonferroni correction seems to compensate any coverage issues caused by the uncertainty behind the estimation of $\omega\left(\cdot\right)$ or by the misspecification of the estimation model. This result suggests that using the feasible set $\widehat{R}_{N}$ in place of the unfeasible set ${R}_{N}$ may work appropriately despite the absence of a formal proof.

\newpage

\section{Additional Empirical Results}\label{AppAddEmpirical}
\setcounter{table}{0}
\renewcommand\thetable{F.\arabic{table}}

\setcounter{figure}{0}
\renewcommand\thefigure{F.\arabic{figure}}

\setcounter{equation}{0}
\renewcommand\theequation{F.\arabic{equation}}

\setcounter{theorem}{0}
\renewcommand\thetheorem{F.\arabic{theorem}}

\setcounter{lemma}{0}
\renewcommand\thelemma{F.\arabic{lemma}}

\setcounter{proposition}{0}
\renewcommand\theproposition{F.\arabic{proposition}}

\setcounter{corollary}{0}
\renewcommand\thecorollary{F.\arabic{corollary}}

\setcounter{assumption}{0}
\renewcommand\theassumption{F.\arabic{assumption}}

{

In the main text, we presented the aggregated results for the probability of causation (Corollary \ref{CorTarget}). To estimate these parameters, we first bound the conditional probability of causation for each stratum (course-city pair). In this appendix, we discuss these conditional parameters, focusing on their heterogeneity and the impact of each additional assumption on their distribution across strata. Since the estimates based on the Probit link function are very similar to the estimates based on the Logit link function (Section \ref{Sestimation}), we focus on the first group of estimates.

Figure \ref{FigLB} shows the distribution of the estimated lower bounds for each stratum and each set of assumptions. First, notice that the lower bound is zero for many strata when we impose Assumptions \ref{ASexogeneity}-\ref{ASmonotonicity} only (Subfigure \ref{FigLB1}). In contrast, the number of strata whose lower bound is zero is much smaller when we impose Assumptions \ref{ASexogeneity}-\ref{ASstochastic} (Subfigure \ref{FigLB3}). Moreover, adding Assumption \ref{ASstochastic} shifts the distribution of estimated lower bounds to the right. These two results illustrate the identifying power of Assumption \ref{ASstochastic} as discussed in Corollary \ref{CorIDprop4}.

\begin{figure}[!htbp]\caption{Estimated Lower Bounds for the Probability of Causation for each Stratum}\label{FigLB}
	\begin{subfigure}[t]{0.47\textwidth}
			\centering
			\includegraphics[width = \textwidth]{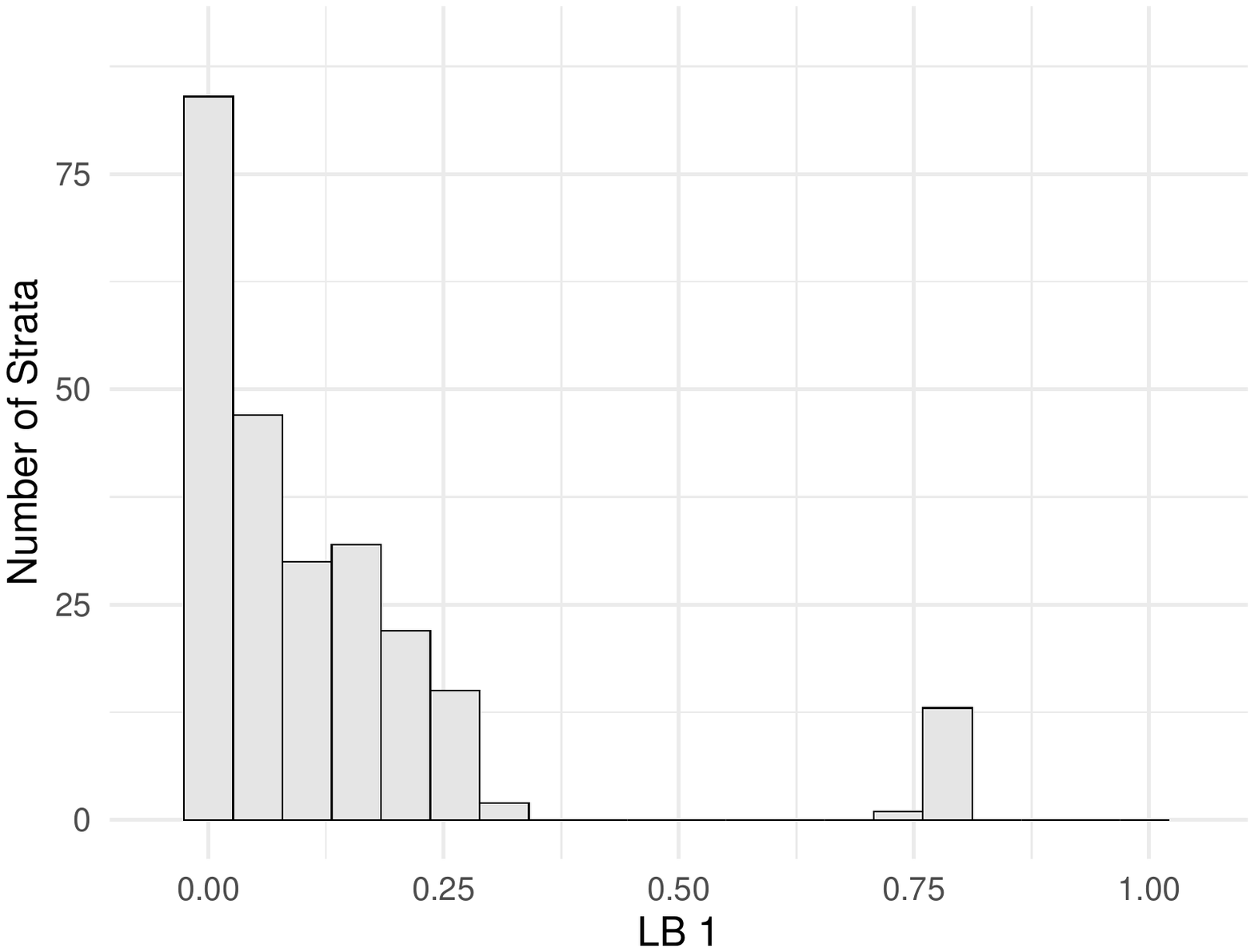}
			\caption{$LB_{1}\left(x\right)$: Assumptions \ref{ASexogeneity}-\ref{ASmonotonicity}}
			\label{FigLB1}
		\end{subfigure}
        \hfill
        \begin{subfigure}[t]{0.47\textwidth}
			\centering
			\includegraphics[width = \textwidth]{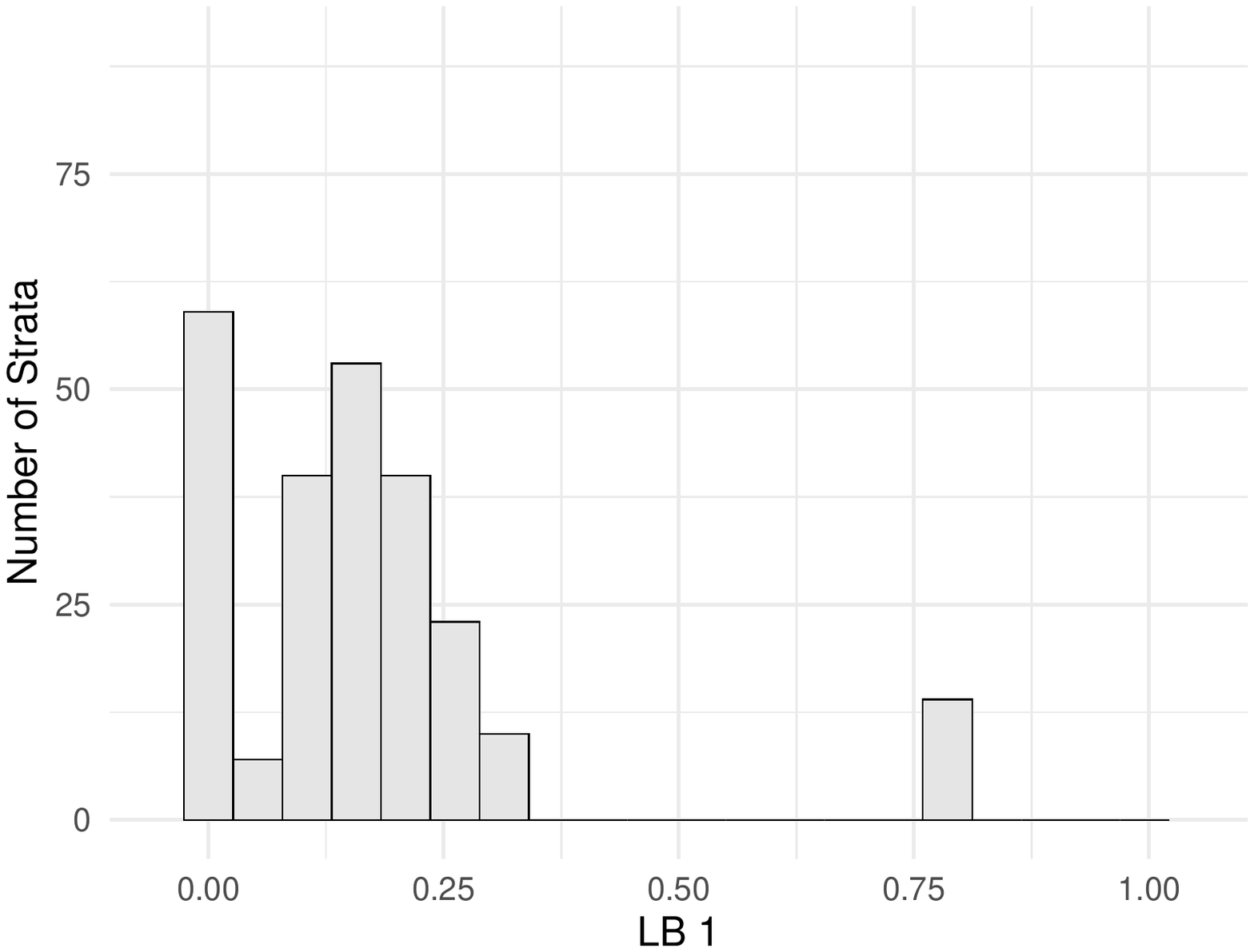}
			\caption{$LB_{3}\left(x\right)$: Assumptions \ref{ASexogeneity}-\ref{ASstochastic}}
			\label{FigLB3}
		\end{subfigure}
	\textit{Notes}: This figure presents frequency histograms of the estimated lower bounds for the probability of causation for each stratum (course-city pair). All bounds were estimated using the Probit link function (Section \ref{Sestimation}). Subfigure \ref{FigLB1} shows the distribution of the lower bounds in Proposition \ref{PROPmonotonicity} while Subfigure \ref{FigLB3} shows the distribution of the lower bounds in Proposition \ref{PROPstochastic}.
\end{figure}

Figure \ref{FigUB} shows the distribution of the estimated upper bounds for each stratum and each set of assumptions. First, notice that the upper bound is one for many strata when we impose Assumptions \ref{ASexogeneity}-\ref{ASmonotonicity} only (Subfigure \ref{FigUB1}). In contrast, the number of strata whose upper bound is one is much smaller when we impose Assumptions \ref{ASexogeneity}-\ref{ASmonotonicityY} (Subfigure \ref{FigUB2}). Moreover, adding Assumption \ref{ASmonotonicityY} shifts the distribution of estimated upper bounds to the left. These two results illustrate the identifying power of Assumption \ref{ASmonotonicityY} as discussed in Corollary \ref{CorIDprop3}.

\begin{figure}[!htbp]\caption{Estimated Upper Bounds for the Probability of Causation for each Stratum}\label{FigUB}
	\begin{subfigure}[t]{0.47\textwidth}
			\centering
			\includegraphics[width = \textwidth]{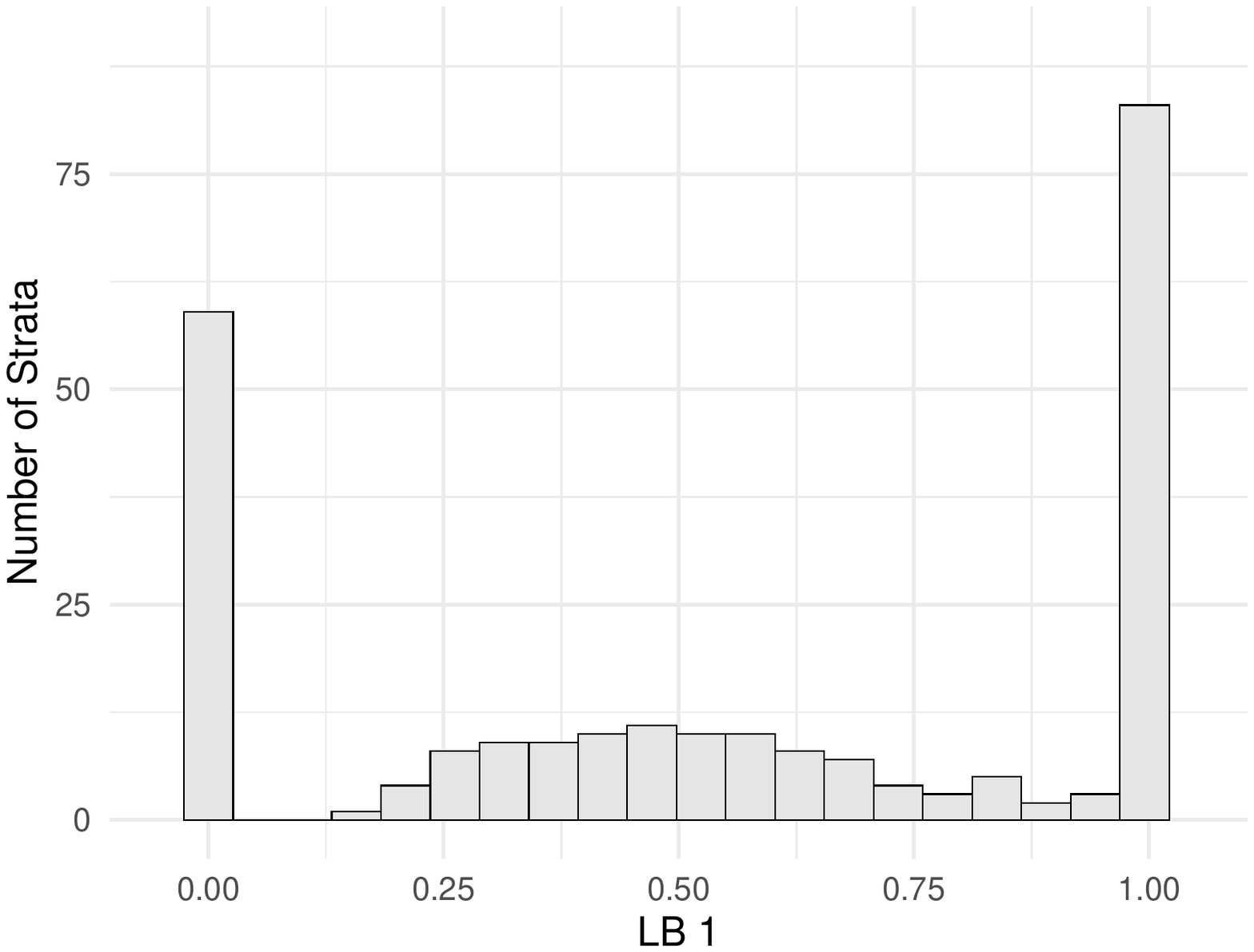}
			\caption{$UB_{1}\left(x\right)$: Assumptions \ref{ASexogeneity}-\ref{ASmonotonicity}}
			\label{FigUB1}
		\end{subfigure}
        \hfill
        \begin{subfigure}[t]{0.47\textwidth}
			\centering
			\includegraphics[width = \textwidth]{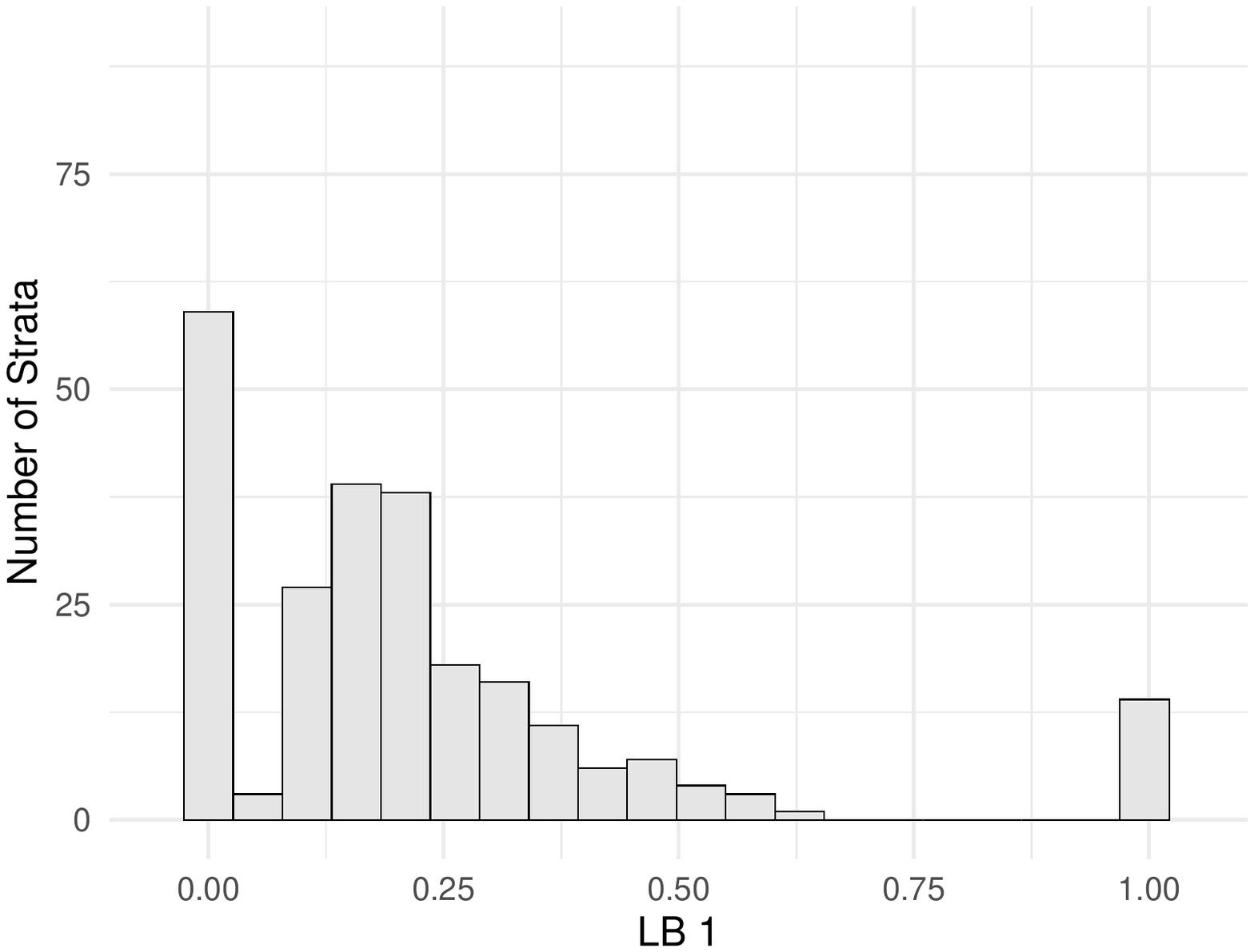}
			\caption{$UB_{2}\left(x\right)$: Assumptions \ref{ASexogeneity}-\ref{ASmonotonicityY}}
			\label{FigUB2}
		\end{subfigure}
	\textit{Notes}: This figure presents frequency histograms of the estimated upper bounds for the probability of causation for each stratum (course-city pair). All bounds were estimated using the Probit link function (Section \ref{Sestimation}). Subfigure \ref{FigUB1} shows the distribution of the upper bounds in Proposition \ref{PROPmonotonicity} while Subfigure \ref{FigUB2} shows the distribution of the upper bounds in Proposition \ref{PROPmonotonicityY}.
\end{figure}

Figure \ref{FigInterval} shows the distribution of the length of the estimated intervals for each stratum and each set of assumptions. Observe that these distributions shift to the left when we impose additional assumptions, i.e., the estimated intervals become shorter. This result illustrates the identifying power of our additional assumptions.

\begin{figure}[!htbp]\caption{Estimated Intervals' Length for each Stratum}\label{FigInterval}
	\begin{subfigure}[t]{0.31\textwidth}
			\centering
			\includegraphics[width = \textwidth]{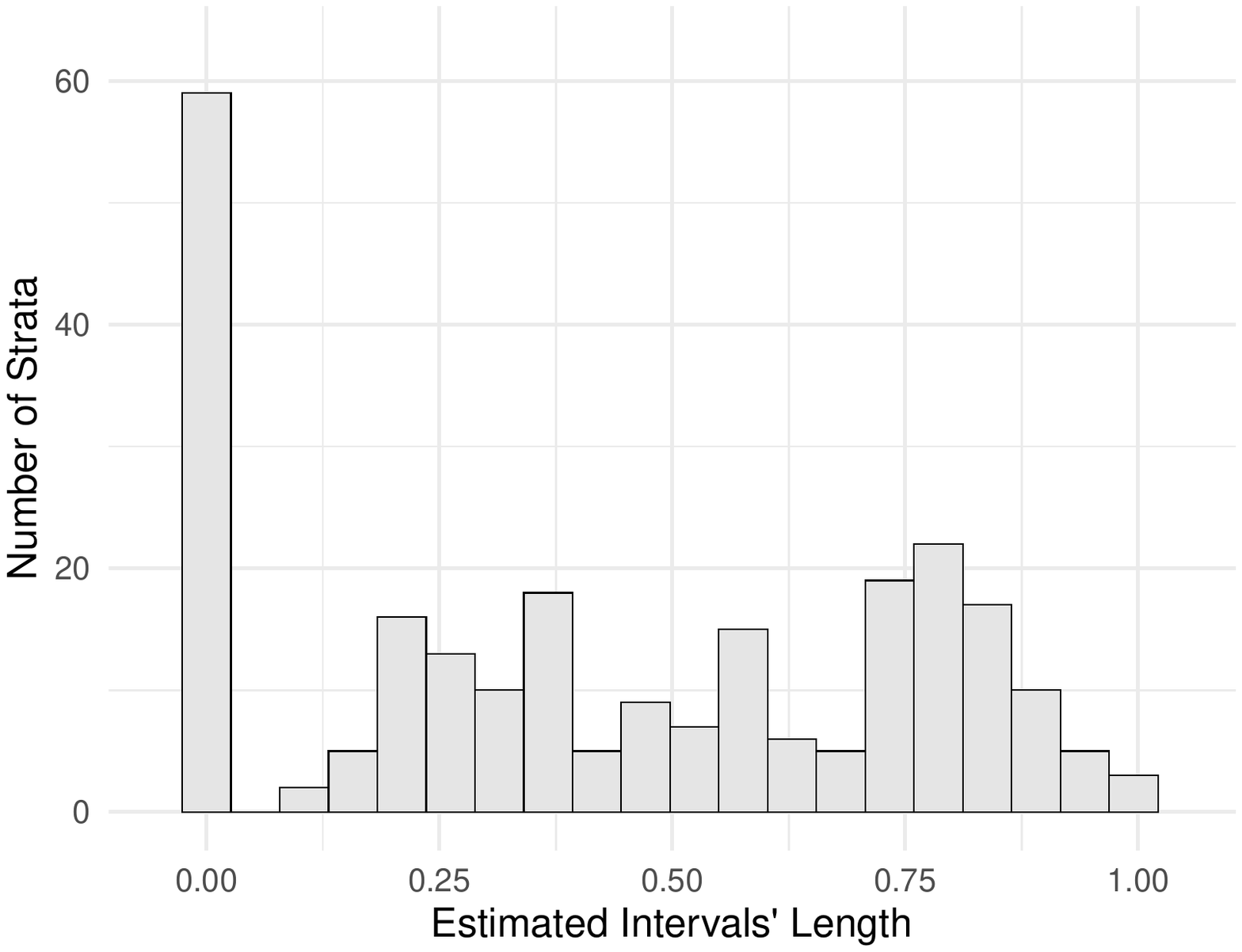}
			\caption{$UB_{1}\left(x\right) - LB_{1}\left(x\right)$: \\ Assumptions \ref{ASexogeneity}-\ref{ASmonotonicity}}
			\label{FigProp2}
		\end{subfigure}
        \hfill
        \begin{subfigure}[t]{0.31\textwidth}
			\centering
			\includegraphics[width = \textwidth]{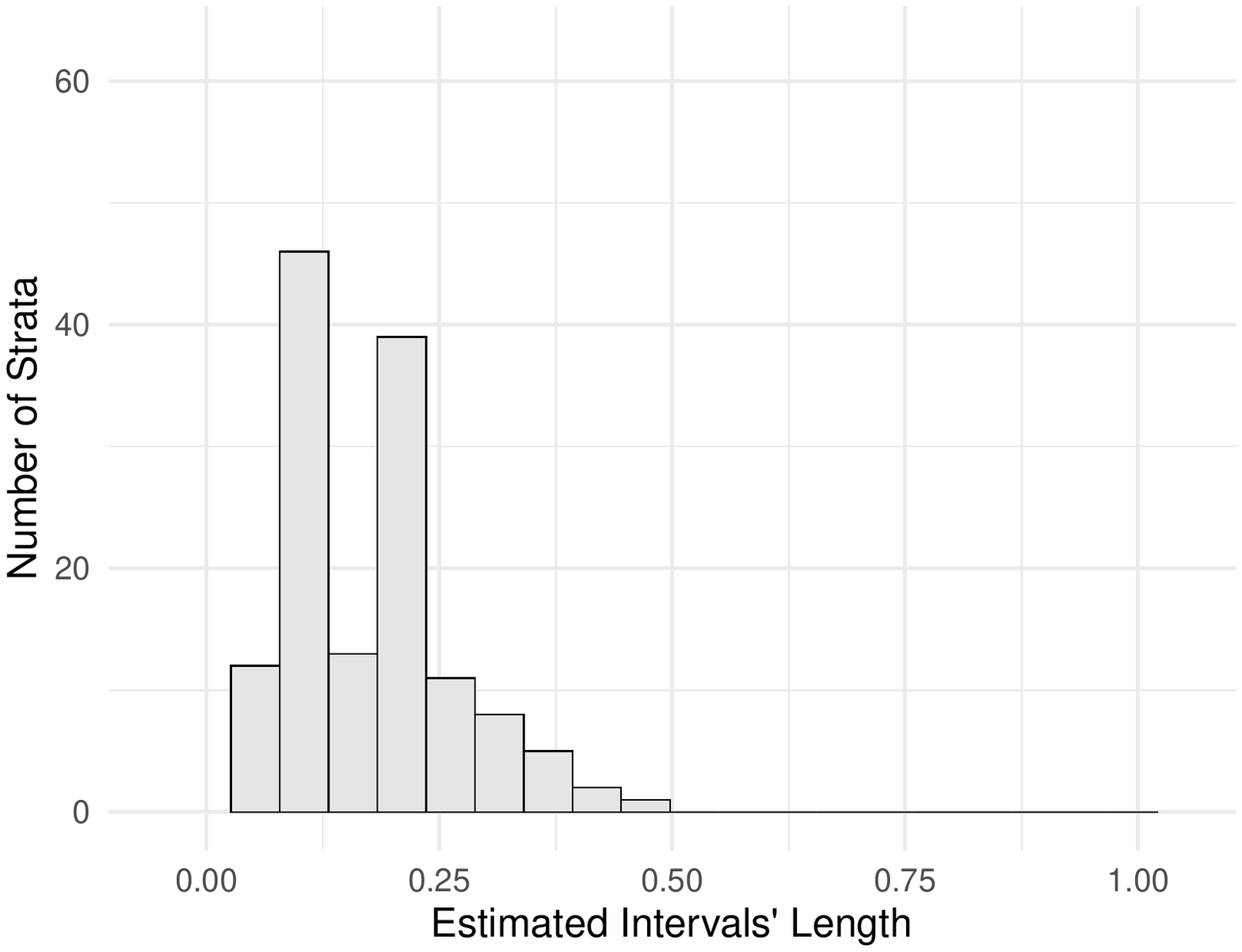}
			\caption{$UB_{2}\left(x\right) - LB_{1}\left(x\right)$: \\ Assumptions \ref{ASexogeneity}-\ref{ASmonotonicityY}}
			\label{FigProp3}
		\end{subfigure}
        \hfill
        \begin{subfigure}[t]{0.31\textwidth}
			\centering
			\includegraphics[width = \textwidth]{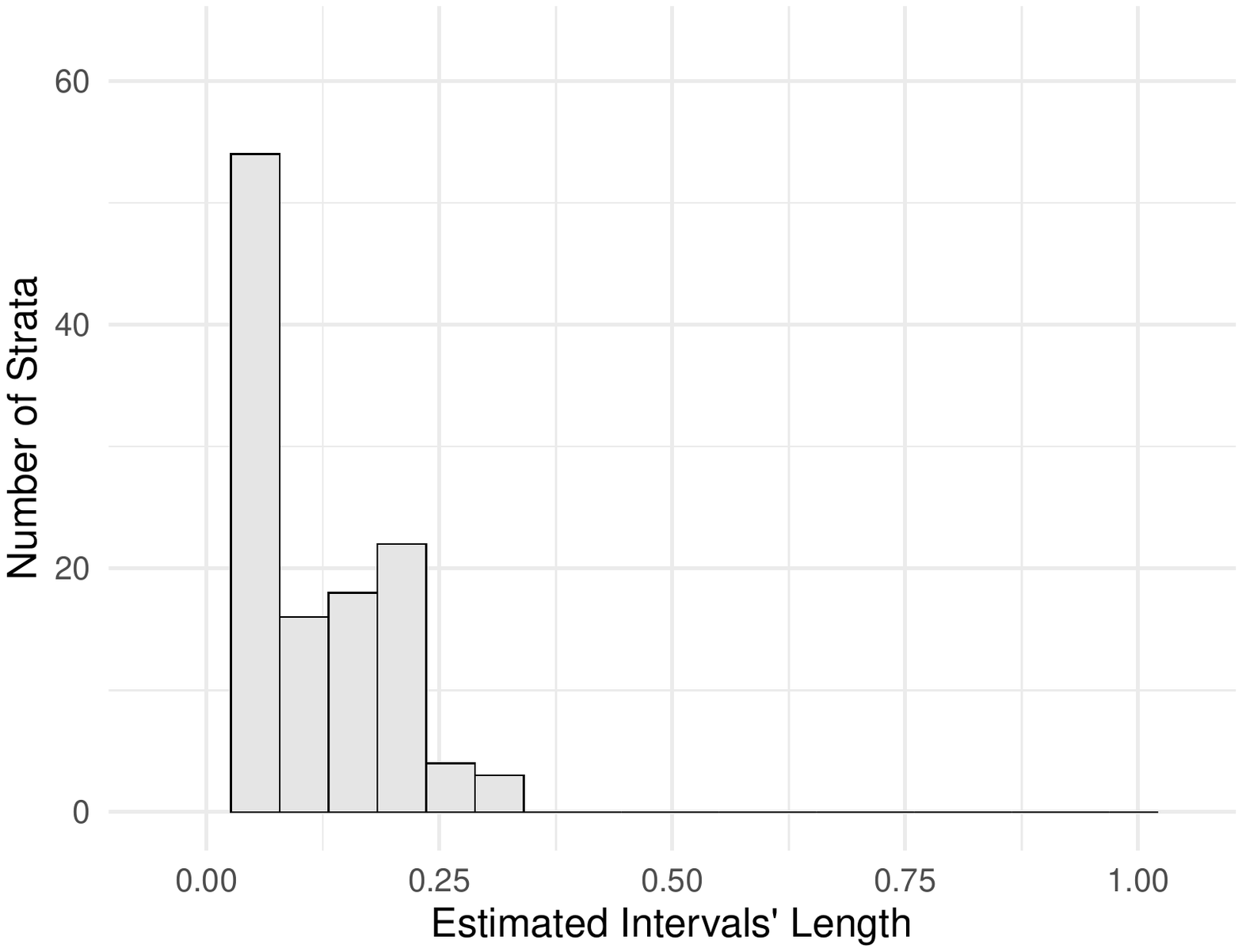}
			\caption{$UB_{2}\left(x\right) - LB_{3}\left(x\right)$: \\ Assumptions \ref{ASexogeneity}-\ref{ASstochastic}}
			\label{FigProp4}
		\end{subfigure}
	\textit{Notes}: This figure presents frequency histograms of the estimated intervals' length for each stratum (course-city pair). All bounds were estimated using the Probit link function (Section \ref{Sestimation}). Subfigure \ref{FigProp2} shows the distribution of the length of the intervals in Proposition \ref{PROPmonotonicity}, Subfigure \ref{FigProp3} shows the distribution of the length of the intervals in Proposition \ref{PROPmonotonicityY}, and Subfigure \ref{FigProp4} shows the distribution of the length of the intervals in Proposition \ref{PROPstochastic}.
\end{figure}

}

\end{document}